\documentclass[aps,prd, twocolumn]{revtex4}
\usepackage{amsmath,amssymb,amsfonts}
\usepackage{graphicx}
\usepackage{enumerate} % advanced enumerate environment
\usepackage{colordvi} % for color text
\usepackage{color} % for color
\usepackage{bm}% bold math
\newcommand{\bfv}{\mbox{\boldmath$v$}}

\newcommand{\bfk}{\mbox{\boldmath$k$}}
\newcommand{\bfp}{\mbox{\boldmath$p$}}
\newcommand{\bfq}{\mbox{\boldmath$q$}}

\newcommand{\sigmav}{\sigma_{\rm v}}
\newcommand{\sigmad}{\sigma_{\rm d}}
\newcommand{\rhom}{\rho_{\rm m}}

\begin{document}
\title{Regularized cosmological power spectrum and correlation function 
in modified gravity models}

\author{Atsushi Taruya}
\affiliation{Yukawa Institute for Theoretical Physics, Kyoto University, Kyoto 606-8502, Japan}
\affiliation{
Kavli Institute for the Physics and Mathematics of the Universe, Todai Institutes for Advanced Study, the University of Tokyo, Kashiwa, Chiba 277-8583, Japan (Kavli IPMU, WPI)}
\author{Takahiro Nishimichi}
\affiliation{Institut d'Astrophysique de Paris, CNRS UMR 7095 and UPMC, 98bis, bd Arago, F-75014 Paris, France}
\author{Francis Bernardeau}
\affiliation{Institut d'Astrophysique de Paris, CNRS UMR 7095 and UPMC, 98bis, bd Arago, F-75014 Paris, France}
\author{Takashi Hiramatsu}
\affiliation{Yukawa Institute for Theoretical Physics, Kyoto University, Kyoto 606-8502, Japan}
\author{Kazuya Koyama}
\affiliation{Institute of Cosmology \& Gravitation, University of Portsmouth, Dennis Sciama Building, Portsmouth, PO1 3FX, United Kingdom}
%\affiliation{Institut de Physique Th\'eorique, CEA, F-91191
%  Gif-sur-Yvette, France\\
%  \ \ CNRS, URA 2306, F-91191, Gif-sur-Yvette, France}
%\affiliation{Research Cener for the Early Universe, Graduate school of Science, The University of Tokyo, Bunkyo-ku, Tokyo 113-0033, Japan}
%
%
%
%%%%%%%%%%%%%%%%%%%%%%%%%%%%%%%%%%%%%%%%%%%%%%%%%%%%%%%%%%%%%%%%%%%%%%%
\date{\today}
\begin{abstract}
Based on the multi-point propagator expansion, 
we present resummed 
perturbative calculations for 
cosmological power spectra and correlation functions 
in the context of modified gravity. 
In a wide class of modified gravity models that have a screening mechanism 
to recover general relativity (GR) 
on small scales, we apply the eikonal approximation to 
derive the governing equation for resummed propagator that 
partly includes the non-perturbative effect in the high-$k$ limit. 
The resultant propagator in the high-$k$ limit 
contains the new corrections arising from the 
screening mechanism as well as the standard exponential damping. 
We explicitly derive 
the expression for new high-$k$ contributions 
in specific modified gravity models, 
and find that in the case of $f(R)$ gravity for a currently 
constrained model parameter, the corrections 
are basically of the sub-leading order and can be neglected. 
Thus, in $f(R)$ gravity, similarly to the GR case, 
we can analytically construct the regularized propagator that reproduces both 
the resummed high-$k$ behavior and the low-$k$ results computed 
with standard perturbation theory, consistently 
taking account of the nonlinear modification of gravity valid at large scales. 
With the regularized multi-point 
propagators, we give predictions for power spectrum and correlation 
function at one-loop order, and compare those with $N$-body simulations 
in $f(R)$ gravity model. 
As an important application, we also discuss the redshift-space distortions 
and compute the anisotropic power spectra and correlation functions. 
\end{abstract}
%%%%%%%%%%%%%%%%%%%%%%%%%%%%%%%%%%%%%%%%%%%%%%%%%%%%%%%%%%%%%%%%%%%%%%%

%\pacs{98.80.-k, 98.62.Py, 98.65.-r}
%\keywords{cosmology, large-scale structure}
\preprint{YITP-14-63}
\maketitle

%%%%%%%%%%%%%%%%%%%%%%%%%%%%%%%%%%%%%%%%%%%%%%%%%%%%%%%%%%%%%%%%%%%%%%%
%%%%%%%%%%%%%%%%%%%%%%%%%%%%%%%%%%%%%%%%%%%%%%%%%%%%%%%%%%%%%%%%%%%%%%%
\section{Introduction}
\label{sec:intro}
%%%%%%%%%%%%%%%%%%%%%%%%%%%%%%%%%%%%%%%%%%%%%%%%%%%%%%%%%%%%%%%%%%%%%%%
%%%%%%%%%%%%%%%%%%%%%%%%%%%%%%%%%%%%%%%%%%%%%%%%%%%%%%%%%%%%%%%%%%%%%%%
%
%
%
The precision observation of large-scale structure of the Universe now 
plays a very crucial role in scrutinizing the standard cosmological model that 
has emerged recently based on the multiple cosmological observations. 
Amongst various cosmological issues, one important subject 
is to clarify the origin and nature of cosmic acceleration, 
first discovered by the distant supernova observations 
\cite{Perlmutter:1998np,Riess:1998cb}.  
The cosmic acceleration may be originated from the 
dark energy, or rather 
it may indicate the breakdown of general relativity on very large 
scales. To observationally explore this, the measurements of both the cosmic 
expansion and the growth of structure are thought to be essential, 
giving us a chance to test gravity on cosmological scales 
or to constrain dark energy equation of state 
(e.g., \cite{2013PhR...530...87W} for review). 
The large-scale structure 
observations with galaxy redshift surveys indeed offer 
an opportunity to measure these two quantities simultaneously.

The key measurements are
the baryon acoustic oscillations (BAO) and redshift-space distortions (RSD), 
imprinted on the large-scale clustering pattern of galaxy distribution. 
With BAO as a standard ruler, we can simultaneously measure the angular diameter distance and Hubble parameter at the distant galaxies through the Alcock-Paczynski effect (e.g., \cite{Alcock_Paczynski:1979,Seo:2003pu,
Blake:2003rh,Glazebrook:2005mb,Shoji:2008xn,Padmanabhan:2008ag}).
On the other hand, RSD caused by 
the peculiar velocity of galaxies induces apparent 
clustering anisotropies, whose 
strength is related to the growth rate of structure formation 
(e.g., \cite{Kaiser:1987qv,Hamilton:1997zq,Peebles:1980,Linder:2007nu}). 
Since both BAO and RSD are now reliably and simultaneously measured 
through the clustering statistics of galaxy distribution (e.g., 
\cite{Eisenstein:2005su,Percival:2007yw,2012MNRAS.426.2719R,Anderson:2013zyy,Beutler:2013yhm} for recent measurements) typically 
on scales close to the linear regime of gravitational evolution, 
the precision estimation of 
power spectrum and/or correlation function is a major priority 
of the ongoing and upcoming galaxy surveys.

With increasing interests in precision measurements, accurate 
theoretical modelings of power spectrum and/or correlation function is 
crucial and is essential to correctly estimate the geometric distances and 
structure growth, taking full account of the nonlinear systematics 
including gravitational clustering and RSD. Development of theoretical 
templates is thus an important research subject, and there have been 
numerous numerical and analytical studies along this line 
\cite{Jeong:2006xd,Crocce:2007dt,Matsubara:2007wj, 2011PhRvD..83h3518M,McDonald:2006hf,Taruya:2007xy, Taruya:2009ir, Bernardeau:2008fa, Pietroni:2008jx,Valageas:2006bi,Taruya:2012ut,Wang:2013hwa,Taruya:2010mx,Reid:2011ar,Carlson:2012bu,Seljak:2011tx,Vlah:2012ni,Okumura:2011pb,Heitmann:2009cu,Lawrence:2009uk}. 
Thanks to these efforts, 
we are now able to discuss the accuracy of theoretical template 
at a percent level. However, one important remark is that the calculation of 
such templates, especially for the prediction of gravitational 
clustering, heavily relies on the underlying theory of gravity. 
So far, general relativity (GR) has been 
implicitly assumed as the underlying theory of gravity in most studies. 
As a consequence, 
although such templates can be employed for consistency tests of GR, their use 
for characterizing or detecting deviation from GR gravity
can be limited.

Further theoretical developments are therefore required in 
a wide context of modified gravity models. While 
a model-independent approach, in which we do not assume gravity 
but rather parametrize it in a fairly generic way 
(e.g., \cite{Hu:2007pj,2013PhRvD..87b4015B,2012PhRvD..86d4015B,Daniel:2008et,2012PhR...513....1C})
is very helpful and should be exploited, most of the 
approaches proposed so far have been restricted to the linear regime. Since the applicable 
range of linear theory calculation is known to be rather narrower at lower redshifts, 
our ability to constrain or test such models is expected to be significantly reduced 
\cite{Taruya:2013quf}.

In this paper, we attempt to extend the  
framework of theoretical templates to deal with modified theories of gravity.  
Here, we specifically examine this issue based on the 
analytical approach with perturbation theory calculations, relevant for the 
measurement of BAO and RSD on large scales.  
Previously, we have presented the basic formalism to treat general 
modified gravity models \cite{Koyama:2009me}, 
and in specific gravity models, 
we have computed power spectra in both real and 
redshift spaces based on the standard perturbation theory (PT) 
\cite{Bernardeau:2001qr}. The 
standard PT is, however, known to produce a poorly convergent series 
expansion, and because of the bad high-$k$ behavior 
(e.g., \cite{Crocce:2005xy,Carlson:2009it,Taruya:2009ir}), 
difficulty arises in computing the correlation function 
through a direct integration of power spectrum.

In the present paper, we shall apply the specific resummed PT scheme 
referred to as the multipoint propagator expansion or $\Gamma$ expansion 
\cite{Bernardeau:2008fa}. 
The building blocks of this PT scheme are the multipoint propagators, 
with which the non-perturbative properties at high-$k$ can be efficiently 
resummed, giving us an improved convergence of the PT expansion. 
In the case of GR, making full use of the analytical properties, 
the regularized propagators, which consistently reproduces 
both the standard PT results at low-$k$ and the expected resummed behaviors  
at high-$k$, have been successfully constructed 
\cite{Bernardeau:2011dp,Bernardeau:2012ux}, and the $\Gamma$ expansion 
has been applied to the predictions of real- and redshift-space power spectra 
and correlation functions, showing a very good agreement with $N$-body 
simulations \cite{Crocce:2012fa,Taruya:2012ut,Taruya:2013my}. 
Clearly, a crucial point for applying this approach to the modified gravity models  
is whether we can systematically construct regularized propagators in a 
semianalytic manner. Here, we specifically show 
that while there appear non-trivial corrections originating 
from the screening mechanism in modified gravity model, 
in the case of $f(R)$ gravity model for a 
currently constrained model parameter, these corrections 
are basically small. Thus, in $f(R)$ gravity, 
the propagator can be constructed in a similar 
manner to the GR case. Then the analytically computed propagators are compared 
with $N$-body simulations, and a good agreement is found. 
With these propagators as building blocks, we will proceed to the calculation 
of the power spectrum and correlation function in both real and redshift 
spaces.

The paper is organized as follows. In Sec.~\ref{sec:BasicEqs_MG}, 
we briefly review the basic formalism to treat perturbations in 
general modified gravity models, and introduce a 
resummed PT scheme based on the multipoint propagator expansion. 
Sec.~\ref{sec:resummed_prop_eikonal} discusses 
the non-perturbative high-$k$ behavior of the propagators 
based on the eikonal approximation, and Sec.~\ref{sec:regularized_pk}  
presents an explicit expression for matter power spectrum 
in terms of the {\it regularized} 
propagators, which satisfy both the expected high-$k$ and low-$k$ 
behaviors. Then, the comparison of PT results with 
$N$-body simulations is made in Sec.~\ref{sec:RegPT}, and  
the applications to the redshift-space observables are discussed 
in Sec.~\ref{sec:RSD}. Finally, the newly 
developed PT calculation is compared with 
standard PT prediction in Sec.~\ref{sec:discussion}, and we summarize our 
findings in Sec.~\ref{sec:conclusion}. 

%
%
%
%
%
%
%%%%%%%%%%%%%%%%%%%%%%%%%%%%%%%%%%%%%%%%%%%%%%%%%%%%%%%%%%%%%%%%%%%%%%%
%%%%%%%%%%%%%%%%%%%%%%%%%%%%%%%%%%%%%%%%%%%%%%%%%%%%%%%%%%%%%%%%%%%%%%%
\section{Basic equations for perturbations}
\label{sec:BasicEqs_MG}
%%%%%%%%%%%%%%%%%%%%%%%%%%%%%%%%%%%%%%%%%%%%%%%%%%%%%%%%%%%%%%%%%%%%%%%
%%%%%%%%%%%%%%%%%%%%%%%%%%%%%%%%%%%%%%%%%%%%%%%%%%%%%%%%%%%%%%%%%%%%%%%

In this section, we begin
by reviewing the framework to treat the evolution of matter
fluctuations in modified gravity models \cite{Koyama:2009me}, and 
present a set of basic equations relevant for the perturbation theory (PT) 
treatment. Then, a resummed PT scheme with multipoint propagator 
expansion \cite{Bernardeau:2008fa} 
is briefly reviewed, and the properties of these multipoint propagators 
are mentioned.

%%%%%%%%%%%%%%%%%%%%%%%%%%%%%%%%%%%%%%%%%%%%%%%%%%%%%%%%%%%%%%%%%%%%%%%
\subsection{Dynamics of matter fluctuations in modified theories of gravity}
\label{subsec:dynamics}
%%%%%%%%%%%%%%%%%%%%%%%%%%%%%%%%%%%%%%%%%%%%%%%%%%%%%%%%%%%%%%%%%%%%%%%

In this paper, we are particularly interested in the evolution 
of matter fluctuations, ignoring a tiny contribution of massive neutrinos. 
Inside the Hubble horizon, the so-called quasi-static approximation 
may be applied, and the time derivatives of 
the perturbed quantities can be neglected compared to the spatial 
derivatives. In GR, based on this approximation, 
we can find the Newtonian correspondence, and 
the standard Poisson equation is recovered. 
On the other hand, in modified theory of gravity, the Poisson equation is 
generically modified due to a new scalar degree of freedom, 
referred to as the scalaron. On large scales, the 
scalaron $\varphi$ mediates the scalar force, and behaves like the Brans-Dicke
scalar field without potential and self-interactions, while it should
acquire some interaction terms on small scales, which play an
important role to recover GR and to evade the solar-system constraints. 
Indeed, there are several known mechanism 
such as chameleon and Vainshtein mechanisms 
(e.g., \cite{Khoury:2003rn,Deffayet:2001uk}), in which the nonlinear
interaction terms naturally arise and eventually become dominant, 
leading to a recovery of GR. As a result, 
the Poisson equation is coupled to the field equation for
scalaron $\varphi$ with self-interaction term. Under 
the quasi-static approximation, we have \cite{Koyama:2009me}
%%%%%%%%%%%%%%%%%%%%%%%%%%%%%%%%%%%%%%%%%%%%%%%%%%%
\begin{align}
&\frac{1}{a}\nabla^2\psi=\frac{\kappa^2}{2}\,\rhom\,\delta
-\frac{1}{2a^2}\nabla^2\varphi,
\label{eq:Poisson_eq}\\
&(3+2\omega_{\rm BD})\frac{1}{a^2}\nabla^2\varphi=-2\kappa^2\rhom\,\delta
-\mathcal{I}(\varphi)
\label{eq:EoM_scalaron}
\end{align}
%%%%%%%%%%%%%%%%%%%%%%%%%%%%%%%%%%%%%%%%%%%%%%%%%%%
with $\kappa^2=8\pi\,G$ and $\omega_{\rm BD}$ being the Brans-Dicke 
parameter. The quantities $\psi$ is the Newton potential, and  
the function $\mathcal{I}$ represents the nonlinear 
self-interaction, which may be generally expanded as
%%%%%%%%%%%%%%%%%%%%%%%%%%%%%%%%%%%%%%%%%%%%%%%%%%%
\begin{align}
&\mathcal{I}(\varphi)=M_1(k)
\nonumber\\
&\quad+\frac{1}{2}\,
\int\frac{d^3\bfk_1 d^3\bfk_2}{(2\pi)^3}\delta_{\rm D}(\bfk-\bfk_{12})\,
M_2(\bfk_1,\bfk_2)\varphi(\bfk_1)\varphi(\bfk_2)
\nonumber\\
&\quad+\frac{1}{3\,!}
\int\frac{d^3\bfk_1 d^3\bfk_2d^3\bfk_3}{(2\pi)^6}\delta_{\rm D}(\bfk-\bfk_{123})\,
\nonumber\\
&\qquad\qquad\quad\times
M_3(\bfk_1,\bfk_2,\bfk_3)\varphi(\bfk_1)\varphi(\bfk_2)\varphi(\bfk_3)
+\cdots.
\label{eq:I_expansion}
\end{align}
%%%%%%%%%%%%%%%%%%%%%%%%%%%%%%%%%%%%%%%%%%%%%%%%%%%

On the other hand, for the matter sector, 
the evolution of matter fluctuations 
is governed by the conservation of energy momentum tensor, 
which would remain unchanged even if the gravity sector is modified. 
Under the single-stream approximation, which is relevant for the scale 
of our interest, the matter fluctuations are
treated as a pressureless fluid flow, whose evolution equations are 
given by \cite{Bernardeau:2001qr}
%%%%%%%%%%%%%%%%%%%%%%%%%%%%%%%%%%%%%%%%%%%%%%%%%%%
\begin{align}
&\frac{\partial \delta}{\partial t} +
\frac{1}{a}\nabla\cdot[(1+\delta)\bfv]=0,
\label{eq:eq_continuity}\\
&\frac{\partial \bfv}{\partial t} + H\,\bfv+
\frac{1}{a}(\bfv\cdot\nabla)\cdot\bfv=-\frac{1}{a}\nabla\psi. 
\label{eq:eq_Euler}
\end{align}
%%%%%%%%%%%%%%%%%%%%%%%%%%%%%%%%%%%%%%%%%%%%%%%%%%%

Eqs.~(\ref{eq:EoM_scalaron})--(\ref{eq:eq_continuity}) are the basic 
equations for perturbations in a general framework 
of modified gravity models. In Fourier space, they  
can be reduced to a more compact form.
Assuming the irrotationality of fluid quantities,
the velocity field is expressed in terms of
the velocity divergence, $\theta=\nabla\cdot\bfv/(aH)$.
Then, we introduce the two-component multiplet 
(e.g.,\cite{Crocce:2005xy}):  
%%%%%%%%%%%%%%%%%%%%%%%%%%%%%%%%%%%%%%%%%%%%%%%%%%%%%%%%%%%%%%%%%%%%%%
\begin{align}
\Psi_a(\bfk;t)=\Bigl(\delta(\bfk;t), \,\,
-\theta(\bfk;t) \Bigr),  
\end{align}
%%%%%%%%%%%%%%%%%%%%%%%%%%%%%%%%%%%%%%%%%%%%%%%%%%%%%%%%%%%%%%%%%%%%%%
where the subscript $a=1,\,2$ selects the density and the velocity 
components of CDM plus baryons. The governing equations for $\Psi_a$ 
become \cite{Koyama:2009me}
%%%%%%%%%%%%%%%%%%%%%%%%%%%%%%%%%%%%%%%%%%%%%%%%%%%%%%%%%%%%%%%%%%%%%%
\begin{widetext}
\begin{eqnarray}
&&\frac{\partial \Psi_a(\bfk;\tau)}{\partial\tau} +
\Omega_{ab}(k;\tau)\,\Psi_b(\bfk;\tau) =
\int\frac{d^3\bfk_1d^3\bfk_2}{(2\pi)^3}\,\delta_{\rm D}(\bfk-\bfk_{12})\,
\gamma_{abc}(\bfk_1,\bfk_2)\,\Psi_b(\bfk_1;\tau)\Psi_c(\bfk_2;\tau)
\nonumber\\
&&\quad\quad\quad\quad\quad\quad
+\delta_{a2}\sum_{n=2}\int \frac{d^3\bfk_1\cdots d^3\bfk_n}{(2\pi)^{3(n-1)}}\,
\delta_{\rm D}(\bfk-\bfk_{1\cdots n})\,
\sigma^{(n)}(\bfk_1,\cdots,\bfk_n;\tau)\,
\Psi_1(\bfk_1;\tau)\cdots\Psi_1(\bfk_n;\tau),
\label{eq:basic_eq}
\end{eqnarray}
\end{widetext}
%%%%%%%%%%%%%%%%%%%%%%%%%%%%%%%%%%%%%%%%%%%%%%%%%%%%%%%%%%%%%%%%%%%%%%
where the time variable $\tau$ is defined by $\tau=\ln a(t)$, and 
$\delta_{ab}$ is the Kronecker delta. Here, we introduced the shortcut 
notations, $\bfk_{12}=\bfk_1+\bfk_2$ and $\bfk_{1\cdots n}=\bfk_1+\cdots+\bfk_n$. 
The matrix $\Omega_{ab}$ is given by
%%%%%%%%%%%%%%%%%%%%%%%%%%%%%%%%%%%%%%%%%%%%%%%%%%%%%%%%%%%%%%%%%%%%%%
\begin{equation}
\Omega_{ab}(k;\tau)=\left(
\begin{array}{cc}
{\displaystyle 0} & \,\,{\displaystyle -1 }\\
\\
{\displaystyle -\frac{\kappa^2}{2}\frac{\rhom}{H^2}
\left[1+
\frac{1}{3}\frac{(k/a)^2}{\Pi(k)}\right] }&
\,\,{\displaystyle 2+\frac{\dot{H}}{H^2}}
\end{array}
\right)
\label{eq:Omega_ab}
\end{equation}
%%%%%%%%%%%%%%%%%%%%%%%%%%%%%%%%%%%%%%%%%%%%%%%%%%%%%%%%%%%%%%%%%%%%%%
with the function $\Pi$ defined by
%%%%%%%%%%%%%%%%%%%%%%%%%%%%%%%%%%%%%%%%%%%%%%%%%%%%%%%%%%%%%%%%%%%%%%
\begin{align}
\Pi(k)=\frac{1}{3}\left\{(3+2\omega_{\rm BD})\frac{k^2}{a^2}+M_1\right\}.
\label{eq:def_Pi}
\end{align}
%%%%%%%%%%%%%%%%%%%%%%%%%%%%%%%%%%%%%%%%%%%%%%%%%%%%%%%%%%%%%%%%%%%%%%
From the $(2,1)$ component of $\Omega_{ab}$, we can define the effective
Newton constant as
%%%%%%%%%%%%%%%%%%%%%%%%%%%%%%%%%%%%%%%%%%%%%%%%%%%%%%%%%%%%%%%%%%%%%%
\begin{equation}
G_{\rm eff} =  G \left[1 + \frac{1}{3} \frac{(k/a)^2}{\Pi(k)}
\right].
\label{eq:effective_G}
\end{equation}
%%%%%%%%%%%%%%%%%%%%%%%%%%%%%%%%%%%%%%%%%%%%%%%%%%%%%%%%%%%%%%%%%%%%%%
Note that in the cases with $M_1$=0, the effective Newton constant is given by
\begin{equation}
G_{\rm eff} = \frac{2 (2+\omega_{\rm BD})}{3 +2 \omega_{\rm BD} }G.
\label{eq:effectiveG}
\end{equation}
%%%%%%%%%%%%%%%%%%%%%%%%%%%%%%%%%%%%%%%%%%%%%%%%%%%%%%%%%%%%%%%%%%%%%%
For a positive $\omega_{\rm BD} >0$, the effective gravitational
constant is larger than GR and the gravitational force is
enhanced. On the other hand, if $M_1 \gg k^2/a^2$, $G_{\rm eff}$ 
becomes $G$.

In Eq.~(\ref{eq:basic_eq}), there appear two types of vertex functions. 
One is the standard vertex function arising from the nonlinearity of the 
fluid flow, $\gamma_{abc}$:
%%%%%%%%%%%%%%%%%%%%%%%%%%%%%%%%%%%%%%%%%%%%%%%%%%%%%%%%%%%%%%%%%%%%%%
%\begin{widetext}
\begin{eqnarray}
\gamma_{abc}(\bfk_1,\bfk_2)=
\left\{
\begin{array}{lcl}
{\displaystyle 
\frac{1}{2}\left\{1+\frac{\bfk_2\cdot\bfk_1}{|\bfk_2|^2}\right\} }
&;& (a,b,c)=(1,1,2), 
\\
\\
{\displaystyle 
\frac{1}{2}\left\{1+\frac{\bfk_1\cdot\bfk_2}{|\bfk_1|^2}\right\} }
&;& (a,b,c)=(1,2,1), 
\\
\\
{\displaystyle 
\frac{(\bfk_1\cdot\bfk_2)|\bfk_1+\bfk_2|^2}{2|\bfk_1|^2|\bfk_2|^2}}
&;& (a,b,c)=(2,2,2), 
\\
\\
{\displaystyle 0 } &;& \mbox{otherwise}.
\end{array}
\right.
\label{eq:def_gamma_abc}
\end{eqnarray}
%\end{widetext}
%%%%%%%%%%%%%%%%%%%%%%%%%%%%%%%%%%%%%%%%%%%%%%%%%%%%%%%%%%%%%%%%%%%%%%
Note the symmetric properties
of the vertex function, $\gamma_{abc}(\bfk_1,\bfk_2)=\gamma_{acb}(\bfk_2,\bfk_1)$. Another vertex function is characterized by the kernel $\sigma^{(n)}$, 
which represents the mode coupling 
of the density fields $\Psi_1$ with velocity-divergence field. This coupling 
comes from the non-linear interaction terms of the scalaron $\varphi$ 
[i.e., Eqs.~(\ref{eq:Poisson_eq}) and (\ref{eq:EoM_scalaron}) through 
(\ref{eq:I_expansion})]. 
The explicit form of the higher-order vertex functions is given by 
(see Appendix \ref{sec:vertex} for derivation): 
%%%%%%%%%%%%%%%%%%%%%%%%%%%%%%%%%%%%%%%%%%%%%%%%%%%%%%%%%%%%%%%%%%%%%%
\begin{widetext}
\begin{align}
&\sigma^{(2)}(\bfk_1,\bfk_2;\tau)= -\frac{1}{12H^2}\,
\left(\frac{\kappa^2\,\rhom}{3}\right)^2 \left(
\frac{k_{12}^2}{a^2} \right)
\frac{M_2(\bfk_1, \bfk_2)}{\Pi(k_{12})\Pi(k_1)\Pi(k_2)}, 
\label{eq:vertex_sigma_2}
\\
&\sigma^{(3)}(\bfk_1,\bfk_2,\bfk_3;\tau)=
-\frac{1}{36 H^2}\,
\left(\frac{\kappa^2 \rhom}{3}\right)^3
\left(\frac{k_{123}^2}{a^2}\right)
\frac{1}
{\Pi(k_{123})\Pi(k_1)\Pi(k_2)\Pi(k_3)}
\left\{M_3(\bfk_1, \bfk_2, \bfk_3)-
\frac{M_2(\bfk_{12},\bfk_3)M_2(\bfk_1,\bfk_2)}{\Pi(k_{12})}
\right\},
\label{eq:vertex_sigma_3}\\
&\sigma^{(4)}(\bfk_1,\bfk_2,\bfk_3,\bfk_4;\tau)=
-\frac{1}{144H^2}\left(\frac{\kappa^2\,\rhom}{3}\right)^4
\left(\frac{k_{1234}^2}{a^2}\right)\frac{1}
{\Pi(k_{1234})\Pi(k_1)\Pi(k_2)\Pi(k_3)\Pi(k_4)}
\nonumber\\
&\qquad\qquad\times\Bigl[M_4(\bfk_1,\bfk_2,\bfk_3,\bfk_4)+
\frac{1}{3}
\Bigl\{
\frac{M_2(\bfk_1,\bfk_2)}{\Pi(k_{12})}
\left(\frac{M_2(\bfk_{12},\bfk_{34})\,\,M_2(\bfk_3,\bfk_4)}{\Pi(k_{34})}
-6\,M_3(\bfk_{12},\bfk_2,\bfk_3)\right)
\nonumber\\
&\qquad\qquad\qquad\qquad\qquad\qquad\qquad\qquad\qquad\quad
+2\,\frac{M_2(\bfk_{123},\bfk_4)}{\Pi(k_{123})}
\Bigl(\frac{M_2(\bfk_{12},\bfk_{13})\,M_2(\bfk_1,\bfk_2)}{\Pi(k_{12})}-M_3(\bfk_1,\bfk_2,\bfk_3)\Bigr)
\,\Bigr\}\,\Bigr].
\label{eq:vertex_sigma_4}
\end{align}
\end{widetext}
%%%%%%%%%%%%%%%%%%%%%%%%%%%%%%%%%%%%%%%%%%%%%%%%%%%%%%%%%%%%%%%%%%%%%%
Note that the expression of vertex functions $\sigma^{(3)}$ and
$\sigma^{(4)}$ 
is not yet symmetrized under the permutation of wave vectors, and 
it has to be symmetrized.

So far the framework to treat perturbations is general, and 
can be applied to any gravity model that satisfies 
the conservation law of the matter sector. As representative examples of 
modified gravity models that can explain the late-time cosmic acceleration, 
we shall below consider the $f(R)$ gravity 
\cite{Hu:2007nk,Starobinsky:2007hu}
and Dvali-Gabadadze-Poratti (DGP) model \cite{Dvali:2000hr}, 
and present the explicit expressions for model-dependent 
parameters $\omega_{\rm BD}$ and coupling functions $\Pi$ and $M_i$. 
While these models are rather specific and have been tightly 
constrained recently by observations, the mechanisms to recover GR on 
small scales 
are typical and a broad class of modified gravity models can fall 
into either of two models. We thus expect that even the PT calculations 
in these specific modified gravity models can give a fairly generic 
view on the deviation of gravity from GR.

%--%--%--%--%--%--%--%--%--%--%--%--%--%--%--%--%--%--%--%--%--%--%--%
\subsubsection{$f(R)$ gravity}
\label{subsubsec:fR_gravity}
%--%--%--%--%--%--%--%--%--%--%--%--%--%--%--%--%--%--%--%--%--%--%--%

The $f(R)$ gravity is a representative modified gravity model for which 
the Einstein-Hilbert action is generalized to include an arbitrary function 
of the scalar curvature $R$:
%%%%%%%%%%%%%%%%%%%%%%%%%%%%%%%%%%%%%%%%%%%%%%%%%%%%%%%%%%%%%%%%%%%%%%%
\begin{align}
S=\int d^4x \sqrt{-g}\left[\frac{R+f(R)}{2\kappa^2}\right]+L_{\rm m}, 
\end{align}
%%%%%%%%%%%%%%%%%%%%%%%%%%%%%%%%%%%%%%%%%%%%%%%%%%%%%%%%%%%%%%%%%%%%%%%
where $L_{\rm m}$ is the Lagrangian for matter sector. This theory is known to 
be equivalent to the Brans-Dicke theory with parameter $\omega_{\rm BD}=0$, but  
due to the nonlinear functional form of $R$, the Brans-Dicke scalar 
can acquire a nontrivial potential.  This can be seen from the trace of the 
modified Einstein equations. In the universe dominated by ordinary matter, 
we have
%%%%%%%%%%%%%%%%%%%%%%%%%%%%%%%%%%%%%%%%%%%%%%%%%%%%%%%%%%%%%%%%%%%%%%%
\begin{align}
3\Box f_R - R + f_R R -2 f= -\kappa^2\,\rhom
\end{align}
%%%%%%%%%%%%%%%%%%%%%%%%%%%%%%%%%%%%%%%%%%%%%%%%%%%%%%%%%%%%%%%%%%%%%%%
where $f_R=df/dR$ and $\Box=\nabla_\mu\nabla^\mu$. 
The field $f_R$ is identified with the scalaron, i.e., 
the extra scalar field, and its perturbations are defined as
%%%%%%%%%%%%%%%%%%%%%%%%%%%%%%%%%%%%%%%%%%%%%%%%%%%%%%%%%%%%%%%%%%%%%%
\begin{equation}
\varphi = \delta f_R\equiv f_R-\overline{f}_R,
\label{eq:scalaron_fR}
\end{equation}
%%%%%%%%%%%%%%%%%%%%%%%%%%%%%%%%%%%%%%%%%%%%%%%%%%%%%%%%%%%%%%%%%%%%%%
where the bar indicates that the quantity is evaluated on the
background universe. Imposing the conditions $| \bar{f}_R | \ll 1$
and $|\bar{f}/\bar{R}| \ll 1$, 
the background expansion can be close to $\Lambda$CDM cosmology, and 
the quasi-static approximation leads to 
%%%%%%%%%%%%%%%%%%%%%%%%%%%%%%%%%%%%%%%%%%%%%%%%%%%%%%%%%%%%%%%%%%%%%%
\begin{equation}
3 \frac{1}{a^2} \nabla^2 \varphi = - \kappa^2 \rhom \delta
+ \delta R,
\quad \delta R \equiv R(f_R)-R(\overline{f}_R), 
\end{equation}
%%%%%%%%%%%%%%%%%%%%%%%%%%%%%%%%%%%%%%%%%%%%%%%%%%%%%%%%%%%%%%%%%%%%%%
The above equation indeed corresponds to Eq.~(\ref{eq:EoM_scalaron}) 
with $\omega_{\rm BD}=0$. Then, expanding $\delta R$ in terms of $\varphi$, 
we obtain the explicit functional form of the coupling functions:
%%%%%%%%%%%%%%%%%%%%%%%%%%%%%%%%%%%%%%%%%%%%%%%%%%%%%%%%%%%%%%%%%%%%%%%
\begin{align}
&M_n(\tau)=\frac{d^n\overline{R}(f_R)}{df_R^n},\quad
\label{eq:coupling_func_fR}
\end{align}
%%%%%%%%%%%%%%%%%%%%%%%%%%%%%%%%%%%%%%%%%%%%%%%%%%%%%%%%%%%%%%%%%%%%%%%
which only depends on time. Then, this gives
%%%%%%%%%%%%%%%%%%%%%%%%%%%%%%%%%%%%%%%%%%%%%%%%%%%%%%%%%%%%%%%%%%%%%%%
\begin{align}
\Pi(k)=\left(\frac{k}{a}\right)^2 +\frac{\overline{R}_{,f}}{3}, 
\end{align}
%%%%%%%%%%%%%%%%%%%%%%%%%%%%%%%%%%%%%%%%%%%%%%%%%%%%%%%%%%%%%%%%%%%%%%%
where we define $\overline{R}_f(\tau)\equiv d\overline{R}(f_R)/df_R$.

In this paper, we will present the results of PT calculations 
in $f(R)$ gravity, and the predictions of propagator, power spectrum, and 
correlation function are compared with $N$-body simulations. 
For this purpose, in this paper, we will below consider the specific 
function of the form:
%%%%%%%%%%%%%%%%%%%%%%%%%%%%%%%%%%%%%%%%%%%%%%%%%%%%%%%%%%%%%%%%%%%%%%%
\begin{align}
f(R)\propto \frac{R}{A R+1},  
\label{eq:fR_HuSawicki}
\end{align}
%%%%%%%%%%%%%%%%%%%%%%%%%%%%%%%%%%%%%%%%%%%%%%%%%%%%%%%%%%%%%%%%%%%%%%%
where $A$ is a dimensional constant of length squared. In particular, 
we are interested in the high curvature limit $A\,R\gg1$, and $f(R)$ 
can be expanded as
%%%%%%%%%%%%%%%%%%%%%%%%%%%%%%%%%%%%%%%%%%%%%%%%%%%%%%%%%%%%%%%%%%%%%%%
\begin{align}
f(R)\simeq -2 \kappa^2 \rho_\Lambda + |f_{R0}|\,\frac{\overline{R}_0^2}{R}.
\label{eq:fR_model_nbody}
\end{align}
%%%%%%%%%%%%%%%%%%%%%%%%%%%%%%%%%%%%%%%%%%%%%%%%%%%%%%%%%%%%%%%%%%%%%%%
Here, $\rho_\Lambda$ is the constant energy density related to $A$. The quantity 
$\overline{R}_0$ is the background curvature at present time, and we  
defined $f_{R0}=\overline{f}_R(\overline{R}_0)$. With the current 
observational constraint $|f_{R0}|\ll1$ 
(e.g., \cite{Marchini:2013oya,Lombriser:2010mp,Yamamoto:2010ie,Okada:2012mn,Schmidt:2009am}, see also \cite{2012PhRvL.109x1301W} for a strong constraint 
from small-scales), 
the background cosmology becomes indistinguishable with $\Lambda$CDM model, but the extra term is still non-negligible for 
the evolution of matter fluctuations, giving rise to a different 
growth history of structure.

%--%--%--%--%--%--%--%--%--%--%--%--%--%--%--%--%--%--%--%--%--%--%--%
\subsubsection{DGP model}
\label{subsubsec:DGP_model}
%--%--%--%--%--%--%--%--%--%--%--%--%--%--%--%--%--%--%--%--%--%--%--%

The DGP braneworld model is another modified gravity model that 
has a screening mechanism.  
The DGP model is the 5D gravity theory with the induced 4D gravity 
on a brane in which we are living. 
Thus, on large scales larger than the characteristic scale $r_c$, 
the gravity becomes $5D$, while 
on small scales, gravity becomes $4D$, but it is not described by GR. 
As a result, the Friedman equation is modified on the brane 
\cite{Dvali:2000hr}:  
%%%%%%%%%%%%%%%%%%%%%%%%%%%%%%%%%%%%%%%%%%%%%%%%%%%%%%%%%%%%%%%%%%%%%%%
\begin{align}
\epsilon \,\frac{H}{r_c}=H^2-\frac{\kappa^2\,\rhom}{3},
\end{align}
%%%%%%%%%%%%%%%%%%%%%%%%%%%%%%%%%%%%%%%%%%%%%%%%%%%%%%%%%%%%%%%%%%%%%%%
where $\epsilon=\pm1$ represents two distinct branches of the solutions 
($\epsilon=1$ is the self-accelerating branch, and $-1$ is called 
the normal branch). 

Notable point in the DGP model is that the GR is recovered via 
the Vainshtein mechanism, by which the scalaron becomes massless, but 
acquires a large second-order derivative interaction. 
The resultant coupling functions become \cite{Koyama:2009me}
%%%%%%%%%%%%%%%%%%%%%%%%%%%%%%%%%%%%%%%%%%%%%%%%%%%%%%%%%%%%%%%%%%%%%%%
\begin{align}
&M_1=0, \quad M_2(\bfk_1,\bfk_2;\tau)=2\,\frac{r_c^2}{a^4}
\bigl\{k_1^2k_2^2-(\bfk_1\cdot\bfk_2)^2\bigr\},  
\nonumber\\
&M_i=0\,\,(i\geq3).
\end{align}
%%%%%%%%%%%%%%%%%%%%%%%%%%%%%%%%%%%%%%%%%%%%%%%%%%%%%%%%%%%%%%%%%%%%%%%
The quasi-static perturbations on 4D brane are described by the Brans-Dicke 
theory, where the Brans-Dicke parameter is given by
%%%%%%%%%%%%%%%%%%%%%%%%%%%%%%%%%%%%%%%%%%%%%%%%%%%%%%%%%%%%%%%%%%%%%%%
\begin{align}
&\omega_{\rm BD}(\tau)=\frac{3}{2}\left\{\beta(\tau)-1\right\};\,\,
\beta(\tau)=1-2\epsilon\,H\,r_c\left(1+\frac{\dot{H}}{3H^2}\right),
\end{align}
%%%%%%%%%%%%%%%%%%%%%%%%%%%%%%%%%%%%%%%%%%%%%%%%%%%%%%%%%%%%%%%%%%%%%%%
with $\dot{H}$ being the cosmic time derivative of the Hubble parameter. 
Then, the function $\Pi$ becomes
%%%%%%%%%%%%%%%%%%%%%%%%%%%%%%%%%%%%%%%%%%%%%%%%%%%%%%%%%%%%%%%%%%%%%%%
\begin{align}
\Pi(k)=\beta(\tau)\left(\frac{k}{a}\right)^2.
\end{align}
%%%%%%%%%%%%%%%%%%%%%%%%%%%%%%%%%%%%%%%%%%%%%%%%%%%%%%%%%%%%%%%%%%%%%%%

%%%%%%%%%%%%%%%%%%%%%%%%%%%%%%%%%%%%%%%%%%%%%%%%%%%%%%%%%%%%%%%%%%%%%%%
\subsection{Multipoint propagator expansion}
\label{subsec:Gamma_expansion}
%%%%%%%%%%%%%%%%%%%%%%%%%%%%%%%%%%%%%%%%%%%%%%%%%%%%%%%%%%%%%%%%%%%%%%%

Provided the basic equations for matter fluctuations, 
a straightforward approach to deal with the nonlinear 
evolution perturbatively is to just expand the perturbed quantities like 
$\Psi_a=\Psi_a^{(1)}+\Psi_a^{(2)}+\cdots$, and to solve the equations 
order by order, regarding the initial field as a small expansion 
parameter. This is the so-called standard PT treatment 
\cite{Bernardeau:2001qr}. As we mentioned in Sec.~\ref{sec:intro}, 
the standard PT is known to produce a poorly convergent series expansion, 
and is difficult to compute the correlation function because of the 
bad UV behavior. 

Alternatively, 
we may first introduce the non-perturbative statistical quantities, and 
expand the statistical quantities for our interest in terms of these. 
The multi-point propagator expansion or the $\Gamma$ expansion is one such 
PT expansion, and is regarded as a resummed PT treatment, in which 
the standard PT expansion is reorganized by the non-perturbative quantities  
\cite{Bernardeau:2008fa}. A key property is that all the 
statistical quantities such as the power spectra and bispectra can be 
reconstructed by an expansion series written solely in terms of 
the multipoint propagators. The multi-point propagator is a fully 
non-perturbative quantity, and with this object, a good 
convergence of the PT expansion is guaranteed. 
This is in marked contrast to the standard PT expansion. Although 
these have been confirmed and checked in the case of GR, we expect them 
to hold even in modified gravity models as long as the deviation from 
GR is small.

The $(p+1)$-point propagator is defined by
%%%%%%%%%%%%%%%%%%%%%%%%%%%%%%%%%%%%%%%%%%%%%%%%%%%%%%%%%%%%%%%%%%%%%%%
\begin{widetext}
\begin{align}
\frac{1}{p!}\,\left\langle\frac{\delta^p \Psi_a(\bfk;\tau)}
{\delta\,\delta_0(\bfk_1)\,\,\cdots\,\delta\,\delta_0(\bfk_p)}\right\rangle=\delta_D(\bfk-\bfk_{1\cdots  p})\,\frac{1}{(2\pi)^{3(p-1)}}\Gamma_a^{(p)}(\bfk_1,\cdots,\bfk_p;\tau),
\label{eq:def_Gamma_p}
\end{align}
\end{widetext}
%%%%%%%%%%%%%%%%%%%%%%%%%%%%%%%%%%%%%%%%%%%%%%%%%%%%%%%%%%%%%%%%%%%%%%%
with $\langle\cdots\rangle$ being the ensemble average.  
Here, $\delta_0$ is the initial density field given at an early 
time $\tau_0$. With the multi-point propagator, 
the power spectra of cosmic fields 
are systematically constructed as follows. 
Defining the power spectra $P_{ab}$ as
%%%%%%%%%%%%%%%%%%%%%%%%%%%%%%%%%%%%%%%%%%%%%%%%%%%%%%%%%%%%%%%%%%%%%%%
\begin{align}
\langle\Psi_a(\bfk;\tau)\Psi_b(\bfk';\tau)\rangle
=(2\pi)^3\delta_{\rm D}(\bfk+\bfk')\,P_{ab}(|\bfk|;\tau), 
\label{eq:def_Pk}
\end{align}
%%%%%%%%%%%%%%%%%%%%%%%%%%%%%%%%%%%%%%%%%%%%%%%%%%%%%%%%%%%%%%%%%%%%%%%
we have \cite{Bernardeau:2008fa}
%%%%%%%%%%%%%%%%%%%%%%%%%%%%%%%%%%%%%%%%%%%%%%%%%%%%%%%%%%%%%%%%%%%%%%%
\begin{align}
&P_{ab}(k;\tau)=\Gamma^{(1)}_a(k;\tau)\Gamma^{(1)}_b(k;\tau) P_0(k) 
\nonumber\\
&\qquad\qquad
+\sum_{n=2}n!\,\int\frac{d^3\bfq_1\cdots d^3\bfq_n}{(2\pi)^{3(n-1)}} \,
\delta_{\rm D}(\bfk-\bfq_{1\cdots n})
\nonumber\\
&\quad\qquad\qquad\times
\Gamma^{(n)}_a(\bfq_1,\cdots,\bfq_n;\tau)\Gamma^{(n)}_b(\bfq_1,\cdots,\bfq_n;\tau)
\nonumber\\ 
&\quad\qquad\qquad \times P_0(q_1)\,\cdots P_0(q_n),
\label{eq:Pk_regpt}
\end{align}
%%%%%%%%%%%%%%%%%%%%%%%%%%%%%%%%%%%%%%%%%%%%%%%%%%%%%%%%%%%%%%%%%%%%%%%
where the quantity $P_0$ is the initial power spectrum defined as
%%%%%%%%%%%%%%%%%%%%%%%%%%%%%%%%%%%%%%%%%%%%%%%%%%%%%%%%%%%%%%%%%%%%%%%
\begin{align}
\langle\delta_0(\bfk)\delta_0(\bfk')\rangle=(2\pi)^3\delta_{\rm D}(\bfk+\bfk')\,
P_0(k).
\end{align}
%%%%%%%%%%%%%%%%%%%%%%%%%%%%%%%%%%%%%%%%%%%%%%%%%%%%%%%%%%%%%%%%%%%%%%%

As it is clear from Eq.~(\ref{eq:Pk_regpt}), 
the nonlinear effects in the power spectrum are wholly encapsulated in 
the multi-point propagators, and thus the construction 
of the propagators keeping their non-perturbative properties 
is quite essential in the analytic treatment of PT. Therefore, 
subsequent sections are devoted to the discussion on how to analytically 
construct the propagators in the context of modified gravity.  
Section \ref{sec:resummed_prop_eikonal} discusses 
the non-perturbative high-$k$ behavior of the propagators 
based on the eikonal approximation, and Section \ref{sec:regularized_pk}  
presents a consistent construction of the {\it regularized} 
propagators which satisfies both the expected high-$k$ and low-$k$ 
behaviors.

%%%%%%%%%%%%%%%%%%%%%%%%%%%%%%%%%%%%%%%%%%%%%%%%%%%%%%%%%%%%%%%%%%%%%%%
%%%%%%%%%%%%%%%%%%%%%%%%%%%%%%%%%%%%%%%%%%%%%%%%%%%%%%%%%%%%%%%%%%%%%%%
\section{Resummed linear propagator with eikonal approximation}
\label{sec:resummed_prop_eikonal}
%%%%%%%%%%%%%%%%%%%%%%%%%%%%%%%%%%%%%%%%%%%%%%%%%%%%%%%%%%%%%%%%%%%%%%%
%%%%%%%%%%%%%%%%%%%%%%%%%%%%%%%%%%%%%%%%%%%%%%%%%%%%%%%%%%%%%%%%%%%%%%%

In this section, we derive the {\it resummed} linear propagator, in 
which the behaviors of the high-$k$ limit is reproduced at the tree-level 
calculation as a result of resummation. This resummed linear propagator 
will be used to systematically construct the multi-point propagator 
in next section. Here, following Ref.~\cite{Bernardeau:2011vy}, we apply the 
eikonal approximation to the perturbation equations 
(\ref{eq:basic_eq}). The eikonal approximation enables us to derive 
the effective evolution equation for short-wave 
fluctuations under the influence of long-wave modes, which are
regarded as external random background. With this treatment, 
if we neglect the non-linear mode couplings, the fluid equations can 
be rewritten as linear equations embedded in an external random medium.

%%--%--%--%--%--%--%--%--%--%--%--%--%--%--%--%--%--%--%--%--%--%--%--%
\subsection{Eikonal approximation}
\label{subsec:Eikonal}
%%--%--%--%--%--%--%--%--%--%--%--%--%--%--%--%--%--%--%--%--%--%--%--%

To be more explicit, consider first the non-linear mode coupling 
in Eq.~(\ref{eq:basic_eq}) associated with standard vertex function, 
$\gamma_{abc}(\bfk_1,\bfk_2)$.  
Through the relation $\bfk=\bfk_1+\bfk_2$, the contribution of the 
non-linear coupling can be split into two different cases: 
the one coming from coupling two modes of very different amplitudes,
$k_1\ll k_2$ or $k_2\ll k_1$, 
and the one coming from coupling two modes of comparable amplitudes. 
In the first case, the small wave modes ought to be much smaller than 
$\bfk$. Let us denote these small modes by $\bfq$, and divide the 
domain of integral into soft and hard domains. 
Then, the coupling term may be rewritten as \cite{Bernardeau:2011vy}
%%%%%%%%%%%%%%%%%%%%%%%%%%%%%%%%%%%%%%%%%%%%%%%%%%%%%%%%%%%%%%%%%%%%%%
\begin{widetext}
\begin{align}
&\int\frac{d^3\bfk_1d^3\bfk_2}{(2\pi)^3}
\delta_{\rm D}(\bfk-\bfk_{12}) \gamma_{abc}(\bfk_1,\bfk_2)
\Psi_b(\bfk_1)\Psi_c(\bfk_2)
\nonumber\\
&\qquad\qquad\qquad=
\Xi(\bfk;\tau)\Psi_a(\bfk;\tau)
+\int_{\mathcal{H}}\frac{d^3\bfk_1d^3\bfk_2}{(2\pi)^3}
\delta_{\rm D}(\bfk-\bfk_{12}) \gamma_{abc}(\bfk_1,\bfk_2)
\Psi_b(\bfk_1)\Psi_c(\bfk_2)
\end{align}
\end{widetext}
%%%%%%%%%%%%%%%%%%%%%%%%%%%%%%%%%%%%%%%%%%%%%%%%%%%%%%%%%%%%%%%%%%%%%%
where the first term at the right-hand side represents the 
contribution from the soft domain, taking the limit,  
$k\gg q$. The expression for the function $\Xi$ becomes
%%%%%%%%%%%%%%%%%%%%%%%%%%%%%%%%%%%%%%%%%%%%%%%%%%%%%%%%%%%%%%%%%%%%%%
\begin{align}
\Xi(\bfk;\tau)=\int_{\mathcal{S}}
\frac{d^3\bfq}{(2\pi)^3}\,\left(\frac{\bfk\cdot\bfq}{q^2}\right)\,
\theta(\bfq;\tau), 
\label{eq:Xi}
\end{align}
%%%%%%%%%%%%%%%%%%%%%%%%%%%%%%%%%%%%%%%%%%%%%%%%%%%%%%%%%%%%%%%%%%%%%%
where the subscript $_\mathcal{S}$ implies that the integral is restricted to the soft domain.

Similarly, the mode coupling arising from the non-linear interaction
of the scalaron [i.e., the second term in RHS of Eq.~(\ref{eq:basic_eq})] 
can be split into two domains: the soft domain in which one of the modes is 
much larger than others, and the hard domain in which there is 
no particularly larger mode than others. We obtain
%%%%%%%%%%%%%%%%%%%%%%%%%%%%%%%%%%%%%%%%%%%%%%%%%%%%%%%%%%%%%%%%%%%%%%
\begin{widetext}
\begin{align}
&\delta_{a2}\int\frac{d^3\bfk_1\cdots d^3\bfk_n}{(2\pi)^{3(n-1)}}
\delta_{\rm D}(\bfk-\bfk_{1\cdots n}) 
\sigma^{(n)}(\bfk_1,\cdots,\bfk_n)
\Psi_1(\bfk_1)\cdots\Psi_1(\bfk_n)
\nonumber\\
&\qquad\qquad\qquad
=\omega_{ab}^{(n)}(\bfk;\tau)\,\Psi_b(\bfk;\tau)
+\delta_{a2}\int_{\mathcal{H}}\frac{d^3\bfk_1\cdots d^3\bfk_n}{(2\pi)^{3(n-1)}}
\delta_{\rm D}(\bfk-\bfk_{1\cdots n}) 
\sigma^{(n)}(\bfk_1,\cdots,\bfk_n)
\Psi_1(\bfk_1)\cdots\Psi_1(\bfk_n).
\end{align}
\end{widetext}
%%%%%%%%%%%%%%%%%%%%%%%%%%%%%%%%%%%%%%%%%%%%%%%%%%%%%%%%%%%%%%%%%%%%%%
Here, the matrix $\omega_{ab}^{(n)}$ includes the contribution from the 
soft domain, and the non-vanishing contribution appears only 
in $\omega_{21}^{(n)}$. Let us rewrite it with 
%%%%%%%%%%%%%%%%%%%%%%%%%%%%%%%%%%%%%%%%%%%%%%%%%%%%%%%%%%%%%%%%%%%%%%
\begin{align}
\omega_{21}^{(n)}(\bfk;\tau)=\frac{\kappa^2}{2}\frac{\rhom}{H^2}\,
\frac{1}{3}\frac{(k/a)^2}{\Pi(k)}\Delta^{(n)}(k;\tau). 
\label{eq:omega_21}
\end{align}
%%%%%%%%%%%%%%%%%%%%%%%%%%%%%%%%%%%%%%%%%%%%%%%%%%%%%%%%%%%%%%%%%%%%%%
Here, the function $\Delta^{(n)}$ represents the sum of all possible 
combinations of the soft/hard domains of the integral. 
We find that the non-vanishing contribution of $\Delta^{(n)}$ 
leads to the modification of the effective Newton 
constant given in Eq.~(\ref{eq:effective_G}), 
as a result of the screening mechanism in modified gravity: 
%%%%%%%%%%%%%%%%%%%%%%%%%%%%%%%%%%%%%%%%%%%%%%%%%%%%%%%%%%%%%%%%%%%%%%
\begin{align}
G_{\rm eff}\to G\left[1+\frac{1}{3}\frac{(k/a)^2}{\Pi(k)}
\left\{1+\sum_{n=2}\Delta^{(n)}(k;\tau)\right\}\right].
\label{eq:effective_G_n}
\end{align}
%%%%%%%%%%%%%%%%%%%%%%%%%%%%%%%%%%%%%%%%%%%%%%%%%%%%%%%%%%%%%%%%%%%%%%
The explicit expression for $\sigma^{(n)}$ is given by 
%%%%%%%%%%%%%%%%%%%%%%%%%%%%%%%%%%%%%%%%%%%%%%%%%%%%%%%%%%%%%%%%%%%%%%
\begin{align}
&
\Delta^{(n)}(k;\tau)=-\frac{n}{3\,n!}\,\frac{M_n(\tau)}{\Pi(k)}\,
\left(\frac{\kappa^2\rhom}{3}\right)^{n-1}\int_{\mathcal{S}}
\frac{d^3\bfp_1\cdots d^3\bfp_{n-1}}{(2\pi)^{3(n-1)}}\,
\nonumber\\
&
\qquad\times
K_{f(R)}^{(n)}(\bfp_1,\cdots,\bfp_{n-1};\tau)
\frac{\delta(\bfp_1;\tau)\cdots\delta(\bfp_{n-1};\tau)}
{\Pi(p_1)\cdots\Pi(p_{n-1})}
\label{eq:sigma_n_fR}
\end{align}
%%%%%%%%%%%%%%%%%%%%%%%%%%%%%%%%%%%%%%%%%%%%%%%%%%%%%%%%%%%%%%%%%%%%%%
for the $f(R)$ gravity model, and 
%%%%%%%%%%%%%%%%%%%%%%%%%%%%%%%%%%%%%%%%%%%%%%%%%%%%%%%%%%%%%%%%%%%%%%
\begin{align}
&\Delta^{(n)}(k;\tau)=n\left\{-\,\frac{r_c^2}{3\beta(\tau)^2}
\,\frac{\kappa^2\rhom}{3}\right\}^{n-1}
\int_{\mathcal{S}}
\frac{d^3\bfp_1\cdots d^3\bfp_{n-1}}{(2\pi)^{3(n-1)}}\,
\nonumber\\
&
\qquad\times K_{\rm DGP}^{(n)}(\bfp_1,\cdots,\bfp_{n-1})\,
%\nonumber\\
%&\qquad\qquad\times
\delta(\bfp_1;\tau)\cdots\delta(\bfp_{n-1};\tau)
\label{eq:sigma_n_DGP}
\end{align}
%%%%%%%%%%%%%%%%%%%%%%%%%%%%%%%%%%%%%%%%%%%%%%%%%%%%%%%%%%%%%%%%%%%%%%
for the DGP model. Note that the factor $n$ comes from the number of 
possible combinations of the soft/hard domains of the integral. 
The functions $K_{f(R)}^{(n)}$ and $K_{\rm DGP}^{(n)}$ 
are the dimensionless kernels, whose explicit expressions are 
presented in Appendix \ref{sec:kernel}. 
%Note that while the positivity of the kernel is guaranteed in
%the DGP model, it is not always the case for $f(R)$ gravity model. 
In the high-$k$ limit, these contributions are supposed to be 
subdominant compared to that coming from the standard vertex function 
[see Eq.~(\ref{eq:Xi})], and may be treated perturbatively as a
higher-order contribution. This point will be discussed in 
Sec.~\ref{subsec:impact_subleading}.

%%%%%%%%%%%%%%%%%%%%%%%%%%%%%%%%%%%%%%%%%%%%%%%%%%%%%%%%%%%%%%%%%%%%%%%
%%%%%%%%%%%%%%%%%%%%%%%%%%%%%%%%%%%%%%%%%%%%%%%%%%%%%%%%%%%%%%%%%%%%%%%
\subsection{Resummed propagator}
\label{subsec:resummed_prop}
%%%%%%%%%%%%%%%%%%%%%%%%%%%%%%%%%%%%%%%%%%%%%%%%%%%%%%%%%%%%%%%%%%%%%%%
%%%%%%%%%%%%%%%%%%%%%%%%%%%%%%%%%%%%%%%%%%%%%%%%%%%%%%%%%%%%%%%%%%%%%%%

Based on the eikonal approximation in Sec.~\ref{subsec:Eikonal},
we can now reabsorb the effect of the non-linear coupling with 
long-wavelength modes in the linear terms, $\Xi\,\Psi_a$ and 
$\omega_{ab}\,\Psi_b$. As a result, the evolution equation for 
perturbation, Eq.~(\ref{eq:basic_eq}), can be recast as
%%%%%%%%%%%%%%%%%%%%%%%%%%%%%%%%%%%%%%%%%%%%%%%%%%%%%%%%%%%%%%%%%%%%%%
\begin{widetext}
\begin{align}
&\left[\delta_{ab}\left\{\frac{\partial}{\partial\tau} - \Xi(k;\tau)
\right\}+\Omega_{ab}(k;\tau)-
\sum_{n=2}\omega_{ab}^{(n)}(k;\tau)\right]\Psi_{b}(\bfk;\tau)
\nonumber\\
&\qquad\qquad\qquad=
\int_{\mathcal H}
\frac{d^3\bfk_1d^3\bfk_2}{(2\pi)^3}\,\delta_{\rm D}(\bfk-\bfk_{12})\,
\gamma_{abc}(\bfk_1,\bfk_2;\tau)\,\Psi_b(\bfk_1;\tau)\Psi_c(\bfk_2;\tau)
\nonumber\\
&\qquad\quad\quad\quad\quad\quad
+\delta_{b2}\sum_{n=2}\int_{\mathcal H}
\frac{d^3\bfk_1\cdots d^3\bfk_n}{(2\pi)^{3(n-1)}}\,
\delta_{\rm D}(\bfk-\bfk_{1\cdots n})\,
\sigma^{(n)}(\bfk_1,\cdots,\bfk_n;\tau)\,
\Psi_1(\bfk_1;\tau)\cdots\Psi_1(\bfk_n;\tau),
\end{align}
\end{widetext}
%%%%%%%%%%%%%%%%%%%%%%%%%%%%%%%%%%%%%%%%%%%%%%%%%%%%%%%%%%%%%%%%%%%%%%
The solution to this equation can be given in terms of 
the {\it resummed} propagator, $\xi_{ab}$, and it reads
%%%%%%%%%%%%%%%%%%%%%%%%%%%%%%%%%%%%%%%%%%%%%%%%%%%%%%%%%%%%%%%%%%%%%%
\begin{widetext}
\begin{align}
&\Psi_a(\bfk;\tau)=\xi_{ab}(k;\tau,\tau_0)\Psi_b(\bfk;\tau_0)+
\int_{\tau_0}^\tau d\tau'\,\xi_{ab}(k;\tau,\tau')
\nonumber\\
&\quad\quad\qquad\qquad\times \Bigl[
\int_{\mathcal H}
\frac{d^3\bfk_1d^3\bfk_2}{(2\pi)^3}\,\delta_{\rm D}(\bfk-\bfk_{12})\,
\gamma_{bcd}(\bfk_1,\bfk_2;\tau')\,\Psi_c(\bfk_1;\tau')\Psi_d(\bfk_2;\tau')
\nonumber\\
&\qquad\qquad\quad\quad\quad\quad
+\delta_{b2}\sum_{n=2}\int_{\mathcal H}
\frac{d^3\bfk_1\cdots d^3\bfk_n}{(2\pi)^{3(n-1)}}\,
\delta_{\rm D}(\bfk-\bfk_{1\cdots n})\,
\sigma^{(n)}(\bfk_1,\cdots,\bfk_n;\tau')\,
\Psi_1(\bfk_1;\tau')\cdots\Psi_1(\bfk_n;\tau') \,\Bigr],
\label{eq:formal_solu}
\end{align}
\end{widetext}
%%%%%%%%%%%%%%%%%%%%%%%%%%%%%%%%%%%%%%%%%%%%%%%%%%%%%%%%%%%%%%%%%%%%%%
Here, the resummed propagator satisfies 
%%%%%%%%%%%%%%%%%%%%%%%%%%%%%%%%%%%%%%%%%%%%%%%%%%%%%%%%%%%%%%%%%%%%%%
\begin{align}
&\left[\delta_{ab}\left\{\frac{\partial}{\partial\tau} - \Xi(k;\tau)
\right\}+\Omega_{ab}(k;\tau)-\sum_{n=2}\omega_{ab}^{(n)}(k;\tau)\right]
\nonumber\\
&\qquad\qquad\qquad\qquad\qquad\qquad\times\xi_{bc}(\bfk;\tau,\tau')=0
\end{align}
%%%%%%%%%%%%%%%%%%%%%%%%%%%%%%%%%%%%%%%%%%%%%%%%%%%%%%%%%%%%%%%%%%%%%%
with the boundary condition, $\xi_{ab}(\bfk;\tau,\tau)=\delta_{ab}$.

In the absence of the term $\omega_{ab}$, 
the solution of this resummed propagator is expressed in terms of the 
{\it standard} linear propagator of 
Eq.~(\ref{eq:basic_eq}), $g_{ab}$, which satisfies 
$(\partial g_{ac} /\partial \tau)+\Omega_{ab}g_{bc}=0$:
%%%%%%%%%%%%%%%%%%%%%%%%%%%%%%%%%%%%%%%%%%%%%%%%%%%%%%%%%%%%%%%%%%%%%%
\begin{align}
\xi_{ab}(k;\tau,\tau_0)=g_{ab}(k;\tau,\tau_0)\,\exp\left[
\int_{\tau_0}^\tau d\tau' \,\Xi(\bfk,\tau')
\right].
\label{eq:xi0_solution} 
\end{align}
%%%%%%%%%%%%%%%%%%%%%%%%%%%%%%%%%%%%%%%%%%%%%%%%%%%%%%%%%%%%%%%%%%%%%%
In the presence of the asymmetric matrix $\omega_{ab}$, 
no tractable analytic expression is obtained, however, 
assuming that the term $\omega_{ab}$ just gives a sub-dominant contribution
compared to the function $\Xi$, we obtain the approximate expression: 
%%%%%%%%%%%%%%%%%%%%%%%%%%%%%%%%%%%%%%%%%%%%%%%%%%%%%%%%%%%%%%%%%%%%%%
\begin{align}
&\xi_{ab}(k;\tau,\tau_0)=\Bigl\{\,g_{ab}(k;\tau,\tau_0)\,+
\int_{\tau_0}^\tau d\tau' \,
g_{ac}(k;\tau,\tau')
\nonumber\\
&\times\omega_{cd}(k;\tau')g_{db}(k;\tau',\tau_0)\,\Bigr\}
\exp\left[
\int_{\tau_0}^\tau d\tau'' \,\Xi(\bfk,\tau'')\right].
\label{eq:xi1_solution} 
\end{align}
%%%%%%%%%%%%%%%%%%%%%%%%%%%%%%%%%%%%%%%%%%%%%%%%%%%%%%%%%%%%%%%%%%%%%%

The resummed propagator given above can be used to systematically 
compute the multi-point propagators $\Gamma^{(n)}$ defined in 
Eq.~(\ref{eq:def_Gamma_p}), where the non-perturbative 
high-$k$ behaviors have been already encapsulated 
(see Sec.~\ref{sec:regularized_pk}). In GR, the 
correction $\omega_{ab}$ vanishes, and all the multi-point 
propagators are shown to have the exponential damping behaviors 
\cite{Bernardeau:2008fa}. Thus, the non-vanishing contribution of 
$\omega_{ab}$ is a non-trivial result in modified gravity models. 
The 
influence of this on the multi-point propagators will be quantitatively 
estimated in next subsection.

%%--%%--%%--%%--%%--%%--%%--%%--%%--%%--%%--%%--%%--%%--%%--%%--%%--%%%
\subsection{Impact of screening effect on resummed propagator}
\label{subsec:impact_subleading}
%%--%%--%%--%%--%%--%%--%%--%%--%%--%%--%%--%%--%%--%%--%%--%%--%%--%%%

Let us discuss the impact of screening effect on the 
resummed propagator, focusing on the 
new contribution, $\omega_{ab}$. To start with, we define
%%%%%%%%%%%%%%%%%%%%%%%%%%%%%%%%%%%%%%%%%%%%%%%%%%%%%%%%%%%%%%%%%%%%%%%
\begin{align}
G_{ab}(k;\tau,\tau_0)=\big\langle \xi_{ab}(k;\tau,\tau_0)\big\rangle_{\Xi,\omega_{ab}}.
\end{align}
%%%%%%%%%%%%%%%%%%%%%%%%%%%%%%%%%%%%%%%%%%%%%%%%%%%%%%%%%%%%%%%%%%%%%%%
In the cases with the negligible effect of modified gravity at 
an early time $\tau_0\to-\infty$, 
contracting $G_{ab}$ with vector $u_a=(1,1)$ gives 
the two-point propagator $\Gamma_a^{(1)}$ in the high-$k$ limit, 
$G_{ab}u_b\simeq \Gamma_a^{(1)}$. 
To evaluate the impact of the new correction term, we adopt 
Eq.~(\ref{eq:xi1_solution}) and substitute it into the above. We then 
write
%%%%%%%%%%%%%%%%%%%%%%%%%%%%%%%%%%%%%%%%%%%%%%%%%%%%%%%%%%%%%%%%%%%%%%%
\begin{widetext}
\begin{align}
&G_{ab}(k;\tau,\tau_0)=G_{ab,0}(k;\tau,\tau_0) + \delta\,G_{ab}(k;\tau,\tau_0);
\nonumber\\
&\qquad\qquad\qquad
G_{ab,0}(k;\tau)=g_{ab}(k;\tau,\tau_0)\,
\exp\left[-\frac{k^2}{2}\int\frac{dq}{6\pi^2} \,P_0(q) 
\{D_+(q;\tau)-D_+(q;\tau_0)\}^2\right], 
\nonumber\\
&\qquad\qquad\qquad
\delta G_{ab}(k;\tau) = 
\int_{\tau_0}^\tau d\tau' \,
g_{ac}(k;\tau,\tau')g_{db}(k;\tau',\tau_0)\,
\sum_{n=2}\Bigl\langle\omega_{cd}^{(n)}(k;\tau')
\exp\left[\int_{\tau_0}^\tau d\tau'' 
\,\Xi(\bfk,\tau'')\right]\Bigr\rangle_{\Xi,\omega_{ab}}.
\end{align}
\end{widetext}
%%%%%%%%%%%%%%%%%%%%%%%%%%%%%%%%%%%%%%%%%%%%%%%%%%%%%%%%%%%%%%%%%%%%%%%
Using Eq.~(\ref{eq:omega_21}), we recast $\delta G_{ab}$ as
%%%%%%%%%%%%%%%%%%%%%%%%%%%%%%%%%%%%%%%%%%%%%%%%%%%%%%%%%%%%%%%%%%%%%%%
\begin{widetext}
\begin{align}
\delta\,G_{ab}(k;\tau,\tau_0)=\int_{\tau_0}^\tau d\tau' \,
g_{a2}(k;\tau,\tau')g_{1b}(k;\tau',\tau_0)\,
\frac{\kappa^2}{2}\,\frac{\rhom(\tau')}{H^2(\tau')}\,
\frac{1}{3}\frac{(k/a)^2}{\Pi(k;\tau')}\,\sum_{n=2}\,
\left\langle \,\Delta^{(n)}
\exp\left[\int_{\tau_0}^\tau d\tau'' 
\,\Xi(\bfk,\tau'')\right]
\right\rangle.
\label{eq:delta_G_ab}
\end{align}
\end{widetext}
%%%%%%%%%%%%%%%%%%%%%%%%%%%%%%%%%%%%%%%%%%%%%%%%%%%%%%%%%%%%%%%%%%%%%%%
Based on the leading-order calculation in which 
the field $\Psi_a=(\delta,-\theta)$ is treated as linear-order quantity, 
we can explicitly evaluate $\delta G_{ab}$ under the Gaussian initial 
condition. We then find that the contribution from $n=2$ in 
Eq.~(\ref{eq:delta_G_ab}) vanishes in both $f(R)$ gravity and DGP models, 
and the correction from 
$n=3$ can give the leading-order non-vanishing contribution. 
Up to this contribution, the propagator $G_{ab}$ can be recast as
%%%%%%%%%%%%%%%%%%%%%%%%%%%%%%%%%%%%%%%%%%%%%%%%%%%%%%%%%%%%%%%%%%%%%%%
\begin{widetext}
\begin{align}
G_{ab}(k;\tau,\tau_0)\simeq\Bigl[\,g_{ab}(k;\tau,\tau_0)+\delta 
g_{ab}^{(3)}(k;\tau,\tau_0)\,\Bigr]
\exp\left[-\frac{k^2}{2}\int\frac{dq}{6\pi^2} \,P_0(q) 
\{D_+(q;\tau)-D_+(q;\tau_0)\}^2\right]. 
\label{eq:g_ab_d_gab}
\end{align}
\end{widetext}
%%%%%%%%%%%%%%%%%%%%%%%%%%%%%%%%%%%%%%%%%%%%%%%%%%%%%%%%%%%%%%%%%%%%%%%
The correction $\delta g^{(3)}_{ab}$ represents
the first non-vanishing contributions from $n=3$ of $\delta G_{ab}$. 
The explicit expression for $\delta g_{ab}^{(3)}$ is presented in 
Appendix~\ref{sec:sub-leading} for the $f(R)$ gravity and DGP models 
[Eqs.~(\ref{eq:dg^(3)_ab_fR}) (\ref{eq:dg^(3)_ab_DGP})].

%%%%%%%%%%%%%%%%%%%%%%%%%%%%%%%%%%%%%%%%%%%%%%%%%%%%%%%%%%%%%%%%%%%%%%%
\begin{figure*}[t]
%\hspace*{-0.8cm}
\includegraphics[width=8cm]{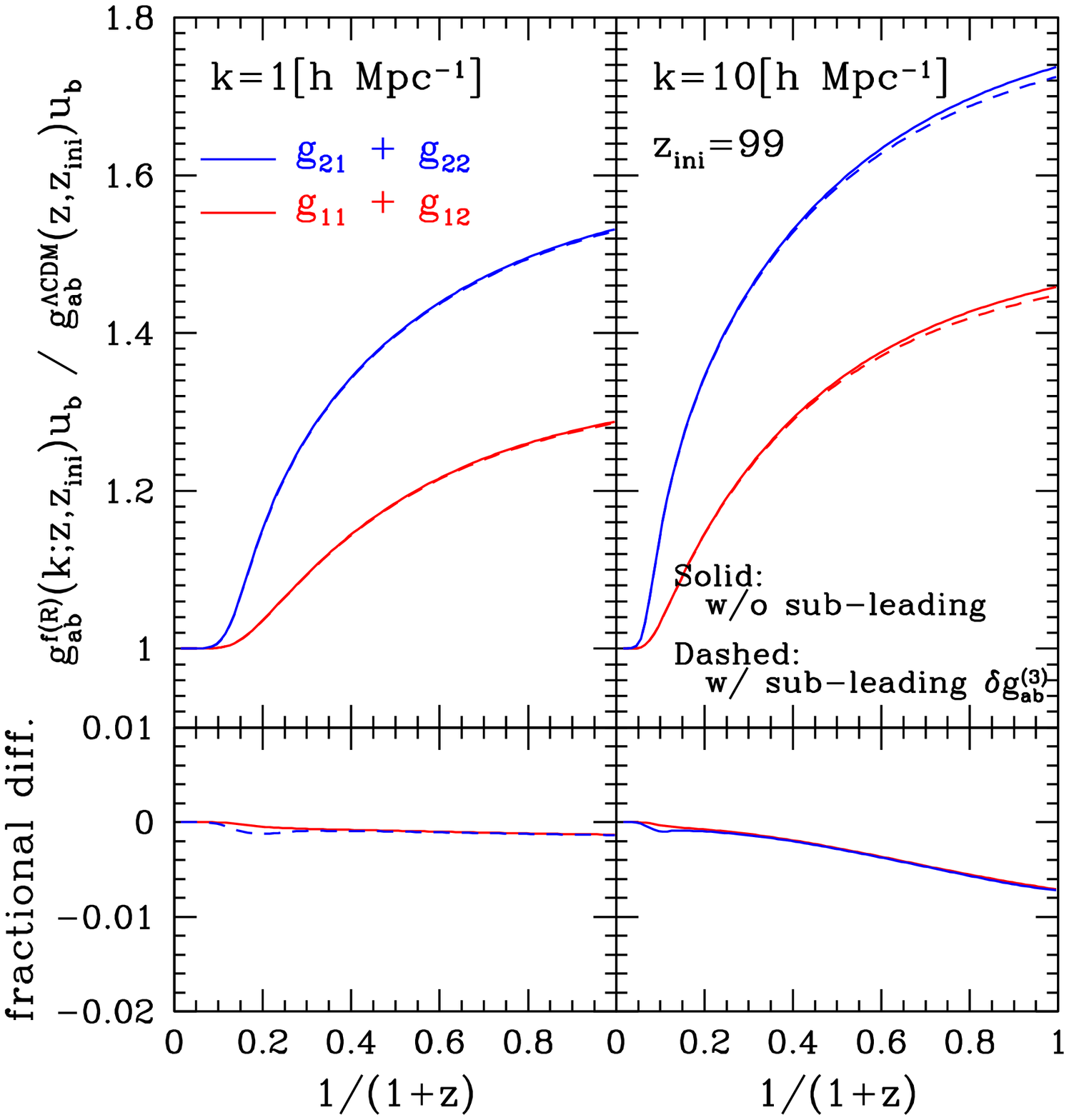}
\includegraphics[width=8cm]{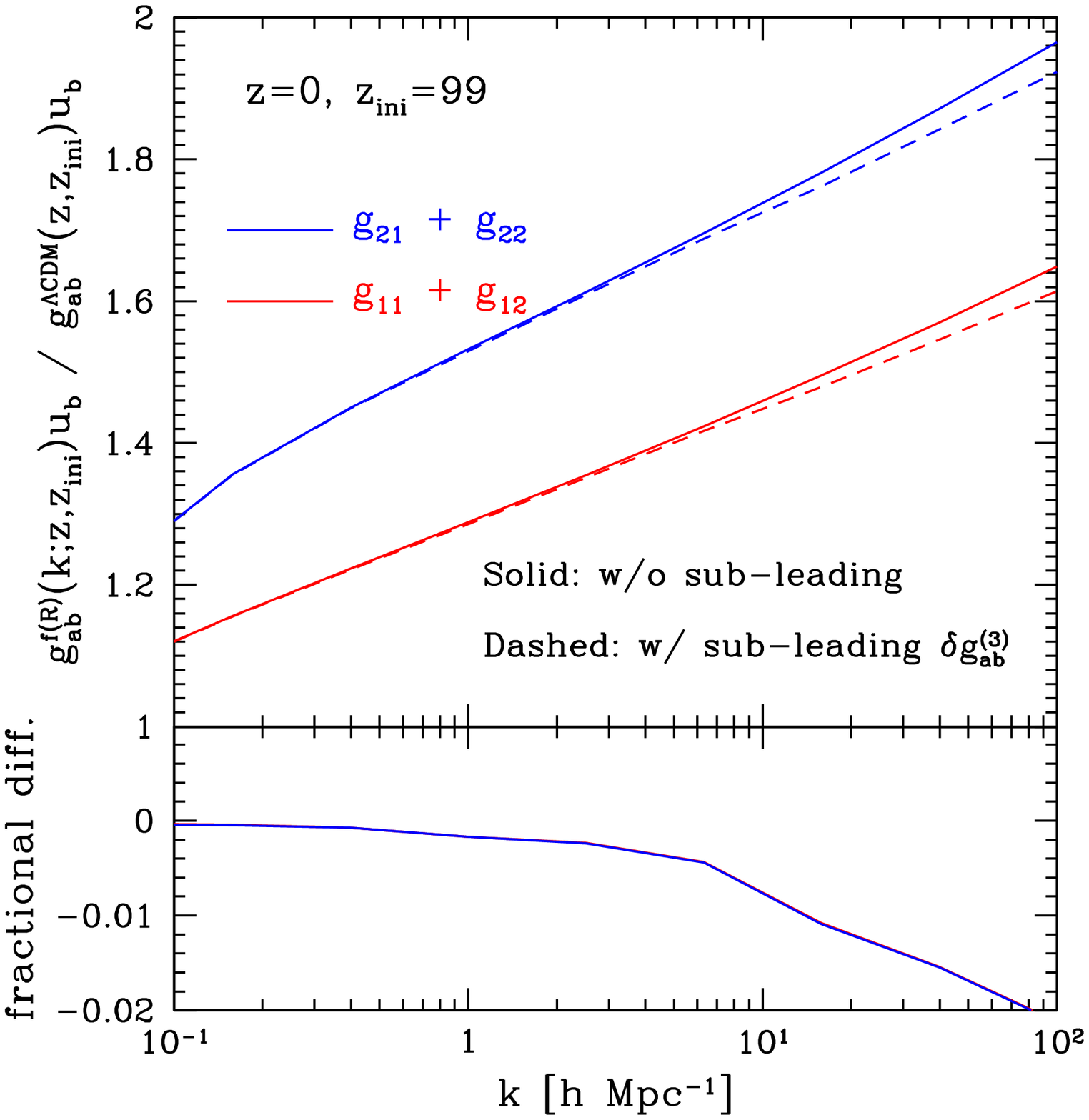}

\vspace*{-0.5cm}

\caption{Resummed linear propagator in $f(R)$ gravity model. 
Left panel shows the  
redshift evolution of the propagator at $k=1\,h$Mpc$^{-1}$ (left) 
and $10\,h$Mpc$^{-1}$ (right), while right panel plots the scale dependence 
of the propagator at $z=0$. Top panels plot the ratio of 
propagator in $f(R)$ gravity to that in GR for 
density (red) or velocity-divergence fields (blue). 
Dashed and solid lines respectively represent the result with and without 
new correction arising from the screening effect, i.e.,  
$[g_{ab}+\delta g^{(3)}_{ab}]^{\rm f(R)}u_b/g_{ab}^{\Lambda{\rm CDM}}u_b$ and
$g_{ab}^{\rm f(R)}u_b/g_{ab}^{\Lambda{\rm CDM}}u_b$. On the other hand, to see 
the size of the correction, 
bottom panels show the fractional difference between the 
propagators with and without the correction,  i.e., 
$[\delta g^{(3)}_{ab}]^{\rm f(R)}u_b/g_{ab}^{\rm f(R)}u_b$. For all panels, 
we assume $f(R)$ gravity of the functional form in 
Eq.~(\ref{eq:fR_model_nbody}), and 
the model parameter is set to $|f_{R,0}|=10^{-4}$.    
In computing the propagators, 
we set the initial redshift to $z_{\rm ini}=99$, and 
adopt the cosmological parameters;   
$\Omega_{\rm m}=0.24$, $\Omega_\Lambda=0.76$, 
$\Omega_{\rm b}=0.0481$, $h=0.73$, $n_s=0.961$, 
$\sigma_8=0.801$~\cite{Taruya:2013quf}. 
\label{fig:gab_sub_ak}}
\end{figure*}
%%%%%%%%%%%%%%%%%%%%%%%%%%%%%%%%%%%%%%%%%%%%%%%%%%%%%%%%%%%%%%%%%%%%%%%

To see quantitatively the impact of sub-leading correction, 
we here consider the $f(R)$ gravity of the functional form in 
Eq.~(\ref{eq:fR_model_nbody}), and 
compute the correction $\delta g_{ab}^{(3)}$. The results are then 
compared with the leading-order term, $g_{ab}$. Fig.~\ref{fig:gab_sub_ak}
show the ratio of propagator in $f(R)$ gravity to that 
in GR ($\Lambda$CDM), without and with the new correction term, i.e., 
$g_{ab}^{\rm f(R)}u_b /g_{ab}^{\rm GR}u_b$ and 
$[g_{ab}+\delta g_{ab}^{(3)}]^{\rm f(R)}u_b/g_{ab}^{\rm GR}u_b$,  where the vector 
$u_b$ is defined by $u_b=(1,1)$. Note that if the effect of modified gravity 
is neglected at $\tau_0$, the combination $g_{ab}u_b$ just gives 
$g_{ab}u_b=(D_+$, $dD_+/d\tau)$, where $D_+$ is the linear growth factor. 
Left panel shows the time evolution of the propagator 
at specific wavenumbers $k=1\,h$Mpc$^{-1}$ (left) and $10\,h$Mpc$^{-1}$ 
(right), while in right panel, we plot the scale dependence 
of the propagator at $z=0$. In all cases, 
the model parameter of $f(R)$ gravity is set to $|f_{R,0}|=10^{-4}$. 
As we see from Fig.~\ref{fig:gab_sub_ak}, 
the new correction $\delta g_{ab}^{(3)}$ can 
give a negative contribution, and it 
basically suppresses the amplitude of the propagators. That is, with 
the new correction $\omega_{ab}$, the propagator in $f(R)$ tends to 
approach the 
one in GR ($\Lambda$CDM), as depicted in dashed lines, and the effect
can become larger at smaller scales. However, 
the  correction itself is very small, and for a 
currently constrained value of the model parameter, 
$|f_{R,0}|\lesssim10^{-4}$, we can safely ignore it at least at the scales of 
our interest.

Note that a negligible contribution of the new correction found above may 
not be always guaranteed in general modified gravity models. Conservatively, 
it would be true only for the models with chameleon-type screening 
mechanisms. Indeed, it is shown analytically and numerically that 
the effect of screening is quite efficient in DGP model, and the model  
can recover the standard GR predictions at relatively larger scales 
(e.g., \cite{2009PhRvD..80j4006S,2009PhRvD..80j4005C,Falck:2014jwa}). 
In this respect, the impact of the high-$k$ correction is non-trivial 
for models with Vainshtein-type mechanisms, and it may give a non-negligible 
contribution to the resummed propagator. We will postpone this issue in a 
separate paper.

%%%%%%%%%%%%%%%%%%%%%%%%%%%%%%%%%%%%%%%%%%%%%%%%%%%%%%%%%%%%%%%%%%%%%%%
%%%%%%%%%%%%%%%%%%%%%%%%%%%%%%%%%%%%%%%%%%%%%%%%%%%%%%%%%%%%%%%%%%%%%%%
\section{Cosmological power spectrum from regularized $\Gamma$ expansion}
\label{sec:regularized_pk}
%%%%%%%%%%%%%%%%%%%%%%%%%%%%%%%%%%%%%%%%%%%%%%%%%%%%%%%%%%%%%%%%%%%%%%%
%%%%%%%%%%%%%%%%%%%%%%%%%%%%%%%%%%%%%%%%%%%%%%%%%%%%%%%%%%%%%%%%%%%%%%%

In this section, based on the resummed propagators, 
we construct the multi-point propagators that consistently reproduce 
the expected high-$k$ and low-$k$ behaviors. Using these {\it regularized}  
propagators, we then give the analytic expression of the real-space power 
spectrum at one-loop order.

%%%%%%%%%%%%%%%%%%%%%%%%%%%%%%%%%%%%%%%%%%%%%%%%%%%%%%%%%%%%%%%%%%%%%%%
%%--%--%--%--%--%--%--%--%--%--%--%--%--%--%--%--%--%--%--%--%--%--%--%
%\subsection{Regularized power spectrum from resummed propagator}
%%--%--%--%--%--%--%--%--%--%--%--%--%--%--%--%--%--%--%--%--%--%--%--%
%%%%%%%%%%%%%%%%%%%%%%%%%%%%%%%%%%%%%%%%%%%%%%%%%%%%%%%%%%%%%%%%%%%%%%%

Let us first derive the expression of the mutli-point propagators in 
which the non-perturbative properties in the high-$k$ limit are effectively 
incorporated. To see the dominant growing-mode contribtuions, 
we may set the 
initial condition, $\Psi_a(\bfk,\tau_0)=\delta_0(\bfk)u_a$ at 
a very early time, $\tau_0\to-\infty$, and  
substitute this into the formal solution,  
Eq.~(\ref{eq:formal_solu}). Then, through 
the definition (\ref{eq:def_Gamma_p}), 
the leading-order expression for the {\it resummed} two- and three-point 
propagators, $\Gamma^{(1)}$ and $\Gamma^{(2)}$, is obtained by taking the 
functional derivative of the formal solution once and twice, respectively. 
The resultant expression ignoring the sub-dominant contribution 
$\omega_{ab}$ becomes similar to those found in the GR case:
%%%%%%%%%%%%%%%%%%%%%%%%%%%%%%%%%%%%%%%%%%%%%%%%%%%%%%%%%%%%%%%%%%%%%%%
\begin{align}
&\Gamma_a^{(1)}(k;\tau)=\left\{g_{a1}(k;\tau,\tau_0)+g_{a2}(k;\tau,\tau_0)\right\}\,
\nonumber\\
&\qquad\qquad\times
\Bigl\langle\exp\left[\int_{\tau_0}^\tau d\tau'\, 
\Xi(\bfk,\tau')\right]\Bigr\rangle,
\label{eq:Gamma1_tree}\\
&\Gamma_a^{(2)}(\bfk_1,\bfk_2;\tau)=
\int_{\tau_0}^\tau d\tau'\,g_{ad}(k;\tau,\tau')
\nonumber\\
&\quad\times
\Bigl[\left\{\gamma_{def}(\bfk_1,\bfk_2;\tau')+
\delta_{d2}\delta_{e1}\delta_{f1}\,\sigma^{(2)}(\bfk_1,\bfk_2;\tau')\right\}
\nonumber\\
&\quad\times
g_{eb}(\bfk_1;\tau',\tau_0)g_{fc}(\bfk_2;\tau',\tau_0)
\Bigr]_{\mathcal{H}}u_bu_c
\nonumber\\
&\qquad\qquad\times
\Bigl\langle\exp\left[\int_{\tau_0}^\tau d\tau'\, 
\Xi(\bfk,\tau')\right]\Bigr\rangle.
\label{eq:Gamma2_tree}
\end{align}
%%%%%%%%%%%%%%%%%%%%%%%%%%%%%%%%%%%%%%%%%%%%%%%%%%%%%%%%%%%%%%%%%%%%%%%
The ensemble average is taken over 
the realizations of the field $\Xi(\bfk)$. 
From the explicit expression 
of $\Xi$ [see Eq.~(\ref{eq:Xi})], the ensemble average of the exponential 
factor becomes
%%%%%%%%%%%%%%%%%%%%%%%%%%%%%%%%%%%%%%%%%%%%%%%%%%%%%%%%%%%%%%%%%%%%%%%
\begin{align}
&\Bigl\langle\exp\left[\int_{\tau_0}^\tau d\tau'\, 
\Xi(\bfk,\tau')\right]\Bigr\rangle
\nonumber\\
&\quad=
\exp\left[-\frac{k^2}{2}\int\frac{dq}{6\pi^2} \,P_0(q) 
\{D_+(q;\tau)-D_+(q;\tau_0)\}^2\right], 
\label{eq:exp_d}
\end{align}
%%%%%%%%%%%%%%%%%%%%%%%%%%%%%%%%%%%%%%%%%%%%%%%%%%%%%%%%%%%%%%%%%%%%%%%
where $P_0$ and $D_+$ are the power spectrum of initial density field 
$\delta_0$, and the linear growth factor, respectively. 
In deriving the expression, 
we assumed Gaussian initial condition, and 
used the fact that the linear velocity-divergence field $\theta$ is 
expressed as $\theta(k;\tau)=\{dD_+(k;\tau)/d\tau\}\,\delta_0(k)$. 
%%%%%%%%%%%%%%%%%%%%%%%%%%%  TABLE  %%%%%%%%%%%%%%%%%%%%%%%%%%%%%%%
\begin{table*}[tb]
\caption{\label{tab:cosmo_params} Cosmological parameters used for 
PT calculations and $N$-body simulations}
\label{tab:nbody_parameters} 
\begin{ruledtabular}
\begin{tabular}{l|cccccccccc}
Name & $L_{\rm box}$\,[$h^{-1}$\,Mpc] & \# of particles & $z_{\rm ini}$ & \# of realizations &
$\Omega_{\rm m}$ & $\Omega_{\Lambda}$ & $\Omega_{\rm b}$ & $h$ & $n_s$ & $\sigma_8$ \\
\hline
\verb|wmap9| & 1,024 & $1,024^3$ & 49 & 1 & 0.281 &  0.719  & 0.0464 & 0.697 & 0.971 & 0.851 \\
\end{tabular}
\end{ruledtabular}
\end{table*}
%%%%%%%%%%%%%%%%%%%%%%%%%%%  TABLE  %%%%%%%%%%%%%%%%%%%%%%%%%%%%%%%

For the late-time evolution dominated by the growing-mode, 
the integral in front of the exponential factor in Eq.~(\ref{eq:Gamma2_tree}) can be reduced
to the second-order standard PT kernel, $F_a^{(2)}$, sometimes referred 
to as $(F_2, G_2)$ (e.g., \cite{Bernardeau:2001qr,Crocce:2005xy}). 
Thus, taking the limit $\tau_0\to-\infty$, we obtain the simplified expression for $\Gamma^{(n)}$:
%%%%%%%%%%%%%%%%%%%%%%%%%%%%%%%%%%%%%%%%%%%%%%%%%%%%%%%%%%%%%%%%%%%%%%%
\begin{align}
&\Gamma_a^{(1)}(k;\tau)=D_a(k;\tau)\,e^{-k^2\sigmad^2(\tau)/2},
\label{eq:Gamma1reg_tree}
\\
&\Gamma_a^{(2)}(\bfk_1,\bfk_2;\tau)=F^{(2)}_a(\bfk_1,\bfk_2;\tau)\,
e^{-k^2\sigmad^2(\tau)/2},
\label{eq:Gamma2reg_tree}
\end{align}
%%%%%%%%%%%%%%%%%%%%%%%%%%%%%%%%%%%%%%%%%%%%%%%%%%%%%%%%%%%%%%%%%%%%%%%
where we define $D_a\equiv g_{a1}+g_{a2}$, which represents 
the linear growth factor $D_+$ and its time derivative $dD_+/d\tau$, 
$D_a=(D_+,\,dD_+/d\tau)$. 
The explicit calculation of the kernel $F_a^{(2)}$ 
taking account of the effect of modified gravity is described in Appendix of 
Ref.~\cite{Taruya:2013quf} (see Eqs.~[A17][A18] of their paper). 
The quantity $\sigmad^2$ is the dispersion of displacement field defined by 
%%%%%%%%%%%%%%%%%%%%%%%%%%%%%%%%%%%%%%%%%%%%%%%%%%%%%%%%%%%%%%%%%%%%%%%
\begin{align}
\sigmad^2=\int\frac{dq}{6\pi^2}\,P_0(q)\,\{D_+(q;\tau)\}^2.
\label{eq:def_sigmad2}
\end{align}
%%%%%%%%%%%%%%%%%%%%%%%%%%%%%%%%%%%%%%%%%%%%%%%%%%%%%%%%%%%%%%%%%%%%%%%

With the two- and three-point propagators given above, 
Eq.~(\ref{eq:Pk_regpt}) truncating at $n=2$ can give
the so-called one-loop power spectrum. However, a naive use of 
Eqs.~(\ref{eq:Gamma1reg_tree}) and (\ref{eq:Gamma2reg_tree}) 
may lead to a small flaw in the PT calculation in a sense that 
one cannot reproduce the standard PT results at low-$k$. 
To reproduce the standard PT result, 
the higher-order correction needs to be included consistently 
in the prediction of propagators. 
Ref.~\cite{Bernardeau:2011dp} has proposed a novel {\it regularization} scheme
for propagators that allows us to interpolate the standard PT result 
and the expected resummed behavior at high-$k$. With this 
regularized treatment, the power spectrum at one-loop order is expressed as
%%%%%%%%%%%%%%%%%%%%%%%%%%%%%%%%%%%%%%%%%%%%%%%%%%%%%%%%%%%%%%%%%%%%%%%
\begin{widetext}
\begin{align}
P_{ab}(k;\tau)=\Gamma_{a,{\rm reg}}^{(1)}(k;\tau)\Gamma_{b,{\rm reg}}^{(1)}(k;\tau)\,
P_0(k)+ 2\int\frac{d^3\bfq}{(2\pi)^3}\,
\Gamma^{(2)}_{a,{\rm reg}}(\bfq,\bfk-\bfq;\tau)
\Gamma^{(2)}_{b,{\rm reg}}(\bfq,\bfk-\bfq;\tau)\,P_0(q)P_0(|\bfk-\bfq|)
\label{eq:reg_pk}
\end{align}
\end{widetext}
%%%%%%%%%%%%%%%%%%%%%%%%%%%%%%%%%%%%%%%%%%%%%%%%%%%%%%%%%%%%%%%%%%%%%%%
with the regularized propagators given by
%%%%%%%%%%%%%%%%%%%%%%%%%%%%%%%%%%%%%%%%%%%%%%%%%%%%%%%%%%%%%%%%%%%%%%%
\begin{align}
&\Gamma^{(1)}_{a,{\rm reg}}(k;\tau)=
\left[D_a(k;\tau)\left\{1+\frac{k^2\sigmad^2}{2}\right\}+
\overline{\Gamma}_{a,{\rm 1\mbox{-}loop}}^{(1)}(k;\tau)\right]
\nonumber\\
&\quad\qquad\qquad\times e^{-k^2\sigmad^2/2}
\label{eq:G1reg}
\\
&\Gamma^{(2)}_{a,{\rm reg}}(\bfq,\bfk-\bfq;\tau)=F_a^{(2)}(\bfq,\bfk-\bfq;\tau)\,
e^{-k^2\sigmad^2/2}.
\label{eq:G2reg}
\end{align}
%%%%%%%%%%%%%%%%%%%%%%%%%%%%%%%%%%%%%%%%%%%%%%%%%%%%%%%%%%%%%%%%%%%%%%%
Hereafter, we call this regularized PT treatment RegPT.

For the PT calculation at one-loop order, 
the regularized three-point propagator $\Gamma^{(2)}_{a,{\rm reg}}$ 
is identical to the one given in Eq.~(\ref{eq:Gamma2reg_tree}), and only 
the two-point propagaor $\Gamma^{(1)}_{a,{\rm reg}}$ gets some corrections.  
The function $\overline{\Gamma}^{(1)}_{a,{\rm 1\mbox{-}loop}}$ represents 
the one-loop correction to the two-point propagator computed with standard 
PT. In this paper, to compute it in the $f(R)$ gravity 
model below, we will use the numerical PT scheme developed by 
Ref.~\cite{2009PhRvD..79j3526H}\footnote{The numerical PT scheme in 
\cite{2009PhRvD..79j3526H} solves the moment equations coupled with 
the propagator $G_{ab}$. For the perturbative treatment, this scheme 
reproduces the standard PT results at one-loop order. From the 
standard PT result of $G_{ab}$, we can obtain
$\overline{\Gamma}_{a,1\mbox{-}{\rm loop}}^{(1)}$ through the relation
$(G_{ab}-g_{ab})u_b=\overline{\Gamma}_{a,1\mbox{-}{\rm loop}}^{(1)}$, where 
$g_{ab}$ is the linear propagator and $u_{a}=(1,1)$. }. 
In the low-$k$ limit $k\sigmad\ll1$, 
the exponential factor in Eq.~(\ref{eq:G1reg}) can be expanded 
and we recover the standard PT results, i.e., $\Gamma^{(1)}_{a,{\rm reg}}\simeq
D_a+\overline{\Gamma}_{a,{\rm 1\mbox{-}loop}}^{(1)}+\mathcal{O}(k^4\sigmad^4)$. 
Further, it is known in the GR case that  the function 
$\overline{\Gamma}^{(1)}_{a,{\rm 1\mbox{-}loop}}$ 
behaves like $\overline{\Gamma}^{(1)}_{a,{\rm 1\mbox{-}loop}}\to
-(k^2\sigmad^2/2)D_a$ in the high-$k$ limit, and thus 
Eq.~(\ref{eq:G1reg}) reproduces Eq.~(\ref{eq:Gamma1reg_tree}). 
In general, the latter property does not necessarily hold 
in modified gravity models. Rather, in the presence of the screening mechanism, 
it will differ from the one in the GR case. In this respect, 
the proposition given in Eq.~(\ref{eq:G1reg}) includes a small flaw, and 
may produce an error in the prediction of propagator. Nevertheless, 
we will see in next section that the final impact of this effect is 
negligible and does not seriously affect the prediction of power spectrum.

%%%%%%%%%%%%%%%%%%%%%%%%%%%%%%%%%%%%%%%%%%%%%%%%%%%%%%%%%%%%%%%%%%%%%%%
%%%%%%%%%%%%%%%%%%%%%%%%%%%%%%%%%%%%%%%%%%%%%%%%%%%%%%%%%%%%%%%%%%%%%%%
\section{Comparison with $N$-body simulations}
\label{sec:RegPT}
%%%%%%%%%%%%%%%%%%%%%%%%%%%%%%%%%%%%%%%%%%%%%%%%%%%%%%%%%%%%%%%%%%%%%%%
%%%%%%%%%%%%%%%%%%%%%%%%%%%%%%%%%%%%%%%%%%%%%%%%%%%%%%%%%%%%%%%%%%%%%%%

We are now in a position to compare the PT predictions with $N$-body 
simulations. We use the simulation data set kindly provided by Baojiu Li. 
The data set of $N$-body simulations were created by the $N$-body 
code, {\tt ECOSMOG} \cite{Li:2011vk}, 
which is a modified version of the mesh-based $N$-body code, 
{\tt RAMSES} \cite{Teyssier:2001cp}. With this code, the simulation 
data were created in both GR and $f(R)$ gravity adopting the functional 
form in Eq.~(\ref{eq:fR_model_nbody}). 
The cosmological paramters used in the $N$-body simulations are 
determined by nine-year WMAP results \cite{Hinshaw:2012aka}, and 
the initial conditions were generated by {\tt mpgraphic} 
\cite{Prunet:2008lr} 
at redshift $z_{\rm ini}=49$, assuming the Gaussianity of initial density 
field. Hereafter we refer the simulation 
data to {\tt wmap9}. Basic parameters of $N$-body simulations are 
summarized in Table \ref{tab:nbody_parameters}. 
In the analysis presented below, we 
consider GR and $f(R)$ gravity with $|f_{R0}|=10^{-4}$, and   
use the output data at $z=0$, $0.5$, $1$, and $2$.

%%%%%%%%%%%%%%%%%%%%%%%%%%%%%%%%%%%%%%%%%%%%%%%%%%%%%%%%%%%%%%%%%%%%%%%
\begin{figure*}[t]
%\hspace*{-0.8cm}
\includegraphics[width=8cm]{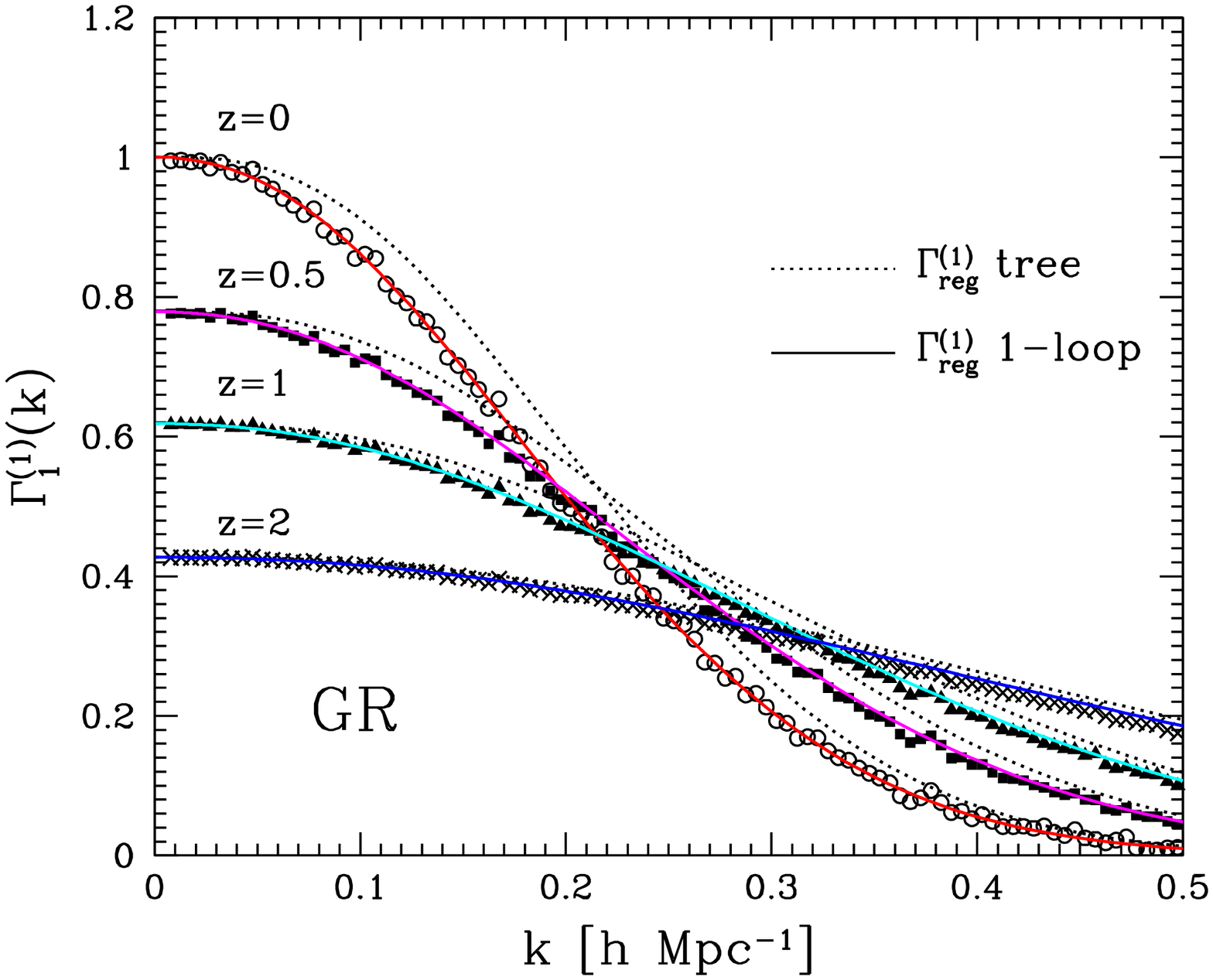}
\includegraphics[width=8cm]{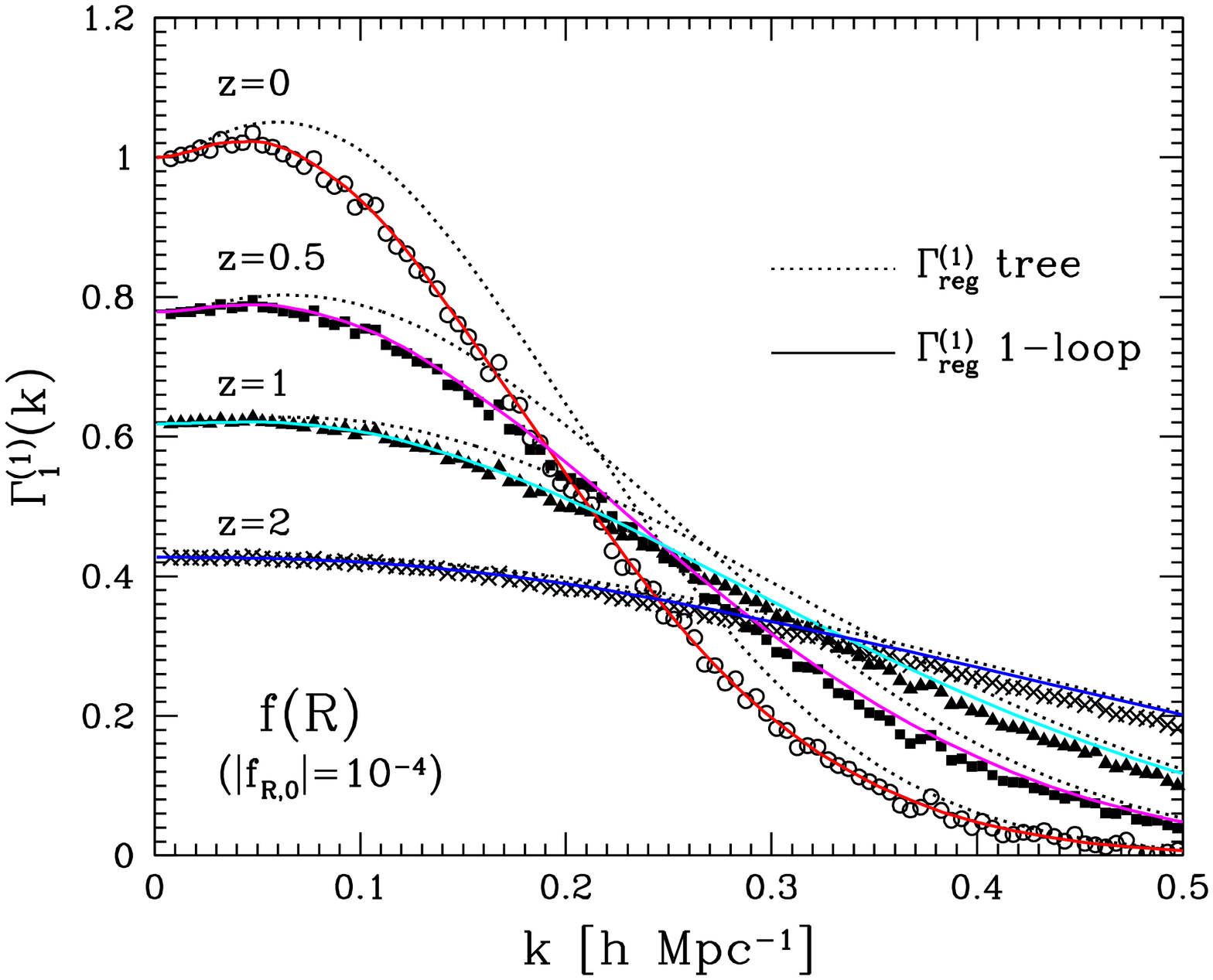}

\vspace*{-1.5cm}

\hspace*{-0.5cm}
\includegraphics[width=8cm]{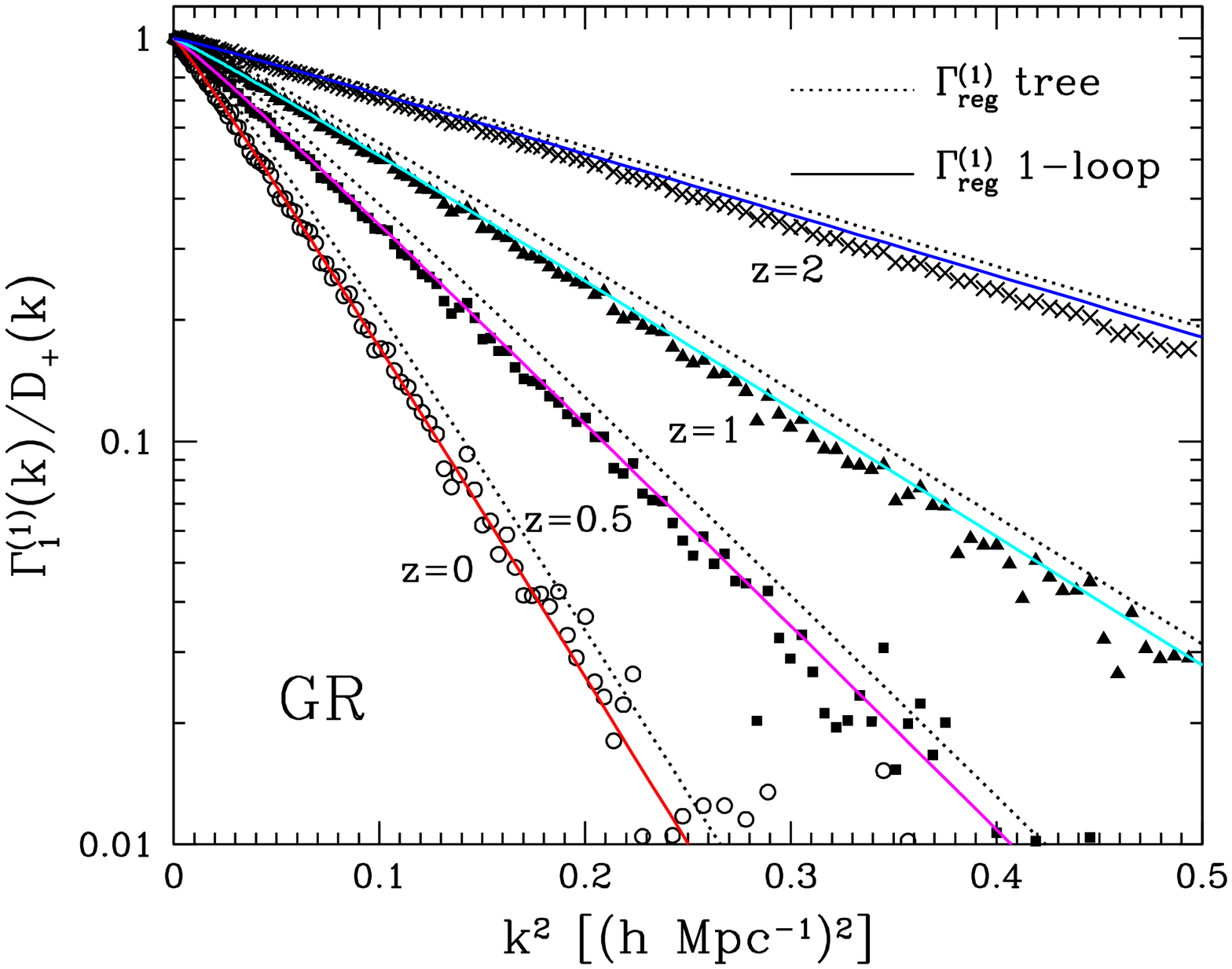}
\includegraphics[width=8cm]{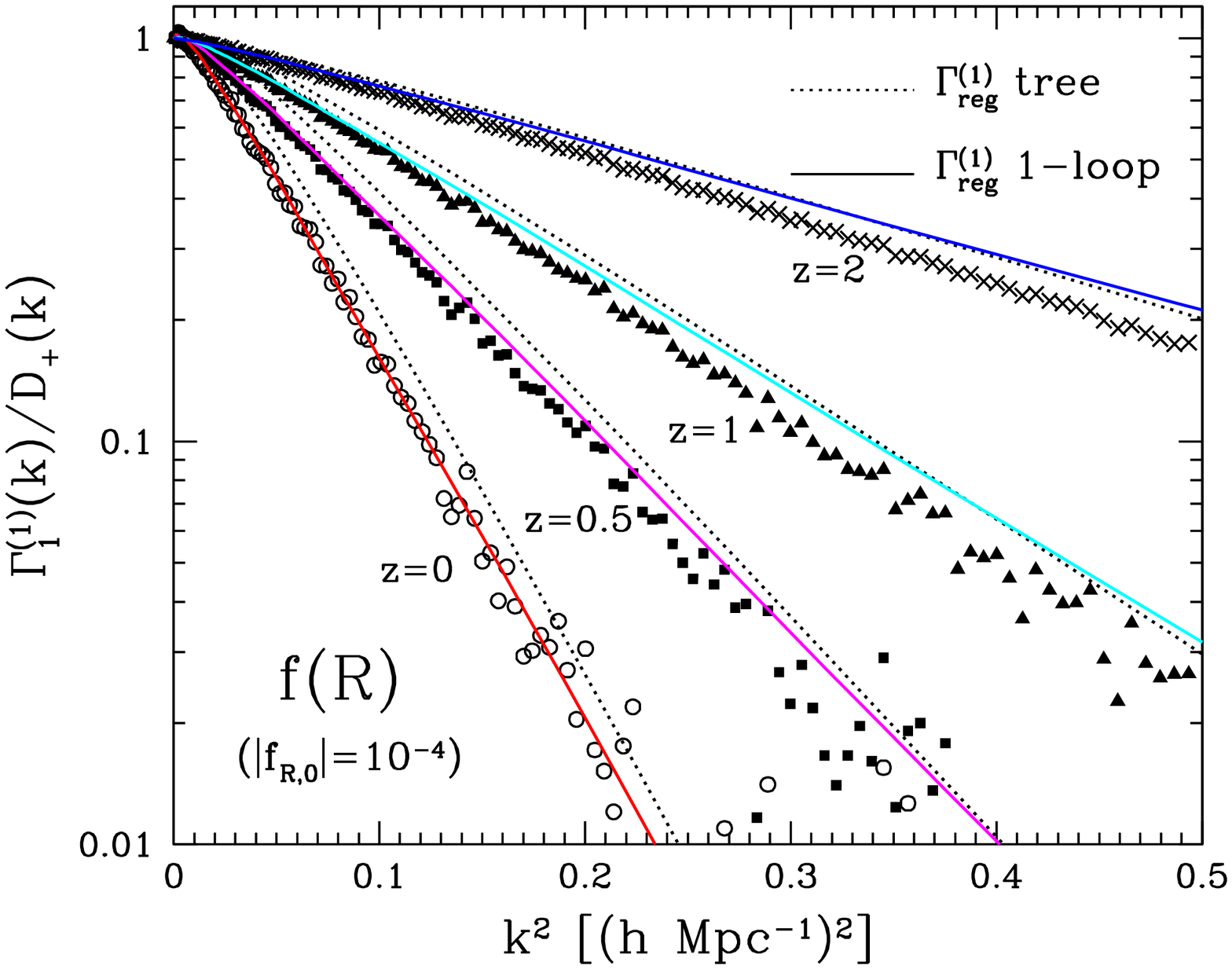}

\vspace*{-0.5cm}

\caption{Two-point propagator of density field, $\Gamma^{(1)}_1(k)$, measured in $N$-body simulations at $z=0$, $0.5$, $1$, and $2$. Left and right panels respectively shows the results in GR and $f(R)$ gravity with $|f_{R,0}|=10^{-4}$. In top panels, the propagators are normally plotted as function of wavenumber. On the other hand, to clearly show the high-$k$ limit behaviors, 
bottom panels plot the normalized propagators $\Gamma^{(1)}/D_+$ as 
function of $k^2$ in semi-log scale. 
In each panel, solid and dotted lines are the regularized propagators at 
tree level and one-loop order, respectively 
[Eqs.~(\ref{eq:Gamma1reg_tree}) and (\ref{eq:G1reg})]. 
\label{fig:Gamma1_nbody}}
\end{figure*}
%%%%%%%%%%%%%%%%%%%%%%%%%%%%%%%%%%%%%%%%%%%%%%%%%%%%%%%%%%%%%%%%%%%%%%%

%%--%%--%%--%%--%%--%%--%%--%%--%%--%%--%%--%%--%%--%%--%%--%%--%%--%%%
\subsection{Propagator}
\label{subsec:Gamma1_nbody}
%%--%%--%%--%%--%%--%%--%%--%%--%%--%%--%%--%%--%%--%%--%%--%%--%%--%%%

Since the power spectrum calculation with $\Gamma$ expansion heavily relies on 
the prescription for propagators, let us first check their behavior in 
$N$-body simulations. 
Fig.~\ref{fig:Gamma1_nbody} plots the measured results of the two-point 
propagator for the density field in $N$-body simulations, $\Gamma_1^{(1)}$ 
in the cases of GR (left) and $f(R)$ gravity (right). 
Since the initial conditions in $N$-body simulations are Gaussian, 
the propagators for density field are easily measured by taking the 
cross correlation between the evolved and initial density fields and 
dividing it by the linear power spectrum used for initial condition 
generator \cite{Crocce:2005xz}. 
Top panels show the propagators plotted in linear scales. 
In bottom panels, to clearly see the damping behaviors at small scales,  
the propagators are divided by the linear growth factor, $\Gamma_1^{(1)}/D_+$, 
and are plotted as function of wavenumber squared $k^2$ 
in semi-logarithmic scales.

As we see from bottom panels, the measured propagators exhibit 
the exponential damping behaviors in both GR and $f(R)$ gravity. 
The results are then in a good agreement with the theoretical predictions 
depicted as solid lines, 
which represent the regularized propagators at one-loop order, 
$\Gamma^{(1)}_{\rm reg}$ [Eqs.~(\ref{eq:G1reg})]. For reference, we also plot the 
tree-level prediction given in Eq.~(\ref{eq:Gamma1reg_tree}), 
which degrades the 
agreement with $N$-body simulations, as expected from previous 
studies in GR. 
Note here that we do not indicate the error in $N$-body simulations, 
since the plotted results are the ratio of measured values, 
and the cosmic variance cancels out at the leading order. Only with one 
realization data, we could not properly estimate the higher-order 
cosmic variance error. 
Nevertheless, the reasonable agreement with prediction 
implies that the propagators 
were reliably estimated in $N$-body simulation, and measured results seem 
robust against numerical systematics.

A closer look at bottom panels, however,  reveals a small discrepancy between 
predictions and simulations. This is rather manifest 
at higher redshifts in both GR and $f(R)$ cases. 
Since both the one-loop and tree-level predictions become 
closer at higher redshifts, the discrepancy would not be ascribed to 
the breakdown of PT treatment. Rather, we suspect a small systematic 
error in the $N$-body simulations. 
A part of the reasons may come from the fact 
that the initial conditions were generated with 
the Zel'dovich dynamics, which is known to produce 
a transient phenomenon due to the non-vanishing decaying mode 
\cite{Crocce:2006ve}. 
Another reason may be the lack of force resolution. 
Generally, simulations with insufficient 
force resolution lead to the incorrect displacements of particles, and thus 
the cross correlation between the evolved density fields and the linear density 
field is partly suppressed. 
This results in a systematic underestimation of propagators, and with 
the same force resolutions, $N$-body simulation starting at higher redshift 
tends to suffer from this systematics. Our previous 
study reveals that the propagators are more sensitive 
to the force error at high 
redshifts than the power spectrum. See Ref.~\cite{Bernardeau:2012ux} 
for more detailed discussion.

Apart from the tiny systematics at high redshifts, the RegPT treatment of the 
propagators successfully reproduces the overall trend of $N$-body simulations, 
and within the precision of 
agreement, the RegPT calculation 
is expected to give an accurate description for the 
power spectrum and correlation function, which will be compared with $N$-body 
results below. 

%%%%%%%%%%%%%%%%%%%%%%%%%%%%%%%%%%%%%%%%%%%%%%%%%%%%%%%%%%%%%%%%%%%%%%%
\begin{figure*}[t]
%\hspace*{-0.8cm}
\includegraphics[width=8cm]{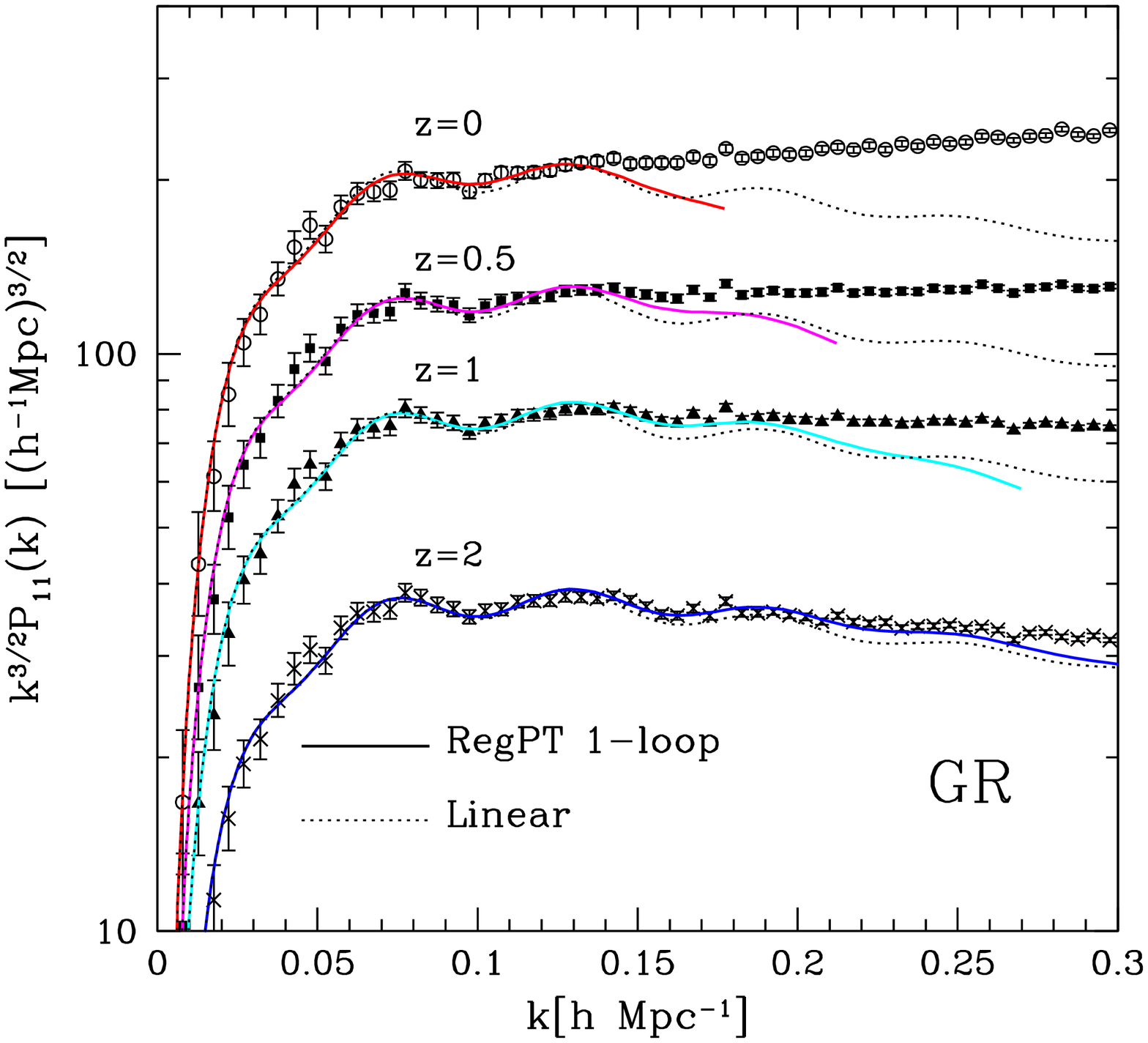}
\includegraphics[width=8cm]{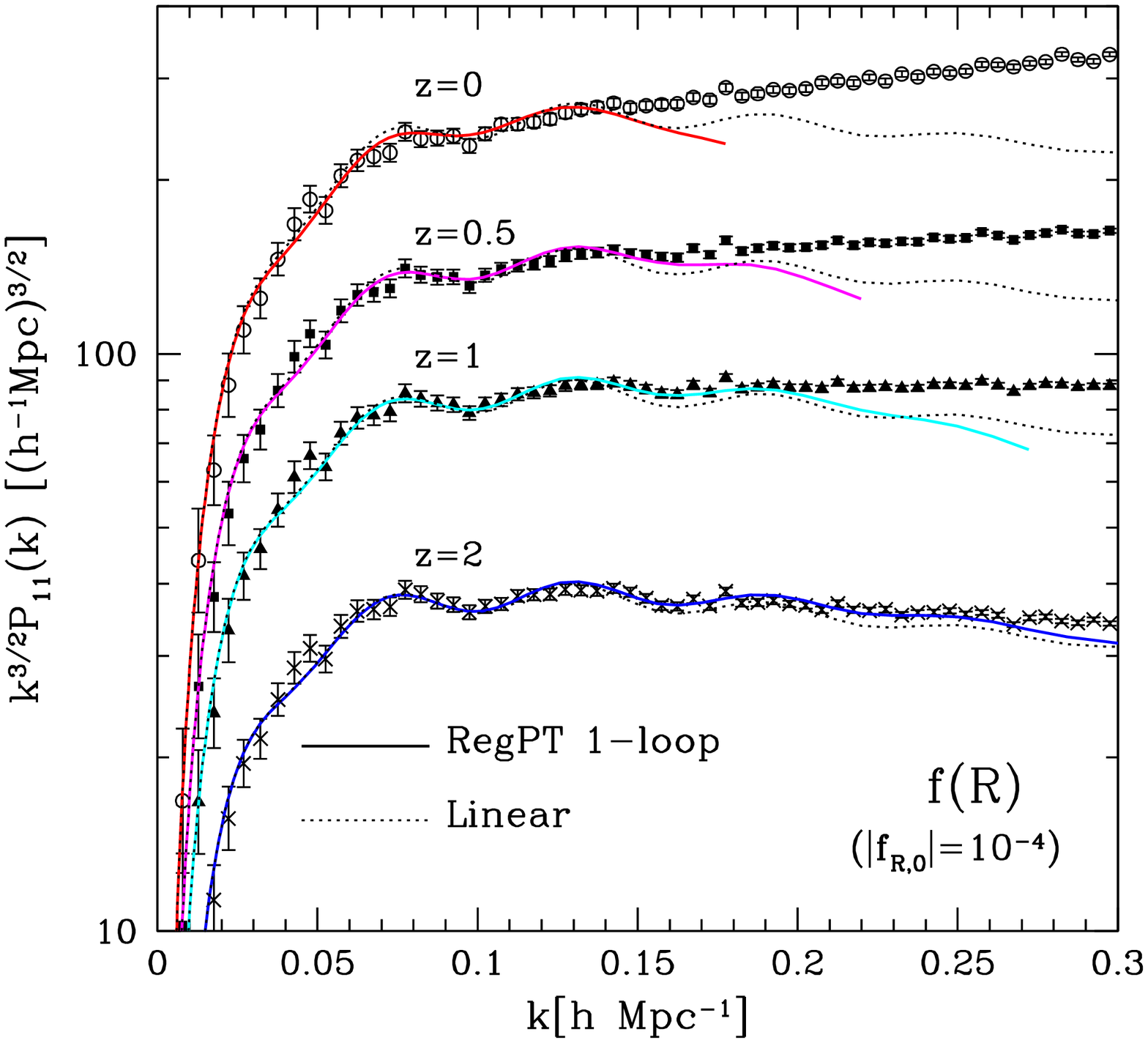}

\caption{Power spectrum of the 
density field in real space multiplied by $k^{3/2}$, 
$k^{3/2}\,P_{11}(k)$, at $z=0$, $0.5$, $1$, and $2$ (from top to bottom). 
Left panel shows the results in GR, while right panel presents the cases in $f(R)$ gravity with $|f_{R,0}|=10^{-4}$. Solid and dotted lines are RegPT 
predictions at one-loop and linear theory predictions, 
respectively. Note that the 
errorbars indicated in $N$-body results are the dispersion of the 
power spectrum amplitude over the modes in each Fourier bin. 
\label{fig:Pkreal_nbody}}
\end{figure*}
%%%%%%%%%%%%%%%%%%%%%%%%%%%%%%%%%%%%%%%%%%%%%%%%%%%%%%%%%%%%%%%%%%%%%%%

%%--%%--%%--%%--%%--%%--%%--%%--%%--%%--%%--%%--%%--%%--%%--%%--%%--%%%
\subsection{Power spectrum}
\label{subsec:pk_nbody}
%%--%%--%%--%%--%%--%%--%%--%%--%%--%%--%%--%%--%%--%%--%%--%%--%%--%%%

We next present the comparison of power spectra between 
$N$-body simulations and PT calculations. 
Fig.~\ref{fig:Pkreal_nbody} presents the power spectra of density 
field multiplied by $k^{3/2}$, i.e., $k^{3/2}\,P_{11}$. The RegPT predictions 
at one-loop order are depicted as solid 
lines, and just for reference, we also plot the linear theory 
predictions in dotted lines.

Because of the scale-dependent linear growth, 
the resultant amplitude of power spectra in $f(R)$ gravity 
becomes relatively larger than that in GR, 
and the differences are manifest at lower redshifts. The RegPT prediction at
one-loop order reproduces the $N$-body results fairly well in both cases 
at the weakly nonlinear scales, where we still clearly see 
the acoustic signature of power spectrum. Although 
the RegPT prediction eventually deviates from the $N$-body result 
at small scales, the range of agreement between $N$-body and PT results 
is almost the same in both GR and $f(R)$ gravity. 
With the resummed PT calculation, 
the nonlinear smearing effect of the BAOs (e.g., \cite{Jeong:2006xd,Crocce:2007dt,Eisenstein:2006nj}), 
which can be seen in the $N$-body results 
even at large scales (e.g., \cite{Seo:2005ys,Nishimichi:2008ry}), is better 
described by the PT results, and the prediction shown here 
is contrasted with the standard 
PT prediction (see Sec.~\ref{sec:discussion}). This point is indeed crucial 
in accurately predicting the shape and location of the baryon acoustic peak 
in the correlation function, which we will discuss below.

%%%%%%%%%%%%%%%%%%%%%%%%%%%%%%%%%%%%%%%%%%%%%%%%%%%%%%%%%%%%%%%%%%%%%%%
\begin{figure*}[t]

\vspace*{-3.0cm}

\includegraphics[width=14cm]{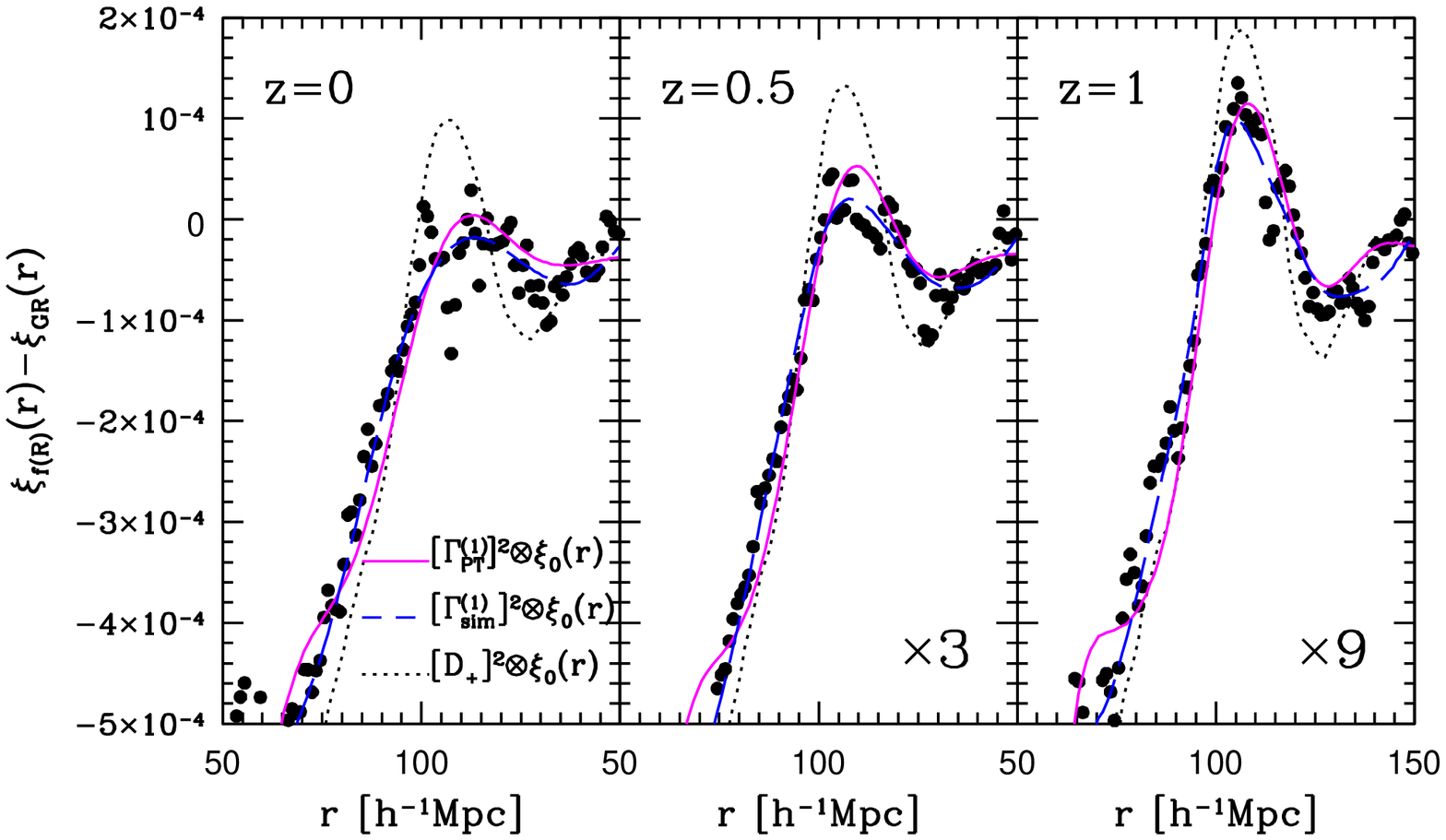}

\vspace*{-2.2cm}

\caption{Difference of the real-space correlation function between $f(R)$ 
gravity with $|f_{R,0}|=10^{-4}$ and GR, 
$\Delta\xi(r)=\xi_{f(R)}(r)-\xi_{\rm GR}(r)$. 
From left to right panels, the results at $z=0$, $0.5$, and $1$ are shown. 
In each panel, filled circles represent the results from $N$-body simulations, 
while the solid, dashed, and dotted lines are estimated from the 
tree-level expression of the correlation function, 
$\xi(r)\simeq[\Gamma^{(1)}]^2\otimes\xi_0(r)$. 
Here, $\xi_0$ indicates the initial 
correlation function for which we computed numerically with 
the random initial data of $N$-body simulation. For the propagator 
$\Gamma^{(1)}$, the regularized one-loop propagator computed analytically 
is used in solid lines, while the dashed lines adopt the 
one directly measured from $N$-body simulations. Finally, the dotted lines 
are obtained with the linear theory prediction, 
just replacing $\Gamma^{(1)}$ with the linear growth factor $D_+$. Note that 
for clarity, the results at $z=0.5$ and $1$ are 
multiplied by the factor $3$ and $9$, respectively. 
\label{eq:xi_diff}}
\end{figure*}
%%%%%%%%%%%%%%%%%%%%%%%%%%%%%%%%%%%%%%%%%%%%%%%%%%%%%%%%%%%%%%%%%%%%%%%

%%--%%--%%--%%--%%--%%--%%--%%--%%--%%--%%--%%--%%--%%--%%--%%--%%--%%%
\subsection{Correlation function}
\label{subsec:xi_nbody}
%%--%%--%%--%%--%%--%%--%%--%%--%%--%%--%%--%%--%%--%%--%%--%%--%%--%%%

The predictions for the correlation function are simply obtained from the power 
spectrum: 
%%%%%%%%%%%%%%%%%%%%%%%%%%%%%%%%%%%%%%%%%%%%%%%%%%%%%%%%%%%%%%%%%%%%%%%
\begin{align}
\xi(r)=\int\frac{dk\,k^2}{2\pi^2}\,P_{11}(k)\frac{\sin(k r)}{kr}.
\end{align}
%%%%%%%%%%%%%%%%%%%%%%%%%%%%%%%%%%%%%%%%%%%%%%%%%%%%%%%%%%%%%%%%%%%%%%%
In the standard PT case, because of the un-regularized UV behavior, 
the above integral cannot be reliably estimated. But now, 
with the RegPT treatment, we are able to evaluate the 
correlation function, which can be directly compared 
with the $N$-body results.

However, only with the single realization 
data, a reliable estimation of the correlation function is rather difficult in 
$N$-body simulations.  
This is because the measured amplitude of the correlation function is 
strongly correlated between different scales. Then, due to the cosmic 
variance error, a small deficit in the 
initial power spectrum in the $N$-body realization, especially at low-$k$,  
can coherently affect the
shape and amplitude of correlation function over the whole scales, 
and the measured result of correlation function 
can drastically differ from what we would 
expect from the {\it true} input power spectrum. 
The proper way to overcome such a problem
is to use a large number of realizations taking ensemble 
averages over a large number of 
different realizations. For the problem we are interested in, however, 
we can still make a meaningful comparison 
with the single realization data by 
combining the $N$-body catalogs in GR and $f(R)$ gravity.
Let us take the difference:
%%%%%%%%%%%%%%%%%%%%%%%%%%%%%%%%%%%%%%%%%%%%%%%%%%%%%%%%%%%%%%%%%%%%%%%
\begin{align}
\Delta\xi(r)=\xi_{f(R)}(r)-\xi_{\rm GR}(r).
\end{align}
%%%%%%%%%%%%%%%%%%%%%%%%%%%%%%%%%%%%%%%%%%%%%%%%%%%%%%%%%%%%%%%%%%%%%%%
Since the two catalogs were created with the same random seed, 
a non-zero value of $\Delta\xi$ implies the systematic 
difference of the dynamics between GR and $f(R)$ gravity. 
On the scales we are interested in, 
the leading-order term in $\Gamma$ expansion is known to play a dominant
role for the nonlinear effect on the correlation function 
(e.g., \cite{Crocce:2007dt,Taruya:2009ir,Taruya:2012ut}). 
Then, from Eq.~(\ref{eq:reg_pk}), the PT prediction gives 
%%%%%%%%%%%%%%%%%%%%%%%%%%%%%%%%%%%%%%%%%%%%%%%%%%%%%%%%%%%%%%%%%%%%%%%
\begin{align}
[\Delta\xi(r)]_{\rm PT}\,\simeq \,
\left([\Gamma_{{\rm reg},f(R)}^{(1)}]^2-[\Gamma_{\rm reg,GR}^{(1)}]^2\right)
\otimes \xi_0(r),
\label{eq:delta_xi_PT}
\end{align}
%%%%%%%%%%%%%%%%%%%%%%%%%%%%%%%%%%%%%%%%%%%%%%%%%%%%%%%%%%%%%%%%%%%%%%%
where the symbol $\otimes$ indicates a convolution. The function $\xi_0$ 
represents the correlation function of the input linear density field,  
which can be computed with the random initial data of $N$-body simulation. 
Thus, plugging the prediction of the regularized two-point propagators 
into the above, the predicted value of $[\Delta\xi]_{\rm PT}$ is directly 
compared with the measured value.

Fig.~\ref{eq:xi_diff} shows the results of the comparison at $z=0$, $0.5$ and 
$1$ (from left to right panels). The measured results 
of $\Delta\xi$ are plotted as filled circles, while the PT predictions 
with the regularized one-loop propagator 
are depicted as solid magenta lines. Note that for clarity, the 
results at $z=0.5$ and $1$ are multiplied by the factor $3$ and $9$, 
respectively. We do not plot here the result at $z=2$, since the differences 
are quite small. The RegPT prediction 
fairly traces the measured result of $\Delta\xi$ quite well, 
and is consistent with the $N$-body estimates of Eq.~(\ref{eq:delta_xi_PT}) 
depicted as blue dashed lines, in which 
we directly use the two-point propagator $\Gamma^{(1)}_{\rm reg}$ measured 
in $N$-body simulations. 
For comparison, we also plot the linear theory prediction (dotted), 
where the two-point propagators in Eq.~(\ref{eq:delta_xi_PT}) are simply 
replaced with the linear growth factors, $D_+$. Clearly, the 
linear theory prediction fails to reproduce the $N$-body trend. 
It is known that the baryon acoustic peak tends to be smeared by the 
nonlinear gravitational growth, and as a result, the location of the 
acoustic peak is slightly shifted (e.g., \cite{Crocce:2007dt,Smith:2007gi}). 
The behavior of $\Delta\xi$ 
seen in both PT predictions and $N$-body results basically follows 
this trend, while a sharp feature in $\Delta\xi$ still remains in 
linear theory predictions. Thus, the PT prediction with $\Gamma$ 
expansion can better describe the correlation function, and even at 
one-loop order it can be used as an accurate theoretical template.

%%%%%%%%%%%%%%%%%%%%%%%%%%%%%%%%%%%%%%%%%%%%%%%%%%%%%%%%%%%%%%%%%%%%%%%
\begin{figure*}[t]
%\hspace*{-0.8cm}
\includegraphics[width=8cm]{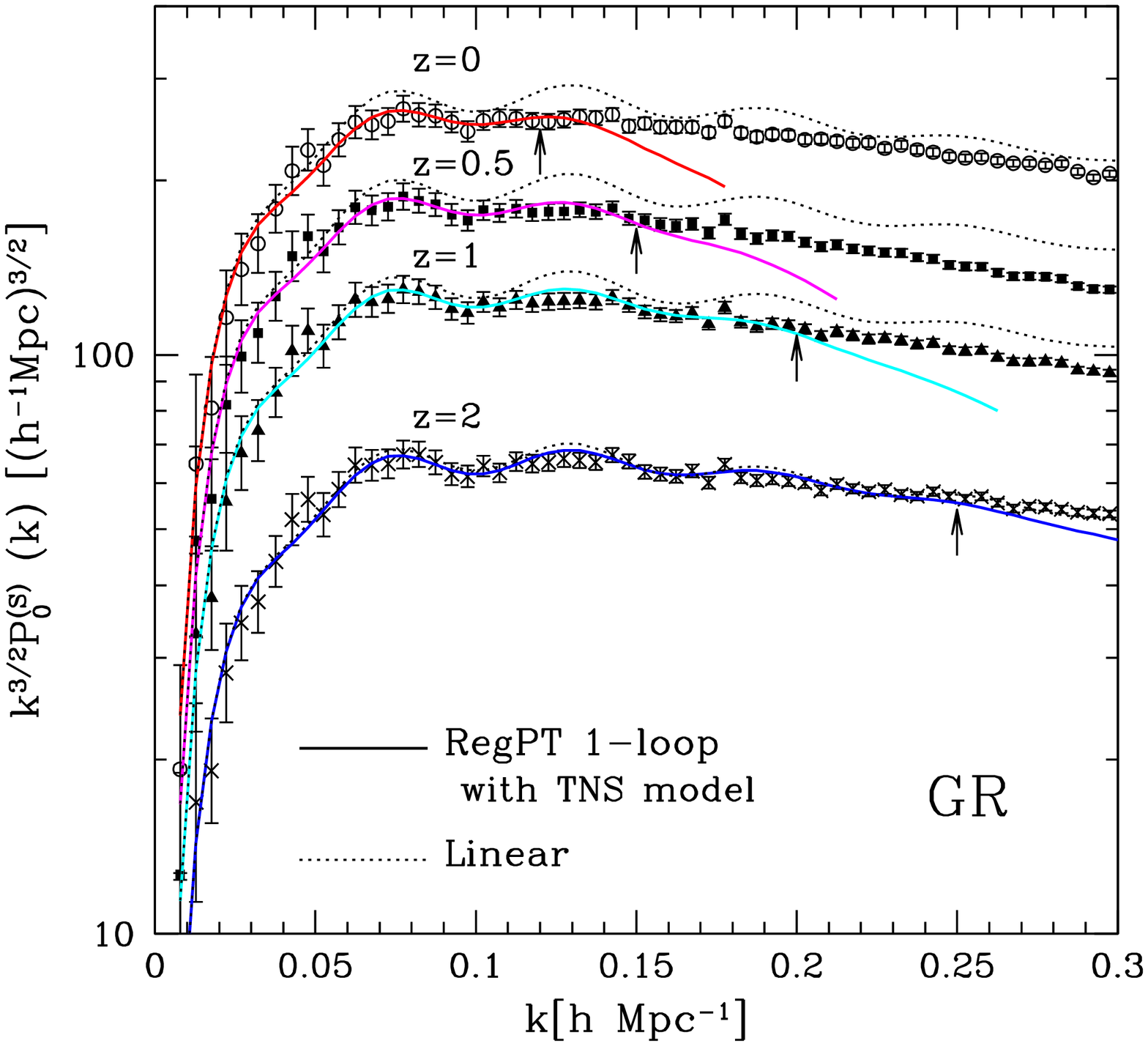}
\includegraphics[width=8cm]{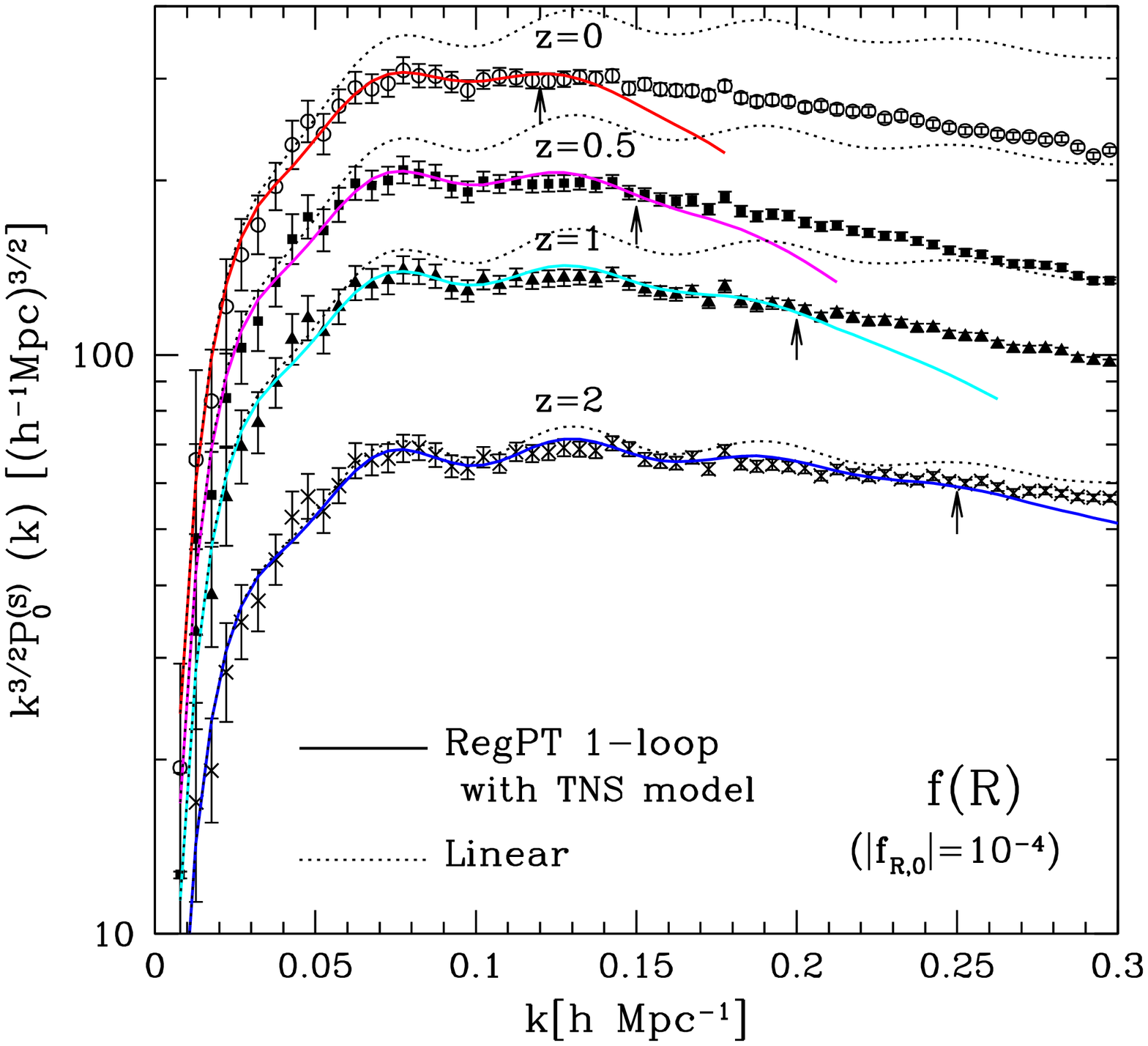}

\vspace*{-0.5cm}

\includegraphics[width=8cm]{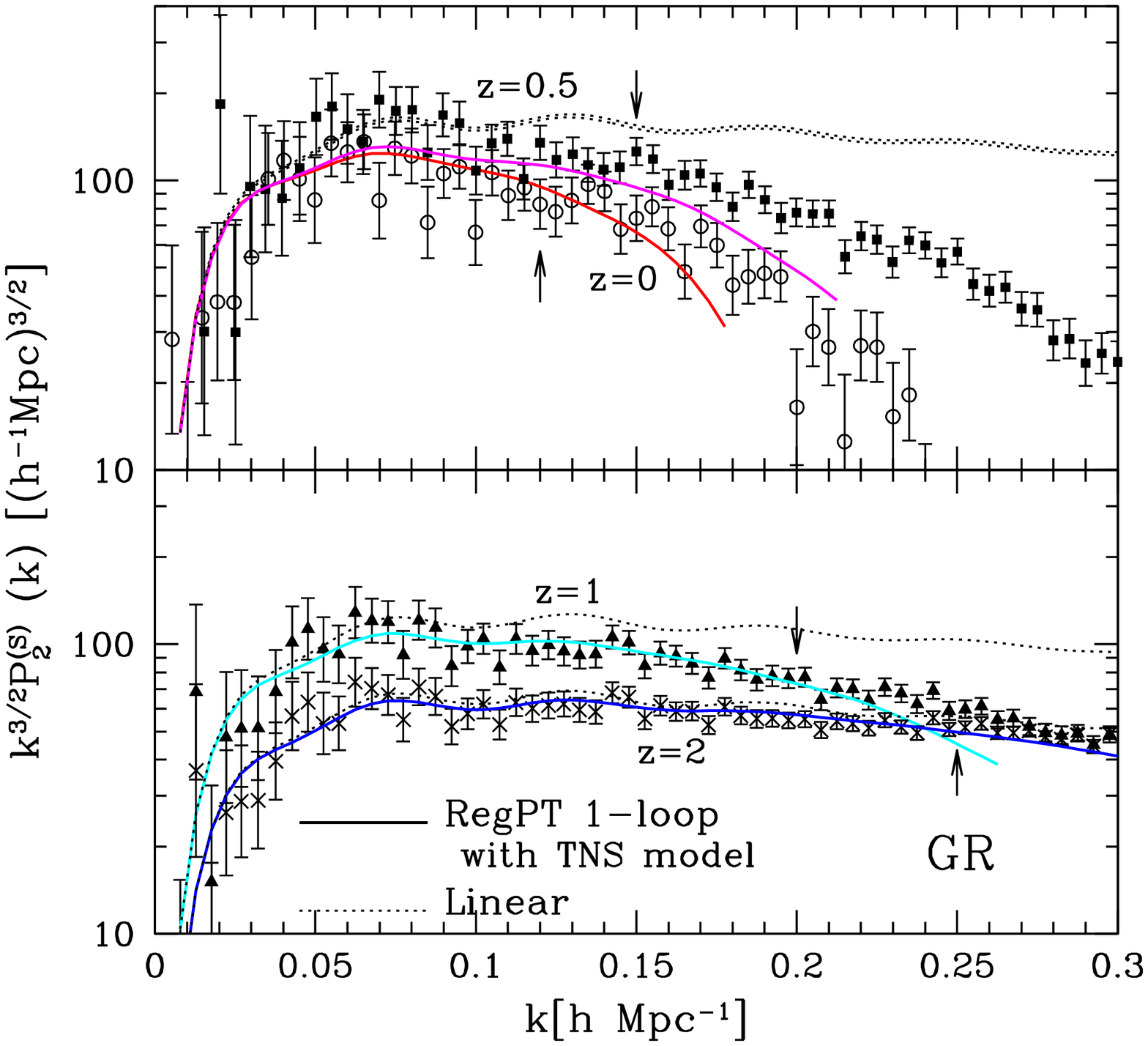}
\includegraphics[width=8cm]{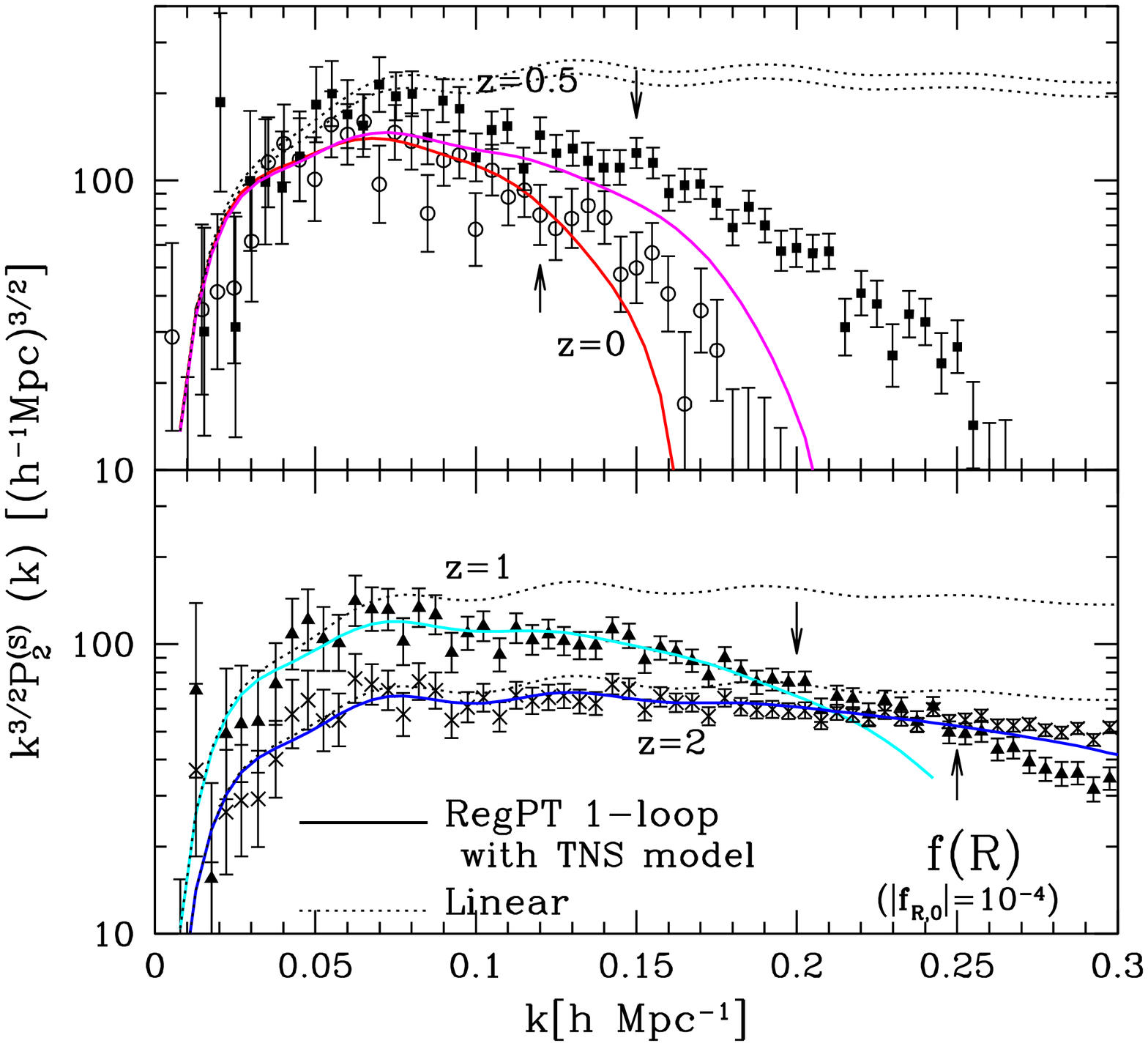}

\caption{Monopole ($\ell=0$, top) and quadrupole ($\ell=2$, bottom) moments of 
redshift-space power spectrum at $z=0$, $0.5$, $1$, and $2$. Left panel shows the 
results in GR, while right panel presents the cases in $f(R)$ gravity with $|f_{R,0}|=10^{-4}$. In each panel, dotted lines represent the linear theory calculations based on the 
Kaiser formula, while the solid lines are the RegPT one-loop results based on the 
TNS model of RSD.
\label{fig:pkred0_pkred2}}
\end{figure*}
%%%%%%%%%%%%%%%%%%%%%%%%%%%%%%%%%%%%%%%%%%%%%%%%%%%%%%%%%%%%%%%%%%%%%%%

%%%%%%%%%%%%%%%%%%%%%%%%%%%%%%%%%%%%%%%%%%%%%%%%%%%%%%%%%%%%%%%%%%%%%%%
\begin{figure*}[t]
%\hspace*{-0.8cm}

\includegraphics[width=10cm]{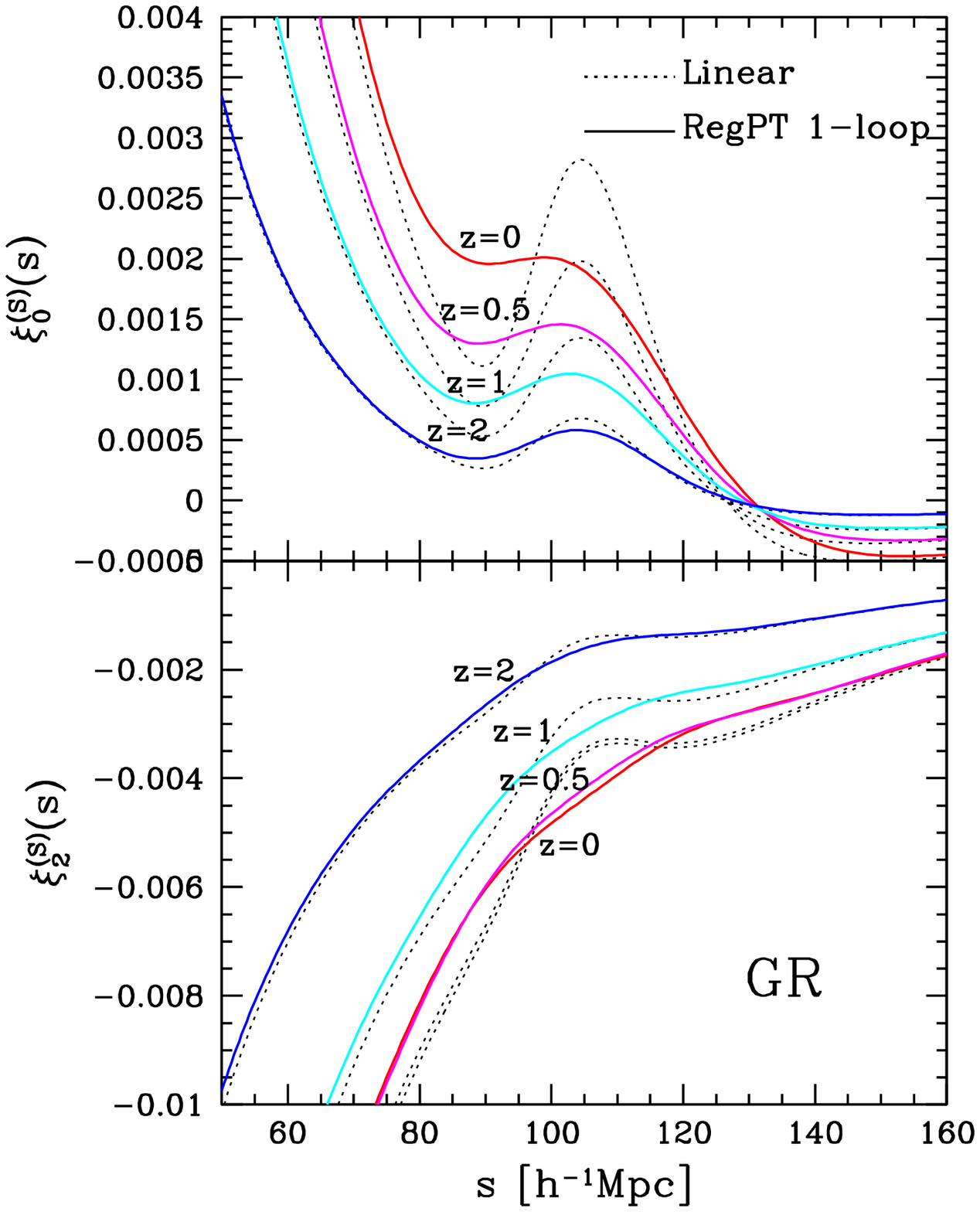}
\hspace*{-2.5cm}
\includegraphics[width=10cm]{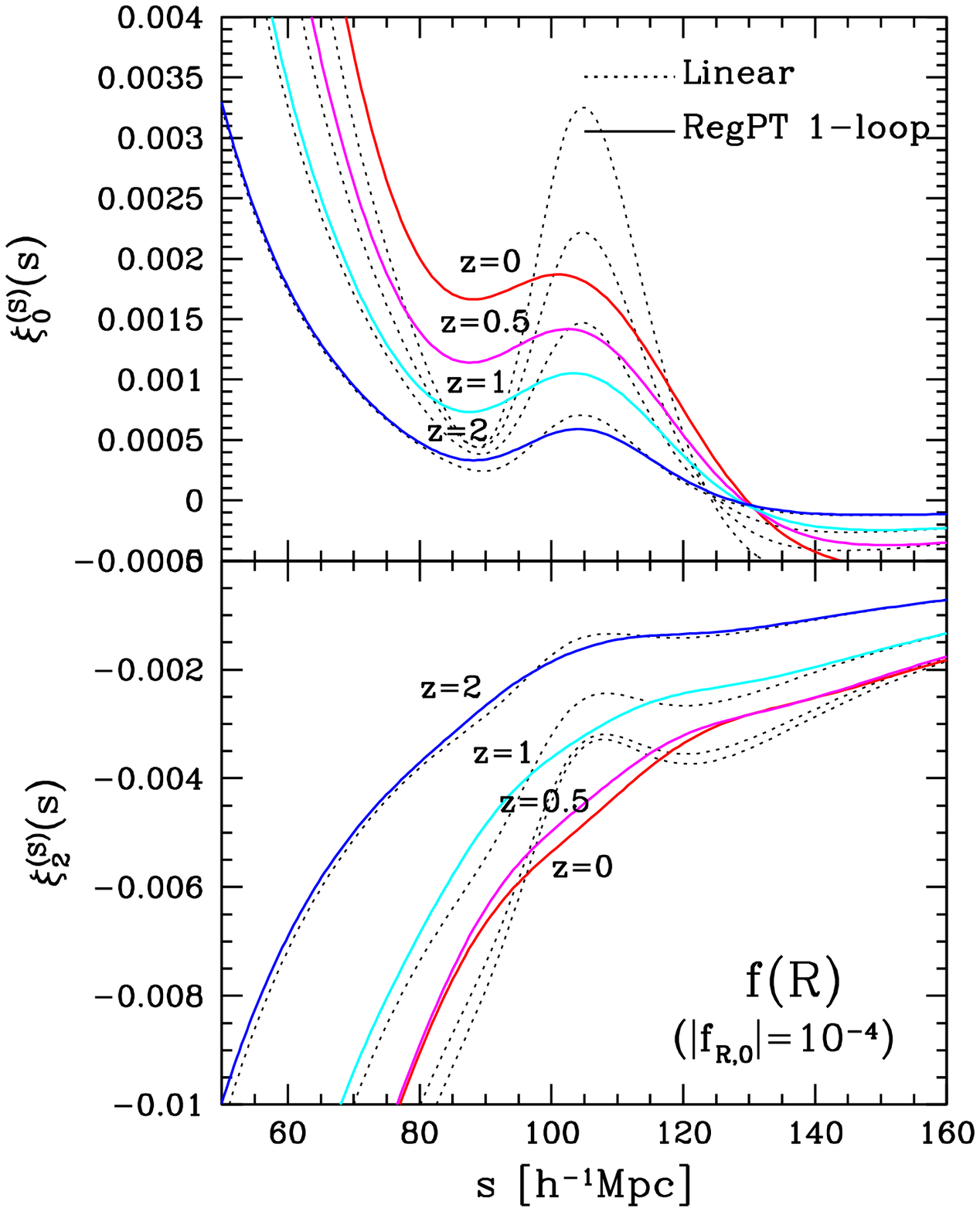}

\vspace*{-0.5cm}

\caption{Prediction of redshift-space correlation functions 
at $z=0$, $0.5$, $1$, and $2$ 
in GR (left) and $f(R)$ gravity  with $|f_{R,0}|=10^{-4}$ (right). 
Based on the TNS model, the monopole ($\ell=0$) and 
quadrupole ($\ell=2$) moments of the correlation function are shown 
in top and bottom panels, respectively. 
Note that in computing the correlation function, we used 
the fitting parameter $\sigmav$ determined from power spectrum 
(Fig.~\ref{fig:pkred0_pkred2}). 
In each panel, solid and dotted lines are the RegPT and linear theory 
predictions. 
\label{fig:xired_TNS}}
\end{figure*}
%%%%%%%%%%%%%%%%%%%%%%%%%%%%%%%%%%%%%%%%%%%%%%%%%%%%%%%%%%%%%%%%%%%%%%%

%%%%%%%%%%%%%%%%%%%%%%%%%%%%%%%%%%%%%%%%%%%%%%%%%%%%%%%%%%%%%%%%%%%%%%%
%%%%%%%%%%%%%%%%%%%%%%%%%%%%%%%%%%%%%%%%%%%%%%%%%%%%%%%%%%%%%%%%%%%%%%%
\section{From real to redshift space}
\label{sec:RSD}
%%%%%%%%%%%%%%%%%%%%%%%%%%%%%%%%%%%%%%%%%%%%%%%%%%%%%%%%%%%%%%%%%%%%%%%
%%%%%%%%%%%%%%%%%%%%%%%%%%%%%%%%%%%%%%%%%%%%%%%%%%%%%%%%%%%%%%%%%%%%%%%

In this section, as an important implication of our resummed PT treatment, we 
examine the prediction of power spectrum and correlation function in 
redshift space.

%%--%%--%%--%%--%%--%%--%%--%%--%%--%%--%%--%%--%%--%%--%%--%%--%%--%%%
\subsection{Model of RSD}
%%--%%--%%--%%--%%--%%--%%--%%--%%--%%--%%--%%--%%--%%--%%--%%--%%--%%%

Since the observed galaxy distribution via spectroscopic measurement 
basically lies in 
the redshift space, the effect of redshift-space distortions (RSD)
has to be properly incorporated into the theoretical template. Modeling 
RSD is, however, a non-trivial issue because of the non-Gaussian 
and nonlinear nature of RSD (e.g., \cite{Scoccimarro:2004tg,Taruya:2010mx,Reid:2011ar,Seljak:2011tx}). In this paper, we will adopt a specific 
PT model of RSD recently proposed by Ref.~\cite{Taruya:2010mx}. This 
model has been shown to explain the redshift-space clustering of dark matter 
in $N$-body simulations \cite{Taruya:2013my,Taruya:2013quf}. 
Combining the prescription for 
galaxy biasing, the model also gives an accurate description of 
observed power spectra and correlation functions, from which a robust 
cosmological constraint was obtained \cite{Nishimichi:2011jm,Ishikawa:2013aea,Oka:2013cba}. Hereafter, 
we call it TNS model, and on the basis of TNS model, we apply  
our RegPT calculation to the prediction of 
redshift-space power spectrum and correlation function in modified gravity 
model.

The TNS model gives a semi-analytic prescription for redshift-space power 
spectrum based on the PT calculations. The functional form of the power 
spectrum looks very similar to the popular and phenomenological streaming 
model, but includes higher-order PT corrections. Denoting the directional 
cosine between observer's line-of-sight and wave vector by $\mu$, we have
%%%%%%%%%%%%%%%%%%%%%%%%%%%%%%%%%%%%%%%%%%%%%%%%%%%%%%%%%%%%%%%%%%%%%%%
\begin{align}
&P^{\rm(S)}(k,\mu)= D_{\rm FoG}(k\,\mu\,\sigmav)
\nonumber\\
&\qquad\qquad\times 
\left\{P_{\rm Kaiser}(k,\mu)+ A(k,\mu) + B(k,\mu)\right\}, 
\label{eq:TNS}
\end{align}
%%%%%%%%%%%%%%%%%%%%%%%%%%%%%%%%%%%%%%%%%%%%%%%%%%%%%%%%%%%%%%%%%%%%%%%
with the function $P_{\rm Kaiser}$ called the nonlinear Kaiser term:
%%%%%%%%%%%%%%%%%%%%%%%%%%%%%%%%%%%%%%%%%%%%%%%%%%%%%%%%%%%%%%%%%%%%%%%
\begin{align}
&P_{\rm Kaiser}(k,\mu)= P_{11}(k) + 2\mu^2\,P_{12}(k) + \mu^4\,P_{22}(k).
\label{eq:Kaiser}
\end{align}
%%%%%%%%%%%%%%%%%%%%%%%%%%%%%%%%%%%%%%%%%%%%%%%%%%%%%%%%%%%%%%%%%%%%%%%
Here, the function $D_{\rm FoG}$ characterizes the non-perturbative 
damping effect caused by both the coherent and small-scale virialized 
motions, and we here assume 
the Gaussian form: $D_{\rm FoG}=\exp[-(k\mu\sigmav)^2]$. The parameter 
$\sigmav$ is a scale-independent constant, and is determined by 
fitting the prediction to simulation or observation. In this respect, 
Eq.~(\ref{eq:TNS}) may be regarded as semi-empirical model, however,  
a salient feature of the TNS model is the presence of 
the correction terms $A$ and $B$ that account for the nonlinear modulation 
of BAO in redshift space quite well. These two corrections 
represent the mode-coupling between 
density and velocity fields originating from the nonlinear mapping from 
real to redshift spaces. They are expressed as
%%%%%%%%%%%%%%%%%%%%%%%%%%%%%%%%%%%%%%%%%%%%%%%%%%%%%%%%%%%%%%%%%%%%%%%
\begin{align}
&A(k,\mu)= \sum_{n=1}^3\sum_{a,b=1}^2\mu^{2n}\frac{k^3}{(2\pi)^2}
\int_0^\infty dr \int_{-1}^1 dx\,
\nonumber\\
&\quad\times\Big\{A_{ab}^n(r,x)\,B_{2ab}(\bfp,\bfk-\bfp,-\bfk)
\nonumber\\
&\qquad\qquad\qquad\qquad\qquad
+\widetilde{A}_{ab}^n(r,x)B_{2ab}(\bfk-\bfp,\bfp,-\bfk)\Big\},
\label{eq:Aterm}\\
&B(k,\mu)= \sum_{n=1}^4\sum_{a,b=1}^2\mu^{2n}\frac{k^3}{(2\pi)^2}
\nonumber\\
&\times \int dr_0^\infty \int_{-1}^1 dx\,B^n_{ab}(r,x)
\frac{P_{a2}(k\sqrt{1+r^2-2rx})P_{b2}(rx)}{(1+r^2-2rx)^a},
\label{eq:Bterm}
\end{align}
%%%%%%%%%%%%%%%%%%%%%%%%%%%%%%%%%%%%%%%%%%%%%%%%%%%%%%%%%%%%%%%%%%%%%%%
where $r$ and $x$ are the dimensionless variables associated with the 
wave vector, $\bfp$, defined by $r=p/k$ and $x=(\bfk\cdot\bfp)/(kp)$, 
respectively. The function $B_{abc}$ is the bispectrum defined by 
%%%%%%%%%%%%%%%%%%%%%%%%%%%%%%%%%%%%%%%%%%%%%%%%%%%%%%%%%%%%%%%%%%%%%%%
\begin{align}
&\langle\Psi_a(\bfk_1)\Psi_b(\bfk_2)\Psi_c(\bfk_3)\rangle
\nonumber\\
&\qquad =(2\pi)^3\,
\delta_{\rm D}(\bfk_1+\bfk_2+\bfk_3)\,B_{abc}(\bfk_1,\bfk_2,\bfk_3).
\label{eq:def_bispectrum}
\end{align}
%%%%%%%%%%%%%%%%%%%%%%%%%%%%%%%%%%%%%%%%%%%%%%%%%%%%%%%%%%%%%%%%%%%%%%%
The non-vanishing coefficients, 
$A_{ab}^n$, $\widetilde{A}_{ab}^n$, and $B_{ab}^n$ are the same as 
those presented in Appendix A of Ref.~\cite{Taruya:2010mx}, and we use 
them to compute Eqs.~(\ref{eq:Aterm}) and (\ref{eq:Bterm}). 
For the prediction at one-loop order, while we apply the regularized one-loop 
expression in Eq.~(\ref{eq:reg_pk}) to the Kaiser term 
$P_{\rm Kaiser}$, the correction terms 
$A$ and $B$ appear as a next-to-leading order correction, and 
for a consistent treatment, the tree-level calculation 
is sufficient for these two terms. Thus, 
the power spectrum and bispectrum 
in the $A$ and $B$ terms may be evaluated as
%%%%%%%%%%%%%%%%%%%%%%%%%%%%%%%%%%%%%%%%%%%%%%%%%%%%%%%%%%%%%%%%%%%%%%%
\begin{align}
&P_{ab,{\rm tree}}(k)=\Gamma^{(1)}_a(k)\Gamma^{(1)}_b(k)\,P_0(k),
\label{eq:pk_tree}\\
&B_{abc,{\rm tree}}(\bfk_1,\bfk_2,\bfk_3)=2\Gamma^{(2)}_a(\bfk_2,\bfk_3)
\Gamma^{(1)}_b(k_2)\Gamma^{(1)}_c(k_3)
\nonumber\\
&\qquad\qquad\times\,P_0(k_2)P_0(k_3)\,+\,\mbox{(cyc.perm)}, 
\label{eq:bk_tree}
\end{align}
%%%%%%%%%%%%%%%%%%%%%%%%%%%%%%%%%%%%%%%%%%%%%%%%%%%%%%%%%%%%%%%%%%%%%%%
with the propagators $\Gamma_a^{(1)}$ and $\Gamma_a^{(2)}$ being evaluated 
with the tree-level expressions [see Eqs.~(\ref{eq:Gamma1reg_tree}) 
and (\ref{eq:Gamma2reg_tree})].

%%%%%%%%%%%%%%%%%%%%%%%%%%%%%%%%%%%%%%%%%%%%%%%%%%%%%%%%%%%%%%%%%%%%%%%
%%%%%%%%%%%%%%%%%%%%%%%%%%%%%%%%%%%%%%%%%%%%%%%%%%%%%%%%%%%%%%%%%%%%%%%

%%%%%%%%%%%%%%%%%%%%%%%%%%%%%%%%%%%%%%%%%%%%%%%%%%%%%%%%%%%%%%%%%%%%%%%
\begin{table}[ht]
\caption{\label{tab:fitting_RSD} Fitting results of the 
free parameter $\sigmav$ 
for the model of redshift-space power spectra shown in 
Fig.~\ref{fig:pkred0_pkred2}}
\begin{ruledtabular}
\begin{tabular}{c|ll}
&  GR & $f(R)$ \\
$z$ &  $\sigmav$ [$h^{-1}$\,Mpc] & $\sigmav$ [$h^{-1}$\,Mpc] \\
\hline\hline
0   & 5.18 &  6.65   \\
0.5 & 4.76 &  6.03   \\
1   & 3.71 &  4.62   \\
2   & 2.25 &  2.64   \\
\end{tabular}
\end{ruledtabular}
\end{table}
%%%%%%%%%%%%%%%%%%%%%%%%%%%%%%%%%%%%%%%%%%%%%%%%%%%%%%%%%%%%%%%%%%%%%%%

%%--%%--%%--%%--%%--%%--%%--%%--%%--%%--%%--%%--%%--%%--%%--%%--%%--%%%
\subsection{Results}
%%--%%--%%--%%--%%--%%--%%--%%--%%--%%--%%--%%--%%--%%--%%--%%--%%--%%%

Fig.~\ref{fig:pkred0_pkred2} plots the 
monopole ($\ell=0$, top) and quadrupole ($\ell=2$, bottom) moments of
redshift-space power spectrum, multiplied by $k^{3/2}$. 
The multipole moment of power spectrum is defined by
%%%%%%%%%%%%%%%%%%%%%%%%%%%%%%%%%%%%%%%%%%%%%%%%%%%%%%%%%%%%%%%%%%%%%%%
\begin{align}
P_\ell^{\rm(S)}(k) =\frac{2\ell+1}{2}\int_{-1}^1 d\mu\,P^{\rm(S)}(k,\mu)\,
\mathcal{P}_{\ell}(\mu). 
\label{eq:pk_ell}
\end{align}
%%%%%%%%%%%%%%%%%%%%%%%%%%%%%%%%%%%%%%%%%%%%%%%%%%%%%%%%%%%%%%%%%%%%%%%
The PT predictions based on the TNS model are shown in solid lines, and 
just for reference, linear theory results are also shown in dotted lines. 
In plotting the PT results, the free parameter $\sigmav$ is first 
determined by fitting the model predictions to the $N$-body results of 
monopole and quadrupole spectra up to the maximum wavenumber, $k_{\rm max}$. 
In each panel, $k_{\rm max}$ is indicated as vertical arrow, 
below which a percent-level agreement between the PT predictions 
and $N$-body simulation is achieved in real-space power spectrum 
(see Fig.~\ref{fig:Pkreal_nbody}).

Overall, similarly to the real space, 
the PT results show a very good performance in both monopole and quadrupole 
spectra. This is true in both GR and $f(R)$ gravity model. 
A closer look at low-$z$ results of the 
quadrupole reveals a small discrepancy with 
$N$-body results at $k\sim k_{\rm max}$, although 
the measured power spectra are bit noisy and the errorbars are large. This 
partly comes from a flaw in the PT model of RSD, but with the other choice of
the damping function $D_{\rm Fog}$ (Lorentzian form, for instance), 
the prediction would be improved \cite{Taruya:2013my}. 
Rather, one remarkable point of 
the low-$z$ results 
may be that the damping of the power spectrum amplitude relative 
to the linear theory prediction is rather significant in $f(R)$ gravity 
compared to that found in GR gravity. 
Table \ref{tab:fitting_RSD} summarizes the 
fitted results of parameter $\sigmav$, clearly showing a stronger 
suppression of the power spectrum in $f(R)$ gravity. As it has been already 
studied (e.g., \cite{Jennings:2012pt,Taruya:2013quf}), this is 
partly explained by the linear theory. 
In $f(R)$ gravity, 
the structure formation is enhanced at small scales 
due to the presence of the scale-dependent linear growth, 
and it produces a larger peculiar velocity. Although 
the screening mechanism should terminate the enhanced structure growth
at some nonlinear scales, it is still ineffective at the scales of 
our interest.

Finally, Fig.~\ref{fig:xired_TNS} shows the prediction of redshift-space 
correlation function. Based on the results in Fig.~\ref{fig:pkred0_pkred2}, 
monopole and quadrupole moments of the correlation functions are computed 
and are shown in solid lines :
%%%%%%%%%%%%%%%%%%%%%%%%%%%%%%%%%%%%%%%%%%%%%%%%%%%%%%%%%%%%%%%%%%%%%%%
\begin{align}
\xi_{\ell}^{\rm(S)}(s)=
i^\ell \int \frac{dk\,k^2}{2\pi^2}\,
P_\ell^{\rm(S)}(k)\,j_\ell(k s)
\label{eq:xi_ell}
\end{align}
%%%%%%%%%%%%%%%%%%%%%%%%%%%%%%%%%%%%%%%%%%%%%%%%%%%%%%%%%%%%%%%%%%%%%%%
We do not present here the $N$-body results, because only with the 
single realization data, we do not reliably estimate the 
correlation function. While a detailed comparison of the results between 
PT calculation and simulation will be made elsewhere, the predictions 
shown in Fig.~\ref{fig:xired_TNS} look reasonable, and one 
expects that the PT results quantitatively describe the impact of both the 
nonlinear gravity and RSD around baryon acoustic peak. It suggests that 
in $f(R)$ gravity, 
the acoustic peak is significantly smeared out, and the peak amplitude 
is largely suppressed relative to the linear theory prediction. The 
resultant shape and amplitude look rather similar to those in GR. Hence, 
taking full account of these nonlinear effects would be essential
in constraining the modified gravity from the observed correlation functions. 

%%%%%%%%%%%%%%%%%%%%%%%%%%%%%%%%%%%%%%%%%%%%%%%%%%%%%%%%%%%%%%%%%%%%%%%
\begin{figure*}[t]
\includegraphics[width=9.5cm]{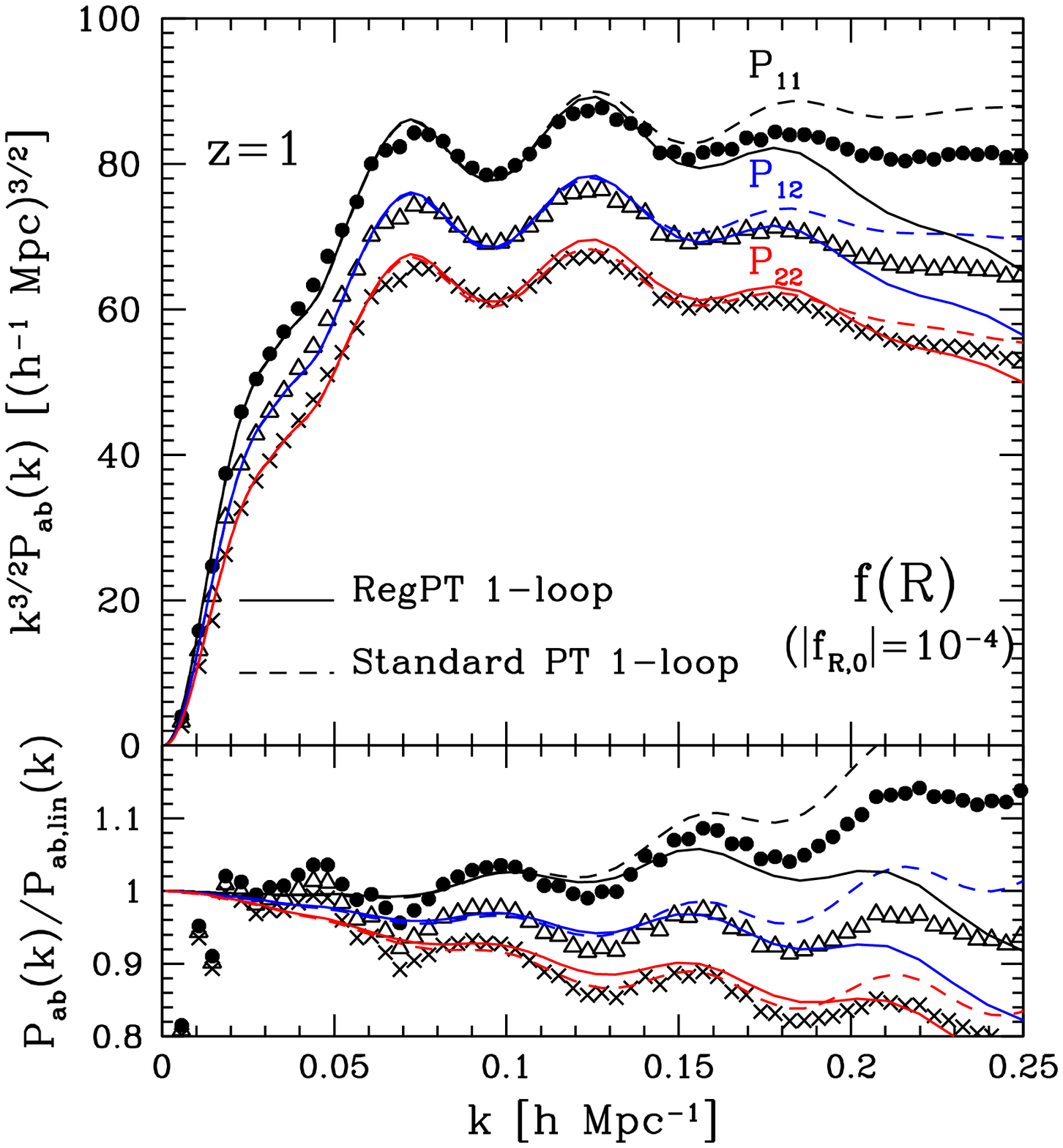}
\hspace*{-0.8cm}
\includegraphics[width=9cm]{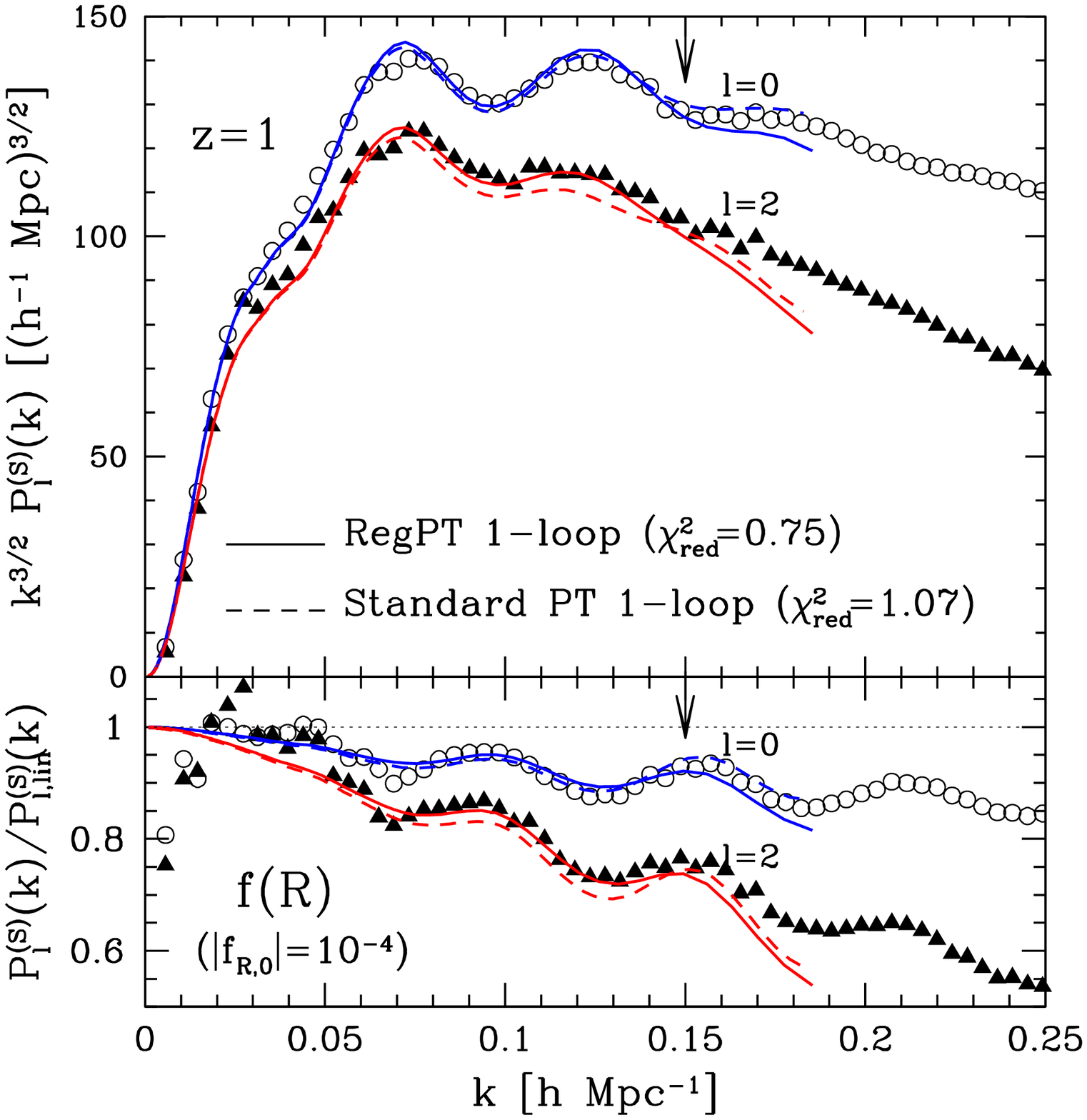}

\caption{Comparison between RegPT and standard PT results at one-loop 
order in real and redshift spaces. 
Left panel plots the auto- and cross-power spectra of density 
and velocity fields 
in real space at $z=1$. The results in $f(R)$ gravity with 
$|f_{R,0}|=10^{-4}$ are shown. Solid and dotted lines are the PT predictions 
computed with RegPT and standard PT, respectively. In right panel, 
redshift-space power spectra at $z=1$ are plotted. 
Based on the TNS model, the monopole ($\ell=0$) and quadrupole ($\ell=2$) power spectra are computed, and the results are compared with $N$-body simulations. 
The solid and dashed lines respectively represent 
the PT predictions computed with RegPT and standard PT.
\label{fig:SPT_vs_RegPT}}
\end{figure*}
%%%%%%%%%%%%%%%%%%%%%%%%%%%%%%%%%%%%%%%%%%%%%%%%%%%%%%%%%%%%%%%%%%%%%%%

%%%%%%%%%%%%%%%%%%%%%%%%%%%%%%%%%%%%%%%%%%%%%%%%%%%%%%%%%%%%%%%%%%%%%%%
%%%%%%%%%%%%%%%%%%%%%%%%%%%%%%%%%%%%%%%%%%%%%%%%%%%%%%%%%%%%%%%%%%%%%%%
\section{Discussion: comparison with standard perturbation theory}
\label{sec:discussion}
%%%%%%%%%%%%%%%%%%%%%%%%%%%%%%%%%%%%%%%%%%%%%%%%%%%%%%%%%%%%%%%%%%%%%%%
%%%%%%%%%%%%%%%%%%%%%%%%%%%%%%%%%%%%%%%%%%%%%%%%%%%%%%%%%%%%%%%%%%%%%%%

While the standard PT treatment fails to compute the correlation 
function, at sufficiently large scales and high redshifts, 
it still provide a reasonably accurate prediction of the power spectrum. 
Indeed, in Ref.~\cite{Taruya:2013quf}, we used the standard PT to 
compute the real- and redshift-space power spectra in $f(R)$ gravity, and 
a good agreement with $N$-body simulation was found at $z=1$. 
It is thus interesting to see how much the standard PT results are 
different from our RegPT treatment.

To compare with the standard PT results, we adopt the same cosmological 
parameters as used in Ref.~\cite{Taruya:2013quf}, and 
the power spectra at $z=1$ are computed with the RegPT treatment 
in both real and redshift spaces. The results are then shown in 
Fig.~\ref{fig:SPT_vs_RegPT}, where the $N$-body results and the standard 
PT predictions taken from Ref.~\cite{Taruya:2013quf} 
are also plotted in symbols and dashed lines, respectively. 
Top panels show the 
power spectra multiplied by $k^{3/2}$, while bottom panels plot 
the ratio of the power spectrum to the linear theory prediction. Note 
that in both PT predictions, the TNS model has been used to compute the 
monopole and quadrupole power spectra in right panel.  

Overall, both of the PT results reproduce the $N$-body 
trend reasonably well at $k\lesssim0.15\,h$\,Mpc$^{-1}$, but 
a closer look at the real-space power spectrum reveals that
the RegPT gives a moderate nonlinear enhancement or suppression of 
the power spectrum amplitude, and it seems to better describe the 
smearing effect on BAO. This trend is also seen in redshift-space
power spectrum. Indeed, the $N$-body results of the quadrupole power 
spectrum is better described by the RegPT prediction. 
To see this quantitatively, we evaluate the $\chi$-squared statistics: 
%%%%%%%%%%%%%%%%%%%%%%%%%%%%%%%%%%%%%%%%%%%%%%%%%%%%%%%%%%%
\begin{align}
\chi_{\rm red}^2=\frac{1}{\nu}\sum_{\ell=0,2}\sum_i
\frac{\left[P_{\ell,{\rm N\mbox{-}body}}^{\rm(S)}(k_i)-
P_{\ell,{\rm PT}}^{\rm(S)}(k_i)\right]^2}{[\Delta P^{\rm(S)}_\ell(k_i)]^2},
\label{eq:chi2_red}
\end{align}
%%%%%%%%%%%%%%%%%%%%%%%%%%%%%%%%%%%%%%%%%%%%%%%%%%%%%%%%%%%
with the quantity $\nu$ being the number of degrees of freedom. 
For simplicity, we here ignore the cross covariance between the monopole 
and quadrupole spectra, which is shown to be fairly small. 
The statistical error $\Delta P^{\rm(S)}_\ell$ is estimated
from the cosmic variance error assuming the hypothetical survey with 
volume $10\,h^{-3}\,$Gpc$^3$ (see Appendix B of Ref.~\cite{Taruya:2010mx} 
for the explicit expressions). With the number of Fourier bins 
below the maximum wavenumber indicated by vertical arrows,  
the resultant numerical values of $\chi^2_{\rm red}$ are shown in the panel, 
clearly showing that the RegPT prediction gives a good performance, and 
reproduce the $N$-body results quite well. Note that the 
$\chi_{\rm red}^2$ value smaller than unity does not implies that the 
model overfits the simulations because the statistical error 
$\Delta P_\ell^{\rm(S)}$ adopted here does not reflect the actual 
error in $N$-body simulations. 
Although the differences of the PT prediction between 
standard PT and RegPT is small at one-loop order, 
a better description of acoustic feature in power spectrum is rather 
crucial in simultaneous measurement of
geometric distances, and the RegPT procedure provides a more accurate 
theoretical template from which one should be able to obtain unbiased 
parameter estimations.

%%%%%%%%%%%%%%%%%%%%%%%%%%%%%%%%%%%%%%%%%%%%%%%%%%%%%%%%%%%%%%%%%%%%%%%
%%%%%%%%%%%%%%%%%%%%%%%%%%%%%%%%%%%%%%%%%%%%%%%%%%%%%%%%%%%%%%%%%%%%%%%
\section{Conclusion}
\label{sec:conclusion}
%%%%%%%%%%%%%%%%%%%%%%%%%%%%%%%%%%%%%%%%%%%%%%%%%%%%%%%%%%%%%%%%%%%%%%%
%%%%%%%%%%%%%%%%%%%%%%%%%%%%%%%%%%%%%%%%%%%%%%%%%%%%%%%%%%%%%%%%%%%%%%%

Future precision observations of large-scale structure enables us to 
not only give a useful cosmological constraint on dark energy but also 
offer a new possibility to probe the theory of gravity itself 
that describes both the dynamics of cosmic expansion and the growth of 
large-scale structure. Toward an unbiased cosmological constraint, 
it is therefore crucial to develop accurate theoretical 
templates of large-scale structure observations. 
In this paper, beyond consistency test of gravity from a precision 
measurement of BAO and RSD, we presented an improved PT prescription 
of the power spectrum and correlation function in modified gravity models.

Based on the resummed PT scheme with $\Gamma$ expansion, we first 
construct a resummed propagator that 
partly includes the non-perturbative effect in the high-$k$ limit. 
While the resultant propagator in the high-$k$ limit 
contains corrections arising from the 
screening mechanism, we explicitly show that 
in the case of $f(R)$ gravity with a currently constrained model parameter, 
the impact of this term is fairly small.  
Thus, in $f(R)$ gravity, the regularized propagators, 
that reproduce both 
the resummed high-$k$ behavior and the low-$k$ results computed 
with standard PT, are successfully constructed 
in a similar manner to the GR case,  
taking account of the nonlinear modification of gravity valid at large scales. 
Then, the analytically calculated propagators have been 
compared with the measured results in $N$-body simulations. 
At one-loop order, the analytic predictions reproduces the measured 
propagators quite well, and a good agreement with $N$-body simulations 
was also obtained for the analytic power spectrum and correlation 
function expressed in terms of the regularized propagators. Furthermore, 
employing a specific mode of RSD, the improved PT prescription 
has been shown to successfully describe the redshift-space observables. 
For the power spectrum at one-loop order, 
while a performance of our improved PT treatment 
apparently looks very similar to that of the standard PT prediction, 
nonlinear modulation of BAO seen in $N$-body simulations is better described 
by our improved PT predictions, showing that the resummed PT 
developed here is better suited for an unbiased cosmological 
constraints from BAO measurements.

Finally, while the explicit PT calculations presented here have 
focused on a specific modified gravity model, 
the resummed PT prescription developed in this paper, as well as 
the framework to treat PT calculations, are quite general, 
and can be applied to a wide class of modified gravity models. 
This is at least true 
for models with chameleon-type screening mechanisms 
within the current observational bounds. Rather, with 
the improved PT prescription presented here, we will be able to 
put a more stringent constraint on modified gravity from a
precision measurement of BAO and RSD. For instance, 
the $f(R)$ gravity of the 
functional form in Eq.~(\ref{eq:fR_HuSawicki}) or (\ref{eq:fR_model_nbody}) 
is currently constrained to $|f_{R,0}|\lesssim10^{-4}$, 
from the RSD measurement with SDSS DR7 luminous red galaxy sample 
\cite{Yamamoto:2010ie}. 
While this is roughly comparable to the one obtained from cluster 
abundance \cite{2009PhRvD..80h3505S}, 
there are now much larger galaxy samples such as BOSS DR11 CMASS sample, 
from which we will get a much more severe constraint on $f(R)$ gravity.
Note also that the template used in Ref.~\cite{Yamamoto:2010ie} 
is based on the fitting formula, which does not properly account for the 
nonlinear modulation of BAO coming from both the gravity and RSD.   
Hence, with the new PT template and the model of RSD presented here, 
a robust and unbiased cosmological analysis will be made possible. 
In doing this, however, a careful study of the galaxy biasing is needed.    
Indeed, the halo clustering properties in $f(R)$ gravity 
has been found to systematically differ from those in GR 
(e.g., \cite{Schmidt:2008tn}). Since 
even in the GR, halo/galaxy clustering not only has a 
scale-dependent property but also possibly exhibits a biasing on 
the velocity field (e.g., \cite{Desjacques:2009kt, Baldauf:2014fza}), 
a proper account of 
these biasing effects is rather crucial toward an unbiased test of gravity.

\begin{acknowledgments}
We are grateful to Baojiu Li for kindly providing us the $N$-body simulation 
data. This work is supported in part by a Grant-in-Aid for
Scientific Research from the Japan Society for the Promotion of Science 
(No.~23740186 for T.H and No.~24540257 for A.T), in part
by grant ANR-12-BS05-0002 of the French Agence Nationale de la Recherche.
TN is supported by JSPS Postdoctoral Fellowships for Research Abroad. 
KK is supported by the UK Science and Technology Facilities Council 
(STFC) grants ST/K00090/1 and ST/L005573/1. 
T.H. acknowledges a support from MEXT HPCI Strategic Program. 
The numerical calculations in this work were partly carried out on 
Cray XC30 at Center for Computational Astrophysics, CfCA, of National 
Astronomical Observatory of Japan. 
\end{acknowledgments}

\appendix
%%%%%%%%%%%%%%%%%%%%%%%%%%%%%%%%%%%%%%%%%%%%%%%%%%%%%%%%%%%%%%%%%%%%%%%
%%%%%%%%%%%%%%%%%%%%%%%%%%%%%%%%%%%%%%%%%%%%%%%%%%%%%%%%%%%%%%%%%%%%%%%
\section{Explicit expression for the kernels, $K_{f(R)}^{(n)}$ and 
$K_{\rm DGP}^{(n)}$}
\label{sec:kernel}
%%%%%%%%%%%%%%%%%%%%%%%%%%%%%%%%%%%%%%%%%%%%%%%%%%%%%%%%%%%%%%%%%%%%%%%
%%%%%%%%%%%%%%%%%%%%%%%%%%%%%%%%%%%%%%%%%%%%%%%%%%%%%%%%%%%%%%%%%%%%%%%

In this appendix, based on the expressions for the vertex functions 
in Eqs.~(\ref{eq:vertex_sigma_2})-(\ref{eq:vertex_sigma_4}), we present the 
explicit expressions for the kernels in the high-$k$ limit, $K_{f(R)}$ 
and $K_{\rm DGP}$, in Sec.~\ref{subsec:Eikonal}. 

%%--%%--%%--%%--%%--%%--%%--%%--%%--%%--%%--%%--%%--%%--%%--%%--%%--%%%
\subsection{$f(R)$ gravity}
%%--%%--%%--%%--%%--%%--%%--%%--%%--%%--%%--%%--%%--%%--%%--%%--%%--%%%

Recalling the fact that the coupling functions $M_i$ in $f(R)$ gravity 
are all scale-independent, and just given as function of time 
[see Eq.~(\ref{eq:coupling_func_fR})], 
the kernel functions defined in Eq.~(\ref{eq:sigma_n_fR}) become
%%%%%%%%%%%%%%%%%%%%%%%%%%%%%%%%%%%%%%%%%%%%%%%%%%%%%%%%%%%%%%%%%%%%%%%
\begin{align}
&K_{f(R)}^{(2)}=1,\quad
\label{eq:K_fR^2} \\
&K_{f(R)}^{(3)}=1-\frac{M_2^2}{3\,M_3}\,\frac{1}{\Pi(p_{12})},
\label{eq:K_fR^3}\\
&K_{f(R)}^{(4)}=1-\frac{M_2\,M_3}{6\,M_4}\,
\frac{1}{\Pi(p_{123})}
+\Bigl\{\frac{M_2^3}{18\,M_4}\frac{1}{\Pi(p_{123})}
-\frac{M_2\,M_3}{3 M_4}\Bigr\}
\nonumber\\
&\qquad\qquad\qquad\qquad
\times\Bigl\{ \frac{1}{\Pi(p_{12})}
+\frac{1}{\Pi(p_{23})}+\frac{1}{\Pi(p_{31})}
\Bigr\}.
\label{eq:K_fR^4}
\end{align}
%%%%%%%%%%%%%%%%%%%%%%%%%%%%%%%%%%%%%%%%%%%%%%%%%%%%%%%%%%%%%%%%%%%%%%%

%%--%%--%%--%%--%%--%%--%%--%%--%%--%%--%%--%%--%%--%%--%%--%%--%%--%%%
\subsection{DGP}
%%--%%--%%--%%--%%--%%--%%--%%--%%--%%--%%--%%--%%--%%--%%--%%--%%--%%%

Using the relations given in Sec.~\ref{subsubsec:DGP_model}, 
taking the high-$k$ limit gives 
the kernel functions defined in Eq.~(\ref{eq:sigma_n_DGP}) as follows:
%%%%%%%%%%%%%%%%%%%%%%%%%%%%%%%%%%%%%%%%%%%%%%%%%%%%%%%%%%%%%%%%%%%%%%%
%\begin{widetext}
\begin{align}
&K_{\rm DGP}^{(2)}(\bfp_1)=1-\mu_1^2,
\\
&K_{\rm DGP}^{(3)}(\bfp_1,\bfp_2)=\frac{2}{3}\Bigl\{2(1-\mu_1^2)(1-\mu_2^2)
\nonumber\\
&+(1-\mu_{12}^2)(1-\mu_{1,2}^2)\Bigr\},
\\
&K_{\rm DGP}^{(4)}(\bfp_1,\bfp_2,\bfp_3)=\frac{1}{3}\Bigl\{
(1-\mu_{23}^2)(1-\mu_{1}^2)(1-\mu_{2,3}^2)
\nonumber\\
&+(1-\mu_{31}^2)(1-\mu_{2}^2)(1-\mu_{3,1}^2)
\nonumber\\
&+
(1-\mu_{12}^2)(1-\mu_{3}^2)(1-\mu_{1,2}^2)
\Bigr\}
\nonumber\\
&+\frac{1}{6}(1-\mu_3^2)\Bigl\{2(1-\mu_1^2)(1-\mu_2^2)+
(1-\mu_{12}^2)(1-\mu_{1,2}^2)\Bigr\}
\nonumber\\
&
+\frac{1}{6}(1-\mu_1^2)\Bigl\{2(1-\mu_2^2)(1-\mu_3^2)+
(1-\mu_{23}^2)(1-\mu_{2,3}^2)\Bigr\}
\nonumber\\
&
+\frac{1}{6}(1-\mu_2^2)\Bigl\{2(1-\mu_3^2)(1-\mu_1^2)+
(1-\mu_{31}^2)(1-\mu_{3,1}^2)\Bigr\}
\nonumber\\
&
+\frac{1}{6}(1-\mu_{123}^2)\Bigl\{(1-\mu_{12,3}^2)(1-\mu_{12}^2)
\nonumber\\
&+
(1-\mu_{23,1}^2)(1-\mu_{2,3}^2)+
(1-\mu_{31,2}^2)(1-\mu_{3,1}^2)\Bigr\},
\end{align}
%\end{widetext}
%%%%%%%%%%%%%%%%%%%%%%%%%%%%%%%%%%%%%%%%%%%%%%%%%%%%%%%%%%%%%%%%%%%%%%%
where we define $\mu_i=(\bfk\cdot\bfp_i)/(k\,p_i)$, $\mu_{ij}=(\bfk\cdot\bfp_{ij})/(k\,p_{ij})$, and $\mu_{i,j}=(\bfp_i\cdot\bfp_j)/(p_i\,p_j)$.

%%%%%%%%%%%%%%%%%%%%%%%%%%%%%%%%%%%%%%%%%%%%%%%%%%%%%%%%%%%%%%%%%%%%%%%
%%%%%%%%%%%%%%%%%%%%%%%%%%%%%%%%%%%%%%%%%%%%%%%%%%%%%%%%%%%%%%%%%%%%%%%
\section{Vertex function $\sigma^{(n)}$}
\label{sec:vertex}
%%%%%%%%%%%%%%%%%%%%%%%%%%%%%%%%%%%%%%%%%%%%%%%%%%%%%%%%%%%%%%%%%%%%%%%
%%%%%%%%%%%%%%%%%%%%%%%%%%%%%%%%%%%%%%%%%%%%%%%%%%%%%%%%%%%%%%%%%%%%%%%

In this appendix, we derive the expression for the vertex function 
$\sigma^{(n)}$. To do this, we first write down the expression 
for scalaron, $\varphi$, and express it in terms of $\delta$. 
From Eqs.~(\ref{eq:EoM_scalaron}) and (\ref{eq:I_expansion}), we have
%%%%%%%%%%%%%%%%%%%%%%%%%%%%%%%%%%%%%%%%%%%%%%%%%%%%%%%%%%%%%%%%%%%%%%%
\begin{widetext}
\begin{align}
&\varphi(\bfk)=\frac{\kappa^2\,\rhom}{3}\,\frac{\delta(\bfk)}{\Pi(k)}-
\frac{1}{3\Pi(k)}\,\Bigl[
\frac{1}{2}M_2(\bfk_1,\bfk_2)\,\varphi(\bfk_1)\,\varphi(\bfk_2)
+\frac{1}{6}M_3(\bfk_1,\bfk_2,\bfk_3)\,\varphi(\bfk_1)\,\varphi(\bfk_2)\,\varphi(\bfk_3)
\nonumber\\
&\qquad\qquad\qquad\qquad\qquad\qquad\qquad\qquad\qquad\qquad
+\frac{1}{24}M_4(\bfk_1,\bfk_2,\bfk_3,\bfk_4)\,\varphi(\bfk_1)\,\varphi(\bfk_2)\,\varphi(\bfk_3)\,\varphi(\bfk_4)+\cdots
\Bigr]
\label{eq:varphi_scalaron}
\end{align}
\end{widetext}
%%%%%%%%%%%%%%%%%%%%%%%%%%%%%%%%%%%%%%%%%%%%%%%%%%%%%%%%%%%%%%%%%%%%%%%
with the function $\Pi$ defined in Eq.~(\ref{eq:def_Pi}). Here, we introduced 
the short-cut notation that the repeated Fourier arguments, 
$\bfk_1,\cdots,\bfk_n$, are, 
after multiplying by the Dirac $\delta$ function 
$\delta_D(\bfk-\bfk_{1\cdots n})$, integrated over. 
To express $\varphi$ in terms of $\delta$, we perturbatively expand $\varphi$: 
%%%%%%%%%%%%%%%%%%%%%%%%%%%%%%%%%%%%%%%%%%%%%%%%%%%%%%%%%%%%%%%%%%%%%%%
\begin{align}
\varphi=\varphi_1+\varphi_2+\varphi_3+\varphi_4+\cdots.
\end{align}
%%%%%%%%%%%%%%%%%%%%%%%%%%%%%%%%%%%%%%%%%%%%%%%%%%%%%%%%%%%%%%%%%%%%%%%
Then, we have
%%%%%%%%%%%%%%%%%%%%%%%%%%%%%%%%%%%%%%%%%%%%%%%%%%%%%%%%%%%%%%%%%%%%%%%
\begin{align}
&\varphi_2=-\frac{1}{6\Pi}\,M_2\,\varphi_1\,\varphi_1,
\nonumber\\
&\varphi_3=-\frac{1}{6\Pi}\,M_2\,(\varphi_2\,\varphi_1+\varphi_1\,\varphi_2)
-\frac{1}{18\Pi}\,M_3\,\varphi_1\,\varphi_1\,\varphi_1,
\nonumber\\
&\varphi_4=-\frac{1}{6\Pi}\,M_2\,(\varphi_2\,\varphi_2+\varphi_1\,\varphi_3+
\varphi_3\,\varphi_1)
\nonumber\\
&\qquad-\frac{1}{18\Pi}\,M_3\,(\varphi_1\,\varphi_1\,\varphi_2+
\varphi_1\,\varphi_2\,\varphi_1+\varphi_2\,\varphi_1\,\varphi_1)
\nonumber\\
&~~~\quad-\frac{1}{72\Pi}\,M_4\,\varphi_1\,\varphi_1\,\varphi_1\,\varphi_1.
\end{align}
%%%%%%%%%%%%%%%%%%%%%%%%%%%%%%%%%%%%%%%%%%%%%%%%%%%%%%%%%%%%%%%%%%%%%%%
In Eq.~(\ref{eq:varphi_scalaron}), at leading-order, 
$\varphi_1$ is rewritten with $\delta$. 
Substituting this relation into the above, we obtain the perturbative 
expression of $\varphi$ in terms of $\delta$:
%%%%%%%%%%%%%%%%%%%%%%%%%%%%%%%%%%%%%%%%%%%%%%%%%%%%%%%%%%%%%%%%%%%%%%%
\begin{align}
&\varphi_1(\bfk)=\frac{\kappa^2\,\rhom}{3}\,\frac{\delta(\bfk)}{\Pi(k)},
\nonumber\\
&\varphi_2(\bfk)=-\frac{1}{6\,\Pi(k)}\,\left(\frac{\kappa^2\,\rhom}{3}\right)^2
M_2(\bfk_1,\bfk_2)\,\frac{\delta(\bfk_1)\,\delta(\bfk_2)}{\Pi(k_1)\,\Pi(k_2)},
\nonumber\\
&\varphi_3(\bfk)=-\frac{1}{18\,\Pi(k)}\,
\left(\frac{\kappa^2\,\rhom}{3}\right)^3\,\Bigl[
M_3(\bfk_1,\bfk_2,\bfk_3)
\nonumber\\
&\qquad\quad\quad-\frac{M_2(\bfk_{12},\bfk_3)M_2(\bfk_1,\bfk_2)}
{\Pi(k_{12})}\Bigr]\,\frac{\delta(\bfk_1)\,\delta(\bfk_2)\,\delta(\bfk_3)}
{\Pi(k_1)\,\Pi(k_2)\,\Pi(k_3)}
\nonumber\\
&\varphi_4(\bfk)=-\frac{1}{72\Pi(k)}\,
\left(\frac{\kappa^2\,\rhom}{3}\right)^4
\,\Bigl[M_4(\bfk_1,\bfk_2,\bfk_3,\bfk_4)
\nonumber\\
&\qquad
+\frac{1}{3}\frac{M_2(\bfk_1,\bfk_2)}{\Pi(k_{12})}
\Bigl\{\frac{M_2(\bfk_{12},\bfk_{34})M_2(\bfk_3,\bfk_4)}{\Pi(k_{34})}
\nonumber\\
&\qquad-6M_3(\bfk_{12},\bfk_3,\bfk_4)\Bigr\}
+\frac{2}{3}\frac{M_2(\bfk_{123},\bfk_4)}{\Pi(k_{123})}
\nonumber\\
&\quad\quad\times\Bigl\{
\frac{M_2(\bfk_{12},\bfk_3)M_2(\bfk_1,\bfk_2)}{\Pi(k_{12})}
-M_3(\bfk_1,\bfk_2,\bfk_3)\Bigr\}\,\Bigr]
\nonumber\\
&\qquad\qquad\qquad\qquad\quad\times
\frac{\delta(\bfk_1)\,\delta(\bfk_2)\,\delta(\bfk_3)\,\delta(\bfk_4)}
{\Pi(k_1)\,\Pi(k_2)\,\Pi(k_3)\,\Pi(k_4)}.
\end{align}
%%%%%%%%%%%%%%%%%%%%%%%%%%%%%%%%%%%%%%%%%%%%%%%%%%%%%%%%%%%%%%%%%%%%%%%
Note that the kernels of integral given above are not yet symmetrized. 

Provided the expression $\varphi$ in terms of $\delta$, the vertex 
functions $\sigma_{21\cdots1}$ are now read off from the comparison between
Eqs.~(\ref{eq:eq_continuity})-(\ref{eq:EoM_scalaron}) and 
Eq.~(\ref{eq:basic_eq}) to give 
%%%%%%%%%%%%%%%%%%%%%%%%%%%%%%%%%%%%%%%%%%%%%%%%%%%%%%%%%%%%%%%%%%%%%%%
\begin{align}
&\sigma^{(n)}(\bfk_1,\cdots,\bfk_n)=\frac{1}{2}
\left(\frac{k_{12\cdots n}}{a\,H}\right)^2
\nonumber\\
&\qquad\qquad\qquad\times(\mbox{symmetrized kernel of}\,\varphi_n),
\end{align}
%%%%%%%%%%%%%%%%%%%%%%%%%%%%%%%%%%%%%%%%%%%%%%%%%%%%%%%%%%%%%%%%%%%%%%%

%%%%%%%%%%%%%%%%%%%%%%%%%%%%%%%%%%%%%%%%%%%%%%%%%%%%%%%%%%%%%%%%%%%%%%%
%%%%%%%%%%%%%%%%%%%%%%%%%%%%%%%%%%%%%%%%%%%%%%%%%%%%%%%%%%%%%%%%%%%%%%%
\section{Explicit expressions for sub-leading correction in resummed propagator}
\label{sec:sub-leading}
%%%%%%%%%%%%%%%%%%%%%%%%%%%%%%%%%%%%%%%%%%%%%%%%%%%%%%%%%%%%%%%%%%%%%%%
%%%%%%%%%%%%%%%%%%%%%%%%%%%%%%%%%%%%%%%%%%%%%%%%%%%%%%%%%%%%%%%%%%%%%%%

In this Appendix, we derive the explicit expressions for 
the corrections to the propagator arising from 
the modification of gravity, $\delta G_{ab}$ [Eq.~(\ref{eq:delta_G_ab})]. 
To do this, our main task is to calculate the following ensemble average: 
%%%%%%%%%%%%%%%%%%%%%%%%%%%%%%%%%%%%%%%%%%%%%%%%%%%%%%%%%%%%%%%%%%%%%%%
\begin{align}
\left\langle\Delta^{(n)}\,\exp\left[\int_{\tau_0}^\tau d\tau'' 
\Xi(\bfk,\tau'')\right]\right\rangle_{\Xi,\omega_{ab}}
\label{eq:sigma_exp_n}
\end{align}
%%%%%%%%%%%%%%%%%%%%%%%%%%%%%%%%%%%%%%%%%%%%%%%%%%%%%%%%%%%%%%%%%%%%%%%
Here, the quantity $\sigma^{(n)}$ represents the $n$-th order corrections 
which contribute to the modification of the effective Newton constant via 
the screening mechanism (see Eq.~[\ref{eq:effective_G_n}]). 
In $f(R)$ gravity, Eq.~(\ref{eq:sigma_n_fR}) with the kernels given by 
Eqs.~(\ref{eq:K_fR^2}) and (\ref{eq:K_fR^3}) lead to 
%%%%%%%%%%%%%%%%%%%%%%%%%%%%%%%%%%%%%%%%%%%%%%%%%%%%%%%%%%%%%%%%%%%%%%%
\begin{align}
&\Delta^{(2)}(k;\tau)=-\frac{1}{3}\frac{\overline{R}_{,ff}(\tau')}{\Pi(k;\tau')}
\left(\frac{\kappa^2\,\rhom}{3}\right)\int\frac{d^3\bfp}{(2\pi)^3}\,
\frac{\delta(\bfp;\tau)}{\Pi(p;\tau)}
\label{eq:sigma_2}\\
&\Delta^{(3)}(k;\tau)=-\frac{1}{6}\frac{\overline{R}_{,fff}(\tau')}{\Pi(k;\tau')}
\left(\frac{\kappa^2\,\rhom}{3}\right)^2
\int\frac{d^3\bfp_1 d^3\bfp_2}{(2\pi)^6}\,
\nonumber\\
&\quad\times
\left\{1-\frac{\overline{R}_{,ff}^2(\tau)}{3\overline{R}_{,fff}(\tau)}
\frac{1}{\Pi(p_{12};\tau)}\right\}
\frac{\delta(\bfp_1;\tau)\delta(\bfp_2;\tau)}{\Pi(p_1;\tau)\Pi(p_2;\tau)}
\label{eq:sigma_3}
\end{align}
%%%%%%%%%%%%%%%%%%%%%%%%%%%%%%%%%%%%%%%%%%%%%%%%%%%%%%%%%%%%%%%%%%%%%%%
Below, based on the linear theory estimate of $\delta$ and $\theta$, 
we evaluate the non-vanishing contributions to 
Eq.~(\ref{eq:sigma_exp_n}).

First consider the $n=2$ case in $f(R)$ gravity model. 
Substituting Eq.~(\ref{eq:sigma_2}) into Eq.~(\ref{eq:sigma_exp_n}), we obtain
%%%%%%%%%%%%%%%%%%%%%%%%%%%%%%%%%%%%%%%%%%%%%%%%%%%%%%%%%%%%%%%%%%%%%%%
\begin{widetext}
\begin{align}
&\Bigg\langle\Delta^{(2)}\exp\left[
\int_{\tau_0}^\tau d\tau'\Xi(\bfk,\tau'')
\right]\Bigg\rangle
\nonumber\\
&\qquad=
-\frac{1}{3}\frac{\overline{R}_{,ff}(\tau')}{\Pi(k;\tau')}
\left(\frac{\kappa^2\,\rhom}{3}\right)
\,\,\Bigl\langle
\int\frac{d^3\bfp}{(2\pi)^3}\frac{\delta(\bfp;\tau')}{\Pi(p;\tau')}
\exp\left[\int_{\tau_0}^\tau d\tau''\int\frac{d^3\bfq}{(2\pi)^3}
\left(\frac{\bfk\cdot\bfq}{q^2}\right)\theta(\bfq;\tau'')\right]
\Bigr\rangle
\nonumber\\
&\qquad
=-\frac{1}{3}\frac{\overline{R}_{,ff}(\tau')}{\Pi(k;\tau')}
\left(\frac{\kappa^2\,\rhom}{3}\right)
\,
\sum_{n=0}\frac{1}{n!}\int\frac{d^3\bfp\,d^3\bfq_1\cdots d^3\bfq_n}{(2\pi)^{3(n+1)}}
\nonumber\\
&\qquad\qquad\qquad\qquad\qquad
\times\int_{\tau_0}^\tau
 d\tau''_1 \cdots d\tau''_n 
\left(\frac{\bfk\cdot\bfq_1}{q_1^2}\right)\cdots
\left(\frac{\bfk\cdot\bfq_n}{q_n^2}\right)
\Bigl\langle\frac{\delta(\bfp;\tau')}{\Pi(p;\tau')}
\theta(\bfq_1;\tau''_1)\cdots\theta(\bfq_n;\tau''_n)\Bigr\rangle
\nonumber\\
&\qquad
=-\frac{1}{3}\frac{\overline{R}_{,ff}(\tau')}{\Pi(k;\tau')}
\left(\frac{\kappa^2\,\rhom}{3}\right)
\,\int_{\tau_0}^\tau d\tau''
\int\frac{d^3\bfp}{(2\pi)^3}\,
\left(-\frac{\bfk\cdot\bfp}{p^2}\right)\,
\frac{P_{\delta\theta}(p;\tau',\tau'')}{\Pi(p;\tau')}\,
\sum_{m=0} \frac{(2m+1)!!}{(2m+1)!}
\nonumber\\
&\qquad\qquad\quad\qquad\qquad\quad\qquad\qquad\quad\quad
\qquad\qquad\quad\times
\left\{-\int_{\tau_0}^\tau d\tau_1''d\tau_2'' \int \frac{d^3\bfq}{(2\pi)^3}
\left(\frac{\bfk\cdot\bfq}{q^2}\right)^2\,P_{\theta\theta}(p;\tau''_1,\tau''_2)
\right\}^m,
\end{align}
\end{widetext}
%%%%%%%%%%%%%%%%%%%%%%%%%%%%%%%%%%%%%%%%%%%%%%%%%%%%%%%%%%%%%%%%%%%%%%%
Here, in the last line, we have assumed 
the Gaussianity of the density and velocity-divergence fields. 
Then, as we see from the integrand, 
the integral over the mode $\bfp$ becomes vanishing because of 
the symmetry, 
and no contribution from $n=2$ can appear in the correction to 
the high-$k$ limit propagator.

Next focus on the $n=3$ in $f(R)$ gravity, where we 
can get the non-vanishing correction. Substituting Eq.~(\ref{eq:sigma_2}) 
into the expression (\ref{eq:sigma_exp_n}),  we have
%%%%%%%%%%%%%%%%%%%%%%%%%%%%%%%%%%%%%%%%%%%%%%%%%%%%%%%%%%%%%%%%%%%%%%%
\begin{widetext}
\begin{align}
&\Bigg\langle\Delta^{(3)}\exp\left[
\int_{\tau_0}^\tau d\tau'\Xi(\bfk,\tau'')
\right]\Bigg\rangle
\nonumber\\
&\qquad
=-\frac{1}{6}\frac{\overline{R}_{,fff}(\tau')}{\Pi(k;\tau')}
\left(\frac{\kappa^2\,\rhom}{3}\right)^2
\int\frac{d^3\bfp_1 d^3\bfp_2}{(2\pi)^6}
\nonumber\\
&\qquad\qquad\qquad\times
\left\{1-\frac{\overline{R}_{,ff}^2(\tau)}{3\overline{R}_{,fff}(\tau)}
\frac{1}{\Pi(p_{12};\tau)}\right\}
\left\langle
\frac{\delta(\bfp_1;\tau)\delta(\bfp_2;\tau)}{\Pi(p_1;\tau)\Pi(p_2;\tau)}
\exp\left[\int_{\tau_0}^\tau d\tau''\int\frac{d^3\bfq}{(2\pi)^3}
\left(\frac{\bfk\cdot\bfq}{q^2}\right)\theta(\bfq;\tau'')\right]
\right\rangle
\nonumber\\
%&\qquad=
%-\frac{1}{6}\frac{\overline{R}_{,fff}(\tau')}{\Pi(k;\tau')}
%\left(\frac{\kappa^2\,\rhom(\tau')}{3}\right)^2
%\int\frac{d^3\bfp_1 d^3\bfp_2}{(2\pi)^6}
%\left\{1-\frac{\overline{R}_{,ff}^2(\tau')}{3\overline{R}_{,fff}(\tau')}
%\frac{1}{\Pi(p_{12};\tau')}\right\}
%\nonumber\\
%&\qquad\qquad\times \sum_{n=0}\frac{1}{n!}\Bigg[
%\,
%\frac{\langle\delta(\bfp_1;\tau')\delta(\bfp_2;\tau')\rangle}
%{\Pi(p_1;\tau')\Pi(p_2;\tau')} \,
%(n-1)!!\left\{-\int_{\tau_0}^\tau d\tau''_1 d\tau''_2
%\int\frac{d^3\bfq}{(2\pi)^3}\left(\frac{\bfk\cdot\bfq}{q^2}\right)^2
%P_{\theta\theta}(q;\tau''_1,\tau''_2)\right\}^{n/2}
%\nonumber\\
%&\qquad\qquad+\frac{n(n-1)}
%{\Pi(p_1;\tau')\Pi(p_2;\tau')} \,
%%\int_{\tau_0}^\tau d\tau''_1 d\tau''_2 \int\frac{d^3\bfr_1 d^3\bfr_2}{(2\pi)^6} 
%\left(\frac{\bfk\cdot\bfr_1}{r_1^2}\right)
%\left(\frac{\bfk\cdot\bfr_2}{r_2^2}\right)
%\langle\delta(\bfp_1;\tau')\theta(\bfr_1;\tau''_1)\rangle
%\langle\delta(\bfp_2;\tau')\theta(\bfr_2;\tau''_2)\rangle
%\nonumber\\
%&\qquad\qquad\qquad\qquad\qquad\qquad\qquad\qquad\qquad\qquad
%\times(n-3)!!\left\{
%-\int_{\tau_0}^\tau dy_1 dy_2 \int \frac{d^3\bfq}{(2\pi)^3}
%\left(\frac{\bfk\cdot\bfq}{q^2}\right)^2\,P_{\theta\theta}(q;y_1,y_2)
%\right\}^{(n-2)/2}
%\Bigg]
%\nonumber\\
&\qquad=-\frac{1}{6}\frac{\overline{R}_{,fff}(\tau')}{\Pi(k;\tau')}
\left(\frac{\kappa^2\,\rhom(\tau')}{3}\right)^2
\nonumber\\
&\qquad\times\Bigg[
\left\{1-\frac{\overline{R}_{,ff}^2(\tau')}{\overline{R}_{,fff}(\tau')\overline{R}_{,f}(\tau')}\right\}\int\frac{d^3\bfp}{(2\pi)^3}
\frac{P_{\delta\delta}(p;\tau')}{\Pi^2(p;\tau')}
\sum_{n=0}\frac{(n-1)!!}{n!}\left\{-\frac{k^2}{3}\int\frac{dq}{2\pi^2}
\left\{D_+(q;\tau)-D_+(q;\tau_0)\right\}^2P_0(q)
\right\}^{n/2}
\nonumber\\
&\qquad\qquad+\int\frac{d^3\bfp_1 d^3\bfp_2}{(2\pi)^6}
\left\{1-\frac{\overline{R}_{,ff}^2(\tau')}{3\overline{R}_{,fff}(\tau')}
\frac{1}{\Pi(p_{12};\tau')}\right\}
\left(\frac{\bfk\cdot\bfp_1}{p_1^2}\right)
\left(\frac{\bfk\cdot\bfp_2}{p_2^2}\right)
\frac{P_0(p_1)P_0(p_2)}{\Pi(p_1;\tau')\Pi(p_2;\tau')}
\nonumber\\
&\qquad\qquad\qquad\qquad\qquad\times
D_+(p_1;\tau')D_+(p_2;\tau')
\left\{D_+(p_1;\tau)-D_+(p_1;\tau_0)\right\}
\left\{D_+(p_2;\tau)-D_+(p_2;\tau_0)\right\}
\nonumber\\
&\qquad\qquad\qquad\qquad\qquad\qquad\qquad\qquad\qquad
\times\sum_{n=0}\frac{(n-3)!!}{(n-2)!}
\left\{-\frac{k^2}{3}
\int\frac{dq}{2\pi^2}\left\{D_+(q;\tau)-D_+(q;\tau_0)\right\}^2P_0(q)
\right\}^{(n-2)/2}\,\Bigg]
\nonumber\\
&\qquad=-\frac{1}{6}\frac{\overline{R}_{,fff}(\tau')}{\Pi(k;\tau')}
\left(\frac{\kappa^2\,\rhom(\tau')}{3}\right)^2
\nonumber\\
&\qquad\times\Bigg[
\left\{1-\frac{\overline{R}_{,ff}^2(\tau')}{\overline{R}_{,fff}(\tau')\overline{R}_{,f}(\tau')}\right\}\int\frac{dp\,p^2}{2\pi^2}
\frac{D_+^2(p;\tau')P_0(p)}{\Pi^2(p;\tau')}
\nonumber\\
&\quad\qquad\qquad+\int\frac{d^3\bfp_1 d^3\bfp_2}{(2\pi)^6}
\left\{1-\frac{\overline{R}_{,ff}^2(\tau')}{3\overline{R}_{,fff}(\tau')}
\frac{1}{\Pi(p_{12};\tau')}\right\}
\left(\frac{\bfk\cdot\bfp_1}{p_1^2}\right)
\left(\frac{\bfk\cdot\bfp_2}{p_2^2}\right)
\frac{P_0(p_1)P_0(p_2)}{\Pi(p_1;\tau')\Pi(p_2;\tau')}
\nonumber\\
&\qquad\qquad\qquad\qquad\qquad\times
D_+(p_1;\tau')D_+(p_2;\tau')
\left\{D_+(p_1;\tau)-D_+(p_1;\tau_0)\right\}
\left\{D_+(p_2;\tau)-D_+(p_2;\tau_0)\right\}
\Bigg]
\nonumber\\
&\qquad\qquad\qquad\qquad\qquad\qquad\qquad\qquad\qquad\qquad
\qquad\qquad
\times\exp\left[
-\frac{k^2}{3}
\int\frac{dq}{2\pi^2}\left\{D_+(q;\tau)-D_+(q;\tau_0)\right\}^2P_0(q)
\right].
\label{eq:sigma_exp_3_ver2}
\end{align}
\end{widetext}
%%%%%%%%%%%%%%%%%%%%%%%%%%%%%%%%%%%%%%%%%%%%%%%%%%%%%%%%%%%%%%%%%%%%%%%
In the above, there appears 
the integral over the two Fourier modes $\bfp_1$ and $\bfp_2$, a part of 
which can be performed analytically. We then have
%%%%%%%%%%%%%%%%%%%%%%%%%%%%%%%%%%%%%%%%%%%%%%%%%%%%%%%%%%%%%%%%%%%%%%%
\begin{widetext}
\begin{align}
&\int\frac{d^3\bfp_1 d^3\bfp_2}{(2\pi)^6}
\left\{1-\frac{\overline{R}_{,ff}^2(\tau')}{3\overline{R}_{,fff}(\tau')}
\frac{1}{\Pi(p_{12};\tau')}\right\}
\left(\frac{\bfk\cdot\bfp_1}{p_1^2}\right)
\left(\frac{\bfk\cdot\bfp_2}{p_2^2}\right)
\frac{P_0(p_1)P_0(p_2)}{\Pi(p_1;\tau')\Pi(p_2;\tau')}
\nonumber\\
&\qquad\qquad\qquad\qquad\qquad\quad\times
D_+(p_1;\tau')D_+(p_2;\tau')
\left\{D_+(p_1;\tau)-D_+(p_1;\tau_0)\right\}
\left\{D_+(p_2;\tau)-D_+(p_2;\tau_0)\right\}
\nonumber\\
%&=-\frac{\overline{R}_{,ff}^2(\tau')}{3\overline{R}_{,fff}(\tau')}
%\int\frac{dp_1dp_2}{(2\pi^2)^2}
%\frac{(p_1p_2)^2\,P_0(p_1)P_0(p_2)}{\Pi(p_1;\tau')\Pi(p_2;\tau')}
%\nonumber\\
%&\qquad\qquad\times
%D_+(p_1;\tau')D_+(p_2;\tau')\left\{D_+(p_1;\tau)-D_+(p_1;\tau_0)\right\}
%\left\{D_+(p_2;\tau)-D_+(p_2;\tau_0)\right\}
%\nonumber\\
%&\times\frac{k^2}{p_1p_2}\int\frac{d\mu_1 d\mu_2 d\phi_1d\phi_2}{(4\pi)^2}
%\frac{\mu_1\mu_2}{\overline{R}_{,f}(\tau')/3+\left[p_1^2+p_2^2+2p_1p_2
%\left\{\cos(\phi_1-\phi_2)\sqrt{(1-\mu_1^2)(1-\mu_2^2)}+\mu_1\mu_2\right\}
%\right]/a^2(\tau')}
%\nonumber\\
&=-\frac{k^2}{36}\frac{a^2(\tau')\,\overline{R}_{,ff}^2(\tau')}
{\overline{R}_{,fff}(\tau')}\int\frac{dp_1dp_2}{(2\pi^2)^2}
\frac{P_0(p_1)P_0(p_2)}{\Pi(p_1;\tau')\Pi(p_2;\tau')}
\nonumber\\
&\qquad\qquad\times
D_+(p_1;\tau')D_+(p_2;\tau')\left\{D_+(p_1;\tau)-D_+(p_1;\tau_0)\right\}
\left\{D_+(p_2;\tau)-D_+(p_2;\tau_0)\right\}
\nonumber\\
&\qquad\qquad\qquad\qquad\times
\Bigg\{2-\frac{\overline{R}_{,f}(\tau')/3+(p_1^2+p_2^2)/a^2(\tau')}
{2p_1p_2/a^2(\tau')}
\log\left[\frac{\overline{R}_{,f}(\tau')/3+(p_1+p_2)^2/a^2(\tau')}
{\overline{R}_{,f}(\tau')/3+(p_1-p_2)^2/a^2(\tau')}\right]\Bigg\}.
\end{align}
\end{widetext}
%%%%%%%%%%%%%%%%%%%%%%%%%%%%%%%%%%%%%%%%%%%%%%%%%%%%%%%%%%%%%%%%%%%%%%%
With the above 
expression and Eq.~(\ref{eq:sigma_exp_3_ver2}), we now obtain
the leading-order non-vanishing corrections to the propagator, 
and the final form of the correction, $\delta g_{ab}^{(3)}$,  
defined in Eq.~(\ref{eq:g_ab_d_gab}), becomes
%%%%%%%%%%%%%%%%%%%%%%%%%%%%%%%%%%%%%%%%%%%%%%%%%%%%%%%%%%%%%%%%%%%%%%%
\begin{widetext}
\begin{align}
&\delta\,g_{ab}^{(3)}(k;\tau,\tau_0)=\int_{\tau_0}^\tau d\tau' \,
g_{a2}(k;\tau,\tau')g_{1b}(k;\tau',\tau_0)\,
\frac{\kappa^2}{2}\,\frac{\rhom(\tau')}{H^2(\tau')}\,
\frac{1}{3}\frac{(k/a)^2}{\Pi(k;\tau')}\,
\left(-\frac{1}{6}\right)
\frac{\overline{R}_{,fff}(\tau')}{\Pi(k;\tau')}\,
\left(\frac{\kappa^2\,\rhom(\tau')}{3}\right)^2
\nonumber\\
&\qquad\times \left[\,
\left\{1-\frac{\overline{R}^2_{,ff}(\tau')}{\overline{R}_{,fff}(\tau')\overline{R}_{,f}(\tau')}\right\}\int\frac{dp\,p^2}{2\pi^2}\frac{D_+^2(p;\tau')\,P_0(p)}{\Pi^2(p;\tau')}
-\frac{k^2}{36}
\frac{a^2(\tau')\overline{R}_{,ff}^2(\tau')}{\overline{R}_{,fff}(\tau')}
\int\frac{dp_1dp_2}{(2\pi)^2}\frac{P_0(p_1)P_0(p_2)}{\Pi(p_1;\tau')\Pi(p_2;\tau')}\right.
\nonumber\\
&\qquad\times
D_+(p_1;\tau')D_+(p_2;\tau')\left\{D_+(p_1;\tau)-D_+(p_1;\tau_0)\right\}
\left\{D_+(p_2;\tau)-D_+(p_2;\tau_0)\right\}
\nonumber\\
&\qquad\times\left.
\left\{2-\frac{\overline{R}_{,f}(\tau')/3+(p_1^2+p_2^2)/a^2(\tau')}{2p_1p_2/a^2(\tau')}\log\left[\frac{\overline{R}_{,f}(\tau')/3+(p_1+p_2)^2/a^2(\tau')}
{\overline{R}_{,f}(\tau')/3+(p_1-p_2)^2/a^2(\tau')}\right]\right\}\,
\right],
\label{eq:dg^(3)_ab_fR}
\end{align}
%%%%%%%%%%%%%%%%%%%%%%%%%%%%%%%%%%%%%%%%%%%%%%%%%%%%%%%%%%%%%%%%%%%%%%%
in $f(R)$ gravity model.

Similarly, in DGP model, the leading-order 
non-vanishing correction to the propagator is shown to appear from $n=3$ 
of Eq.~(\ref{eq:sigma_exp_n}). We here skip the detail of the derivation, and simply present the final result. Straightforward but lengthy 
calculation leads to 
%%%%%%%%%%%%%%%%%%%%%%%%%%%%%%%%%%%%%%%%%%%%%%%%%%%%%%%%%%%%%%%%%%%%%%%
\begin{align}
&\delta\,g_{ab}^{(3)}(k;\tau,\tau_0)=\int_{\tau_0}^\tau d\tau' \,
g_{a2}(k;\tau,\tau')g_{1b}(k;\tau',\tau_0)\,
\frac{\kappa^2}{2}\,\frac{\rhom(\tau')}{H^2(\tau')}\,
\frac{1}{3}\frac{(k/a)^2}{\Pi(k;\tau')}\,
\cdot
\left(-3\right)\left(\frac{r_c^2}{3\beta(\tau')^2}\right)^2
\left(\frac{\kappa^2\,\rhom(\tau')}{3}\right)^2
\nonumber\\
&\qquad\times
\left[
\frac{32}{45}D_+^2(\tau')\int \frac{dp\,p^2}{2\pi^2}\,P_0(p)\,+\,
\left(D_+(\tau')\left\{D_+(\tau)-D_*(\tau_0)\right\}\right)^2\frac{k^2}{3}
\int\frac{dp_1dp_2}{(2\pi)^2}\,P_0(p_1)P_0(p_2)
\right.
\nonumber\\
&\qquad\qquad\qquad\qquad\qquad\qquad\qquad\times\left.\left\{
-\frac{2}{45}(p_1^2+p_2^2)+\frac{(p_1^2-p_2^2)(p_1^2+p_2^2)}{60(p_1p_2)^2}
-\frac{(p_1^2-p_2^2)^4}{120(p_1p_2)^3}\,\log\left|\frac{p_1+p_2}{p_1-p_2}\right|
\right\}
\right].
\label{eq:dg^(3)_ab_DGP}
\end{align}
\end{widetext}
%%%%%%%%%%%%%%%%%%%%%%%%%%%%%%%%%%%%%%%%%%%%%%%%%%%%%%%%%%%%%%%%%%%%%%%

%%%%%%%%%%%%%%%%%%%%%%%%%%%%%%%%%%%%%%%%%%%%%%%%%%%%%%%
%\bibliography{ref}
\bibliographystyle{apsrev}

%%%%%%%%%%%%%%%%%%%%%%%%%%%%%%%%%%%%%%%%%%%%%%%%%%%%%%%

\end{document}